\documentclass[sigconf]{acmart}
\usepackage[acronym]{glossaries}
\usepackage{graphicx}
\usepackage{subcaption}
\usepackage{hyperref}
\usepackage{enumitem}
\usepackage{xspace}
\usepackage[utf8]{inputenc}
\makeglossaries
\AtBeginDocument{%
  }

\copyrightyear{2026}
\acmYear{2026}
\setcopyright{cc}
\setcctype{by}
\acmConference[CHI '26]{Proceedings of the 2026 CHI Conference on Human Factors in Computing Systems}{April 13--17, 2026}{Barcelona, Spain}
\acmBooktitle{Proceedings of the 2026 CHI Conference on Human Factors in Computing Systems (CHI '26), April 13--17, 2026, Barcelona, Spain}
\acmPrice{}
\acmDOI{10.1145/3772318.3790773}
\acmISBN{979-8-4007-2278-3/2026/04}

\begin{document}

\title{\name: Biomechanical Motion Synthesis from Touch Logs}

\author{Michał Patryk Miazga}
\email{miazga@uni-leipzig.de}
\orcid{0009-0003-5579-3036}
\affiliation{%
  \institution{ScaDS.AI, Leipzig University}
  \city{Leipzig}
  \country{Germany}}

\author{Hannah Bussmann}
\email{hannah.bussmann@outlook.de}
\orcid{0009-0005-4698-6009}
\affiliation{%
  \institution{ScaDS.AI, Leipzig University}
  \city{Leipzig}
  \country{Germany}}

\author{Antti Oulasvirta}
\email{antti.oulasvirta@aalto.fi}
\orcid{0000-0002-2498-7837}
\affiliation{%
	\institution{Aalto University
    \& ELLIS Institute Finland}
	\city{Helsinki}
	\country{Finland}}

\author{Patrick Ebel}
\email{ebel@uni-leipzig.de}
\orcid{0000-0002-4437-2821}
\affiliation{%
  \institution{ScaDS.AI, Leipzig University}
  \city{Leipzig}
  \country{Germany}}

\renewcommand{\shortauthors}{Miazga et al.}

\newcommand{\name}{\textsc{Log2Motion}\xspace}
\newcommand{\fix}[1]{\textcolor{red}{ #1}}


\newacronym{AitW}{AitW}{Android in the Wild}
\newacronym{CR}{CR}{Computational Rationality}
\newacronym{HCI}{HCI}{Human-Computer Interaction}
\newacronym{DoF}{DoF}{Degree of Freedom}
\newacronym{DTW}{DTW}{Dynamic Time Warping}
\newacronym{GOMS}{GOMS}{Goals, Operators, Methods, and Selection rules}
\newacronym{KLM}{KLM}{Keystroke-Level Model}
\newacronym{MDP}{MDP}{Markov Decision Process}
\newacronym{POMDP}{POMDP}{Partially Observable Markov Decision Process}
\newacronym{RL}{RL}{Reinforcement Learning}
\newacronym{UCD}{UCD}{User-Centered Design}
\newacronym{UI}{UI}{User Interface}
\newacronym{UX}{UX}{User Experience}
\newacronym{UXD}{UXD}{User Experience Design}
\newacronym{XAI}{XAI}{Explainable AI}
\newacronym{OS}{OS}{Operating System}
\newacronym{PPO}{PPO}{Proximal Policy Optimization}
\newacronym{VR}{VR}{Virtual Reality}

\begin{abstract}
Touch data from mobile devices are collected at scale but reveal little about the interactions that produce them.
While biomechanical simulations can illuminate motor control processes, they have not yet been developed for touch interactions.
To close this gap, we propose a novel computational problem: synthesizing plausible motion directly from logs.
Our key insight is a reinforcement learning-driven musculoskeletal forward simulation that generates biomechanically plausible motion sequences consistent with events recorded in touch logs.
We achieve this by integrating a software emulator into a physics simulator, allowing biomechanical models to manipulate real applications in real-time.
\name produces rich syntheses of user movements from touch logs, including estimates of motion, speed, accuracy, and effort.
We assess the plausibility of generated movements by comparing against human data from a motion capture study and prior findings, and demonstrate \name in a large-scale dataset. 
Biomechanical motion synthesis provides a new way to understand log data, illuminating the ergonomics and motor control underlying touch interactions.

\end{abstract}

\ccsdesc[500]{Computing methodologies~Reinforcement learning}
\ccsdesc[500]{Human-centered computing~Mobile devices}
\ccsdesc[500]{Human-centered computing~User models}

\keywords{Biomechanical Models, User Modeling,  Computational Interaction, Reinforcement Learning, Usage Logs, Android, Motion Synthesis}

\begin{teaserfigure}
\centering
  \includegraphics[width=\textwidth]{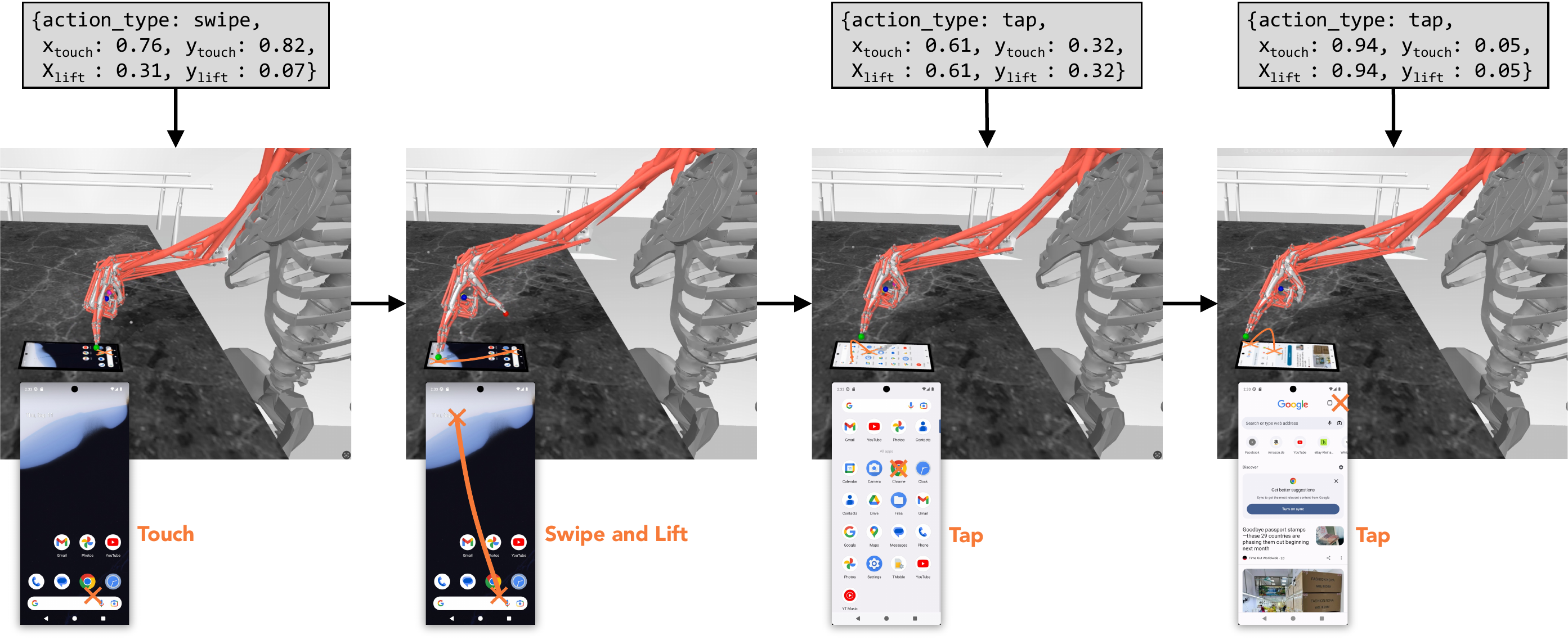}
  \caption{\name bridges interaction logs and the movements that generate them. By integrating a software emulator with a musculoskeletal forward simulation, \name synthesizes plausible human-like interaction motions from logs, enabling the estimation of performance measures such as speed, accuracy, and effort.}
  \label{fig:teaser}
  \Description{A diagram illustrating Log2Motion. On the left, touch events such as swipe and tap are shown as abstract log entries with x–y coordinates for touch and lift. On the right, these logs are linked to synthesized human-like motions generated by a physics-based musculoskeletal simulator. Together, the figure shows how interaction logs can be transformed into plausible movement trajectories for estimating speed, accuracy, and effort.}
\end{teaserfigure}
\maketitle

\section{Introduction}
Despite the vast amounts of user interaction data collected by computer applications, much of it remains underutilized from a \gls{HCI} perspective.
Many organizations produce terabytes or more of log data daily~\cite{leith2021mobile, schmidt_google_2018}.
Yet, despite their scale, these logs provide only a limited view of the interactions that generated them, as they only consist of selections made by the user.
Many approaches have been proposed that analyze~\cite{henze_100000000_2011, dhakal_observations_2018} and visualize such log data~\cite{ebel_exploring_2023, wongsuphasawat_exploring_2012, wongsuphasawat_lifeflow_2011} or use it to train user models for UI evaluation~\cite{tian_what_2022, fang_general-purpose_2024}.
Yet from an \gls{HCI} perspective, it is \emph{not} the tap or swipe events themselves, but rather the \emph{processes} that generated them, that is of primary interest.
%
%
What if we could infer user movements from raw touch logs? Synthesized motion data could enrich real-world logs, for example, revealing that frequent errors arise from excessive finger stretches caused by wide spacing rather than poor labeling. Similarly, designers could estimate key ergonomic and performance metrics such as effort and error probabilities using only a UI prototype that produces logs without requiring expensive user studies.

Previous work on this topic has predominantly focused on the inference of interaction-related performance metrics such as task time, typing speed, or touch accuracy~\cite{weir_user-specific_2012, henze_100000000_2011, mott_cluster_2019}.
However, these approaches can only predict what is already in the logs without access to the generating processes. 
In other words, they do not shed light on how the interaction itself unfolded. 

In accordance with prior work on motion synthesis~\cite{karunratanakul2023guided, petrovich2021action, jiang_synthesis_2019}, we introduce \textit{biomechanical motion synthesis} as a new computational problem in HCI. It describes the generation of plausible human movements from logs using forward biomechanical modeling. The key idea is to synthesize the information in the logs with prior physiological knowledge encoded in a biomechanical model. These physiological constraints meaningfully restrict the space of motion and enable the model to generate muscle activations and movement trajectories that could plausibly have produced the interaction traces in the log.

Biomechanical motion synthesis can effectively complement the insights generated with traditional data-driven methods~\cite{henze_100000000_2011, atterer_knowing_2006}.
By capturing key anatomical and physiological aspects of the human motor system, biomechanical models serve as strong priors for inference. 
%
%
Through forward simulation, these models generate biomechanically realistic motion, enabling accurate predictions of user performance and muscle effort based on musculoskeletal dynamics~\cite{fischer_reinforcement_2021, ikkala_breathing_2022}.
Thus, synthesizing interaction logs with the prior knowledge encoded in the biomechanical model unlocks new design opportunities, offering deeper insights into ergonomics, usability, and user experience~\cite{bachynskyi_is_2014, fischer_sim2vr_2024, ikkala_breathing_2022}.

However, one critical problem remains unsolved.
While there are promising \gls{RL}-based approaches for \emph{forward} biomechanical simulation in HCI, they are not (yet) suitable for synthesizing motion from interaction logs. 
Current methods are typically trained for specific tasks (e.g., pointing), whereas real-world logs reflect a wide range of user behaviors across diverse applications. Furthermore, they have primarily focused on coarse movements of the arm~\cite{fischer_reinforcement_2021, ikkala_breathing_2022}, which limits their relevance for logs produced on mobile devices. 
Synthesizing touch interactions requires accurate modeling of fine motor control in the arm, hand, and fingers.
In addition, current \gls{RL}-based approaches typically operate within isolated physics engines and remain disconnected from real applications.  
To make predictions that are relevant to real-world application design, applications must be meaningfully linked to the biomechanical models that live within physics-based simulations.

To address these gaps, we present \name, a novel method for biomechanical motion synthesis from touch logs using musculoskeletal forward simulation.
\name supports logs containing touch events that can be generated via a software emulator. The logs can stem from real-world mobile interactions or expert demonstrations in UI design tools.
Specifically, \name takes abstract input events from log files (see \autoref{fig:teaser}) and generates plausible motion by learning the underlying muscle activations. From this, we compute (1) movement trajectories, (2) performance measures, and (3) muscle effort.
The method is general and can be applied to any touchscreen scenario involving basic input interactions such as tapping and swiping. In its current form, \name supports index-finger input on a mobile device, whether it is placed on a surface (see \autoref{fig:teaser}) or held in the hand. However, the motor control policies trained in this work can be extended to support other types of interactions.
To this end, \name integrates, for the first time, real mobile applications (specifically, the Android emulator) with a physics simulator that runs a musculoskeletal model~\cite{saul_benchmarking_2015}, whose muscle activation patterns have been experimentally validated for HCI-relevant pointing tasks~\cite{bachynskyi_is_2014}.
To enable the biomechanical model to interact with emulated applications, we introduce a new control policy and training procedure that learn movement policies that accurately reproduce the logged touch events.
This makes it possible to translate usage logs into realistic, dexterous movements within a physics-based environment.

We evaluate \name across multiple dimensions, demonstrating that the simulated movements are biomechanically plausible and human-like. Our analysis reveals close alignment with Fitts’ Law, consistency with human performance data, realistic velocity dynamics, and sensitivity to individual differences in speed–accuracy trade-offs, reflected in variations in error probabilities and muscle effort.
In-depth comparisons with motion capture data show that the synthesized movement trajectories not only qualitatively exhibit key human-like features but also quantitatively fall well within the range of variability observed between human participants.
To demonstrate the generalizability of \name's capabilities to real-world use cases, we use it to synthesize interaction sequences from the \gls{AitW} dataset~\cite{rawles_androidinthewild_2023} and compare them in terms of predicted error probabilities, muscle effort, and movement times. 
Finally, we discuss how biomechanical motion synthesis and \name, in particular, can aid UI design and accessibility evaluation.

In summary, our main contribution is a method for synthesizing user movements from log data using biomechanical forward simulation. 
In solving this, we make three technical contributions:
\begin{itemize}

\item We formulate a motor control problem to simulate dexterous touch interactions. Specifically, we propose a \gls{POMDP} that models finger movements during both discrete (tap) and continuous (swipe) touchscreen interactions. We further extend this formulation to handle continuous movement sequences involving multiple, successive interactions.

\item We design a reward function that aligns simulated movements with logged interaction data and human-like movement patterns, enabling the learning of control policies that accurately synthesize user movements.

\item We integrate a real-time software emulator into a physics simulator that runs a biomechanical forward simulation of the human arm, enabling biomechanical models to directly manipulate real-world applications.

\end{itemize}
\section{Related Work}
HCI research offers a rich set of computational approaches for understanding user behavior. We review key methods through the lens of the central problem addressed in this work: synthesizing biomechanically plausible user motion from logs.

\subsection{Log-Based Analysis of User Interaction Patterns}

Usage logs are a standard data source for large-scale analysis of interaction behavior. In both industry and academia, data science methods have been applied to infer performance metrics~\cite{atterer_knowing_2006, agichtein_improving_2006}, visualize interaction patterns~\cite{liu_patterns_2017, wongsuphasawat_lifeflow_2011, ebel_exploring_2023}, and train machine learning models that predict usability outcomes~\cite{pfeuffer_analysis_2018, ebel_forces_2023}.
However, while effective for large-scale analysis, logs only reveal \emph{what} users did, not \emph{how} they physically carried out interactions or the constraints that shaped them. 
\name can complement this body of work by enabling the inference of physical motion that, given the same usage logs, can be uniquely derived from biomechanical simulation.

\subsection{Forward Models of Task Performance}

\gls{HCI} has a long history of forward models that predict user behavior and performance in compound tasks.
Cognitive models, such as \gls{GOMS}~\cite{john_goms_1996} and the \gls{KLM}~\cite{card_keystroke-level_1980}, deconstruct interactions into a series of goal-directed, rule-based operations. \gls{KLM}, as a simplification of \gls{GOMS}, estimates task execution time as a sum of time spent in lower-level motor and cognitive operators.
\gls{GOMS} adds to this an account of the mediating processes and states in cognition. 
While these methods are not directly suitable for our problem, since they require an expert to define the sequence and do not model physical motion in space, \name builds on the principle of deconstructing interactions in sequential operators.

More recently, RL has been proposed as a way to overcome this problem, with emerging applications across HCI in tasks such as typing~\cite{shi_simulating_2025}, chart reading~\cite{shi_chartist_2025}, and driving~\cite{jokinen_predicting_2025}. 
RL can generate action sequences under the assumption that user behavior is boundedly optimal given internal (cognitive, perceptual, motor) and external (environmental, interface) constraints~\cite{lewis_computational_2014, gershman_computational_2015, oulasvirta_computational_2022}. 
However, these models do not account for the biomechanics that underpin motion. 

\subsection{Biomechanical Forward Simulation in HCI}

In biomechanical forward simulation, a skeletal or musculoskeletal model is actively controlled to generate realistic human motion. While such models have been widely applied in gait analysis~\cite{lee_scalable_2019}, rehabilitation~\cite{van_den_bogert_real-time_2013}, sports science, and ergonomics~\cite{si_realistic_2014}, only recently have applications, for example, evaluating comfort and fatigue in \gls{VR} settings~\cite{evangelista_belo_xrgonomics_2021, jang_modeling_2017}, demonstrated their value for HCI.
Traditionally, these approaches relied on inverse simulation, where motion capture data served as input, and the model worked ``backwards'' to estimate the underlying muscle forces, joint torques, and neural control signals. These inverse models can predict loads and muscle activations for a given trajectory, but cannot generate movement or adapt to new tasks and environments~\cite{bachynskyi_is_2014}. For simulating user interactions, researchers have therefore either handcrafted movement sequences or fitted motion data to the models~\cite{saul_benchmarking_2015, pennestri_virtual_2007}.

Recent work has started to close this gap by introducing RL-based approaches that learn to produce interaction behavior through forward simulation. \citet{ikkala_breathing_2022} presented the \textsc{UitB} framework, which learns muscle control policies based on perceptual feedback in interaction tasks. Their approach successfully learned different interaction methods and reproduced movement characteristics observed in empirical studies of pointing.
To connect the simulation with real applications, \textsc{UitB} was later extended in \textsc{sim2vr}~\cite{fischer_sim2vr_2024}, integrating the framework with VR environments to predict behavior and fatigue in immersive interaction scenarios.
Similarly, deep RL combined with physics-based musculoskeletal models has been used for tasks such as pointing~\cite{fischer_reinforcement_2021}, object manipulation~\cite{berg_sar_2023}, mid-air typing~\cite{hetzel_complex_2021}, or even varying tasks~\cite{chiappa_arnold_2025}. However, most of these approaches did not go beyond relatively coarse mid-air pointing movements.
These works demonstrate that learning muscle-actuated control policies that yield human-like movement is feasible, laying the foundation for using biomechanical forward simulation to model dexterous touch interactions.

\subsection{Summary and Objectives}

To sum up, three problems need to be identified to utilize biomechanical models for motion synthesis.
\textbf{First}, the models need to be able to handle a wide range of sequential actions as expressed in logs. 
In practice, this means taking an approach similar to \gls{KLM} and \gls{GOMS}, where we identify a minimum number of actions that cover the maximum proportion of log events.
\textbf{Second}, to account for touch logs, modeling human dexterity and prehension in hand movements is required~\cite{karakostis_biomechanics_2021}. The challenge lies in the complexity of the overactuated musculoskeletal system, where approximately 600 muscles control more than 250 joints through overlapping and multi-articular connections~\cite{tortora_principles_2018}. While this redundancy improves movement efficiency, it significantly complicates motor control, especially in contact-rich tasks, such as tapping and swiping gestures, where forces must be precisely coordinated~\cite{caggiano_myochallenge_2023}. 
%
\textbf{Third}, existing models operate in isolated physics engines, remaining disconnected from real-world applications. To run simulations, experts need to handcraft the task environment, which is time-consuming and limits the potential of biomechanical forward simulations to actively shape real-world design processes. We build on the idea of \textsc{Sim2VR}~\cite{fischer_sim2vr_2024} to connect applications and physics simulators. 
However, instead of embedding the user within a custom Unity environment, we integrate an OS emulator with the physics simulator and mirror the application state in the same environment that runs the biomechanical model. This eliminates the need for manual scripting or source code modifications that were necessary in systems like \textsc{Sim2VR}.

\section{Method Overview}\label{sec:method}

\begin{figure*}
    \centering
    \includegraphics[width=\textwidth]{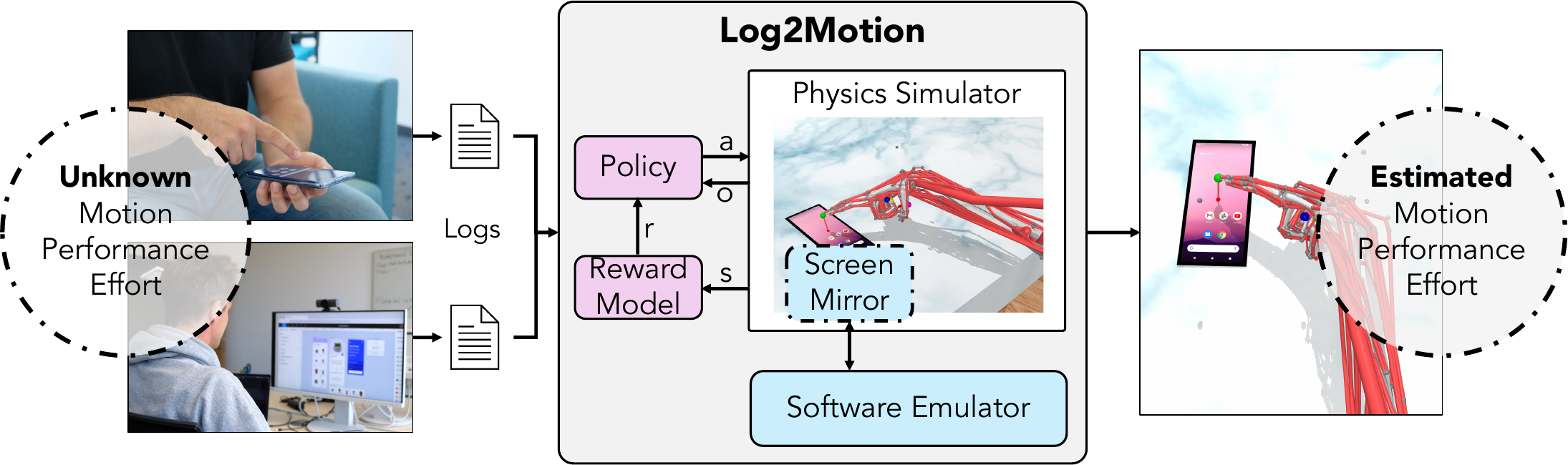}
    \caption{In motion synthesis, a biomechanical body model generates motions that are aligned with sequences of events recorded in a log file. These log files can stem from user interactions or expert demonstrations. \name introduces three key innovations: (1) a problem formulation that defines motion synthesis as a POMDP, (2) an integrated physics simulator that captures both biomechanical and display dynamics, and (3) a reward model that supports learning plausible movement policies for dexterous touch interactions using \gls{RL}.}
    \label{fig:architecture}
    \Description{A schematic sketch of motion synthesis. Log files from user or expert interactions feed into a software emulator and physics simulator, which combine with a reward model to estimate human-like motions. The system produces synthesized trajectories along with performance and effort measures.}
\end{figure*}

We define the problem of biomechanical motion synthesis as learning the muscle activations required to control a biomechanical model so that it generates movements aligned with the interactions recorded in a log file. These synthesized movements must not only match the event sequence (e.g., taps or swipes) but also be biomechanically plausible.
A key practical challenge in achieving this is that musculoskeletal forward simulations are typically isolated from the software environments in which real interactions occur.
As shown in \autoref{fig:architecture}, \name addresses this by coupling a software emulator with a physics-based biomechanical simulation. This connection is facilitated by a Screen Mirror mechanism that renders the emulator’s display state onto a virtual touchscreen and routes touch events back to the emulator. This high-level architecture provides the context for the individual components we describe below.


\name takes as input a log file containing a sequence of touch events from real interactions or expert demonstrations. It integrates the Android Emulator with the MuJoCo physics engine to run biomechanical forward simulations, producing the movement trajectories and the muscle activation patterns that generate them. From these activations, \name estimates motion, effort, and performance. Furthermore, because \name models motor noise, repeated simulations enable statistical estimation of the probability of missing a target and thus the probability of a user making an error.
\name supports any application that responds deterministically to touch events, ensuring reproducible display states across simulations. The current implementation handles tapping and swiping, and additional touch operations can be incorporated following the same principles.

\name presents two critical innovations to solving the motion synthesis problem: 1) an integrated physics simulator with both biomechanics and display dynamics, 2) a problem formulation, including a reward model, that affords learning plausible movement policies for dexterous interaction using \gls{RL}. Although \name focuses on mobile interaction, the approach can, in principle, be applied to any setting where the key elements that we introduce in the following are supported.

\subsection{Usage Logs}

\name is agnostic to the source of the usage logs it receives. In this sense, any usage log that specifies touch events, including their x-y coordinates, suffices.  
We envision these logs stemming mainly from two main sources (see \autoref{fig:architecture}). The first captures actual user input on physical devices in real-world usage. The second source is synthetic logs from UX/UI design tools, which record anticipated user behavior demonstrated by a designer. 
%
In this paper, we demonstrate \name by using logs from the \gls{AitW} dataset~\cite{rawles_androidinthewild_2023}, which contains 715k episodes of human demonstrations of device interactions performed using the Android emulator.

\subsection{Software Emulator}

\name requires a software emulator that can reproduce the display state at each stage of a user interaction represented in a log.
%
%
The emulator must provide access to critical information, such as the coordinates of the touch interaction and the UI element that the model interacted with. This information is necessary to compute error rates (i.e., the model tapping a button different from what was specified in the logs).
Moreover, to enable the training of an RL policy, the emulator needs to respond deterministically to touch events.
For \name, we use the Android Emulator due to its strong developer support and wide adoption. It can run either on a physical mobile device or as a virtual device within the Android Studio environment.

\subsection{Physics Simulator}

For motion synthesis, \name requires a physics simulator that can simultaneously run the Screen Mirror (see below) and the biomechanical model of the user. The simulator must allow headless rendering for training the RL policy. Furthermore, it should be able to manipulate the simulation environment to model different phone sizes and surroundings.
Specifically, \name uses \textit{MuJoCo}~\cite{todorov_mujoco_2012}, a free and open-source physics engine designed for model-based optimization. We selected MuJoCo for its widespread adoption in the robotics research community, its support for biomechanics, and its compatibility with the Gymnasium API for policy learning~\cite{towers_gymnasium_2024}.

\subsection{Biomechanical Body Model}

To synthesize plausible movements, \name relies on a realistic musculoskeletal forward simulation of the human body. The biomechanical model must therefore represent the key parts of the kinematic chain involved in the targeted interactions. It should also allow for free parameters to capture individual differences, such as motor control noise and movement ranges, and be controllable from outside the simulator. This is essential for learning plausible movement policies.

To account for dexterous mobile interactions, \name builds on and extends MyoSuite~\cite{caggiano_myosuite_2022} by extending existing environments and biomechanical models. MyoSuite is a collection of musculoskeletal environments that builds upon validated models from OpenSim~\cite{delp_opensim_2007} and is simulated using the MuJoCo physics engine. MyoSuite integrates with the Gymnasium API~\cite{towers_gymnasium_2024}, which simplifies training \gls{RL} agents to learn the muscle control policies~\cite{caggiano_myosuite_2022}.
Specifically, \name is based on the \textit{myoArm} model~\cite{caggiano_myochallenge_2024-1}, which comprises 63 muscle–tendon units that control arm and finger movements, thereby enabling interaction with the simulated touchscreen device.
To account for noise in the human nervous system~\cite{faisal_noise_2008, harris_signal-dependent_1998}, we add a model of motor noise consisting of two components: a signal-dependent noise term that scales with the magnitude of muscle activation and a constant noise term that represents baseline variability~\cite{van_beers_role_2004}.
Integrating this noise model reflects the empirical observation that repeated human movements between the same points yield slightly different trajectories each time.
We select the index finger as the input finger, as it provides the greatest comfort and results in a low error rate~\cite{colley_exploring_2014}. 
Constraining interactions to the index finger also yields computational benefits in subsequent training processes. However, \name can easily be adopted to account for any other posture (see \autoref{ch:postures}).
To support dexterous interactions, we modified the myoArm environment to operate under \textit{relative muscle control}. Instead of requiring absolute muscle activation values at each time step, the environment interprets inputs as changes relative to the previous activation state. Each muscle’s activation is updated proportionally---either increasing, decreasing, or remaining unchanged---with relative changes clipped to a bounded range of $[-1, 1]$. By adopting relative rather than absolute control, the environment more closely reflects the dynamics of biological motor systems, produces smoother activation patterns, and eases subsequent policy learning~\cite{ikkala_breathing_2022}.

\subsection{Screen Mirror}
%
To synthesize user motion from logs and predict movement, performance, and effort, software emulators, whether running design prototypes or production applications, must be integrated directly with physics-based simulation environments. This integration enables biomechanical models to interact with real-world applications seamlessly, without requiring handcrafted environments or additional conversions.
For the biomechanical model to manipulate an application, three components are required: (1) a 3D asset of the touchscreen device being replicated, (2) bi-directional real-time communication between the virtual screen and the emulator, and (3) a Screen Mirror that renders the application running in the emulator on a virtual screen in the physics simulation.


In our implementation of \name, we extend the method presented by \citet{toyama_androidenv_2021}, which enables bidirectional communication by sending and receiving information to and from the Android device. A 3D model of a mobile device is positioned on a virtual table surface in front of the agent in the physics simulation. The smartphone surface ($66\,\mathrm{mm} \times 148\,\mathrm{mm}$) acts as a contact sensor, capturing touch input with high fidelity. Inspired by real-world capacitive touchscreens, our simulation replicates the grid sensors by detecting contact with the agent’s index finger. We chose a grid spacing of 1.25\,mm, which reflects super-resolution displays and exceeds the typical 3~mm to 5~mm spacing found in most modern devices~\cite {mayer_super-resolution_2021, streli_capcontact_2021}.
An interaction is registered when the index finger touches the target location. Our implementation allows us to log, for each interaction, the touch-down event (initial contact), touch-up event (release of contact), and, accordingly, the total touch duration. In addition to discrete touch events, we implemented support for directional gestures, specifically swiping. A swipe is valid only when the index finger maintains continuous contact with the touchscreen from the starting position to the target position. For each detected interaction, the Screen Mirror sends a signal to the Software Emulator indicating that a touch event has occurred at the corresponding coordinates (x, y) on the screen.

The communication between the Screen Mirror and the Software Emulator is performed in real-time. Registered touch events are transmitted to the Software Emulator, which processes the input and generates the corresponding system-level response. The resulting output frame, representing the visual feedback of the interface, is rendered onto the simulated phone display surface within our environment. The rendering resolution is defined by the Software Emulator, which outputs image frames that reflect the current state of the user interface. These frames are then projected onto the surface of the virtual device. We synchronize the display rendering with the simulation loop to ensure temporal consistency between agent actions and perceptual feedback, with the display refreshing at 10 frames per second by default. 
Within the Screen Mirror, we achieve real-time rendering by modifying and extending the rendering capabilities of both MyoSuite and MuJoCo. Specifically, we define a dedicated display region that precisely matches the dimensions of the target device and explicitly reserve this space within the texture memory in the MuJoCo scene creation file. We then overwrite the associated texture in memory with the latest frame captured from the Software Emulator and enforce a texture refresh to render the updated visual content as a simulation of a mobile display. This feature allows us to modify and refresh textures in real time, providing a flexible framework that smoothly integrates and displays any application running in a software emulator within the MuJoCo physics simulator.





\section{Policy Learning for Biomechanical Motion Synthesis}\label{sec:policylearning}








Learning control policies for biomechanical motion synthesis of arbitrary sequence length is a high-dimensional, multi-objective optimization problem.
Agents must learn to control the myoArm model, which has 27 degrees of freedom, to generate highly accurate interactions that align with log data and reproduce human-like kinematics while accounting for motor noise.
Previous work has largely avoided these challenges by focusing on short-horizon, hand-crafted tasks with limited fidelity. In contrast, our goal is to develop generalizable methods that can be applied to arbitrary sequences of touch interactions.
In solving these challenges, we make two novel contributions to policy learning for biomechanical simulations. First, our reward design and training routines enable accurate and fine-grained movements that have not yet been achieved but are essential to model touch interactions. Second, we decompose continuous movements into motor operators, which enables \name to synthesize arbitrarily long touch sequences accounting for usage logs of any length.

In the following, we introduce the \gls{POMDP} formulation, the decomposition of motion sequences into motor operators, and the reward design and training methods used for policy learning.

\subsection{Problem Formulation} \label{sec:pomdp_formulation}

We define touch interactions on mobile devices as a sequential decision-making process \cite{spaan_partially_2012}, where the agent controls muscle activations to produce coordinated finger and arm movements on the screen. 
Accordingly, we aim to generate finger and arm movements that not only align with the human interactions recorded in usage logs but also adhere to the biomechanical constraints of the human motor system. This requires a validated biomechanical user model to perform gestures and make contact at locations consistent with those recorded during real user interactions.

To address this decision-making problem, we formulate it as a bounded optimality problem within a \gls{POMDP} framework. This formulation allows us to approximate optimal policies representing muscle activation patterns that generate the desired movements through \gls{RL}.
In our formulation, the agent receives partial observations of its internal state and the target and selects muscle activation patterns to move the arm and index finger toward the target. Given an action \( a \in \mathcal{A} \) and the current state \( s \in \mathcal{S} \), the environment transitions to a new state \( s' \in \mathcal{S} \). The agent then receives a reward \( r \in \mathbb{R} \) and an observation \( o \in \mathcal{O} \), which guides its decision-making process. The discount factor $\gamma \in [0, 1)$ determines how much future rewards are prioritized over immediate ones.

The \gls{POMDP} for the muscle control task is as follows:
\begin{itemize}
    \item \textbf{State space} ($\mathcal{S}$): A state $s \in \mathcal{S}$ includes the joint angles ($\boldsymbol{\theta}$), joint velocities ($\dot{\boldsymbol{\theta}}$), and muscle activation levels ($\mathbf{a}_{\text{muscle}}$) of the myoArm model. This represents the internal state of the agent necessary for determining the system dynamics.

    \item \textbf{Observation space \( \mathcal{O} \)}: We design the observation space such that it mimics what information a user might have access to. Accordingly, the agent receives only a subset of the full state space. Specifically, the agent receives proprioceptive, tactile, and visual feedback:
    \begin{itemize}
        \item \textbf{$\mathbf{a}_{\text{muscle}}$:} Muscle activation values.
        \item \textbf{$\mathbf{p}_{\text{finger}}$:} The 3D position of the index finger tip.
        \item \textbf{$\mathbf{v}_{\text{finger}}$:} The velocity of the index finger tip.
        \item \textbf{$\mathbf{p}_{\text{target}}$:} The 3D position of the button (target).
        \item \textbf{$d_{\text{finger-target}}$:} The Euclidean distance between the tip of the index finger and the target location.
        \item \textbf{$w_{\text{target}}$:} Width of the target button.
        \item \textbf{$c_{\text{state}}$:} Contact information indicating whether the index finger tip is in contact with the button, in contact with another object, or not in contact with any object.
        \item \textbf{$f_{\text{muscle}}$:} Current muscle fatigue level across all muscles.
        \item \textbf{$m_{\text{operator}}$:} Type of motor operator (see \autoref{sec:low_level_control_policies}) performed, e.g., touch, swipe.
    \end{itemize}
   \item \textbf{Action space \( \mathcal{A} \)}: The agent’s action is a 63-dimensional continuous vector, where each element represents the activation level of a specific muscle. 
   \item \textbf{Reward function \( R(s, a) \)}:  The reward function provides feedback from the environment and is designed to guide the agent toward accurate, efficient, and dexterous behavior. 

\end{itemize}

\subsection{Reward Design}\label{sec:reward_design}

Our goal is to generate input interactions that are smooth, ergonomic, and comply with empirical findings of human motor behavior. To this end, we design a weighted reward function composed of multiple components, each affecting a specific aspect of the agent's behavior.
This modular structure enables the reward to balance multiple objectives, including motion efficiency, movement speed, and tap accuracy. Inspired by prior work in RL for musculoskeletal forward simulation and reward design~\cite{selder_what_2025, ikkala_breathing_2022, eschmann_reward_2021, grzes_reward_2017}, we select and tune each reward component.
The total reward at each timestep is computed as a weighted sum of the individual components, each encouraging a distinct dimension of the desired behavior:
    \begin{itemize}
        \item \textbf{Task success  \( r_{\text{success}} \):} A positive reward that is given if the agent successfully touches the correct target with the index finger.
        
        \item \textbf{Distance penalty \( r_{\text{dist}} \):} A continuous penalty proportional to the distance between the index finger and the center of the target button. This term encourages directed movements toward the target.
        
        \item \textbf{Muscle activation penalty \( r_{\text{activation}} \):} A muscle activation cost penalizes excessive energy expenditure leading to efficient movements~\cite{cavanagh_efficiency_1985}.
        
        \item \textbf{Error penalty \( r_{\text{error}} \):} A penalty that is given when the agent touches the screen anywhere except the target, discouraging imprecise pointing behavior.
        
        \item \textbf{Jerk penalty \( r_{\text{jerk}} \):}  A penalty for jerky movements~\cite{quinn_modeling_2018}, defined as the rate of change of acceleration over time, capturing abrupt changes in motion. This term promotes smooth movement as seen in humans~\cite{mehrotra_jerk_1997, wann_relation_1988, viviani_minimum-jerk_1995}.
        
        \item \textbf{Failure termination \( r_{\text{termination}} \):} A penalty that is applied if the agent fails to touch the target within the specified time limit. In this case, the episode ends, and the agent receives a penalty equal in magnitude to the success reward.
    \[
    r_{\text{timeout}} = -R_{\text{success}}
    \]
    \end{itemize}

\subsection{Reward with Terminal Constraints} \label{sec:reward_terminal_constrains}
        
Building on the reward design outlined in the previous section, we formulate the reward function to constrain the cumulative return such that intermediate shaping rewards or penalties never dominate the terminal outcome. At the final timestep \( T \), the agent receives a fixed terminal reward defined as:

\[
R_T =
\begin{cases}
+1000, & \text{if the task is successfully completed}, \\
-1000, & \text{if the episode ends in failure or timeout}.
\end{cases}
\]

By anchoring the overall return to this terminal reward, the agent is guided to focus on completing the task correctly and establishes the maximum and minimum bounds of the cumulative return, constraining the value function \( V^\pi(s) \in [-1000, +1000] \). Meanwhile, intermediate shaping rewards \( r_t \), such as distance, efficiency, or smoothness, serve only to accelerate learning and refine behavior without altering the final goal.












\paragraph{Intermediate Reward}
We scale and normalize each intermediate reward, shaping the agent's actions consistently across objectives. By normalizing and scaling each intermediate reward, we ensure that no single component dominates the others, for example, preventing a distance penalty from overshadowing a muscle activation penalty.
To implement the main reward function, we define each component with its associated weight or fixed value, as well as the valid range of values:  
\begin{itemize}
    \item \textbf{$r_{\text{dist}}$:}  \quad Weight: $w_{\text{dist}} =1$ \quad Range: $[-1, 0]$  
    \item \textbf{$r_{\text{activation}}$:}  \quad Weight: $w_{\text{activation}} = 10$ \quad Range: $[-1, 0]$  
    \item \textbf{$r_{\text{error}}$:}  \quad Weight: $w_{\text{error}} = 10 $  \quad Discrete: $-1$ or $0$ 
    \item \textbf{$r_{\text{jerk}}$:}  \quad Weight: $w_{\text{jerk}} = 1$ \quad Range: $[-1, 0]$  

\end{itemize}

This formulation prevents the agent from exploiting intermediate rewards at the expense of task completion. Since the terminal reward always dominates the return, the agent is encouraged to complete the task efficiently and avoid failures decisively.
Compared to previous work~\cite{fischer_sim2vr_2024, ikkala_breathing_2022}, all intermediate rewards are scaled to respect these bounds, ensuring that they never exceed the terminal limits, providing stable and consistent training.

\paragraph{Reward Design for Different Motor Operators}
The reward implementation for different motor operators (see next subsection) reflects the distinct objectives of the agent in each case. 
%
%
For the \textbf{tapping motor operators}, the agent is penalized for incorrect contacts to achieve a speed–accuracy trade-off. Specifically, the scale of the \textit{Error penalty} $r_{\text{error}}$ is adjusted to balance the trade-off between quickly completing the task and maintaining precise fingertip contact with the target. The weight can be set as $w_{\text{error}} \in \{10, 20, 30\}$,
corresponding respectively to \textit{fast}, \textit{normal}, and \textit{accurate} motor operators.
%
For the \textit{swiping} operator, the agent is not penalized for incorrect touches. Swiping is modeled as a directional movement task with a defined start and end region. Accordingly, the distance penalty \( r_{\text{dist}} \) first penalizes the fingertip’s distance to the screen to encourage establishing and maintaining screen contact. Once contact is made, it penalizes the Euclidean distance to the end region. The termination penalty \( r_{\text{termination}} \) is applied whenever contact is broken after the swipe begins. As of now, start and end regions define targeted swipes, but swipes that require less precision (e.g., scrolling through long text) could be captured by enlarging these regions to represent broader directional intent rather than strict accuracy.

\subsection{Motor Operators} \label{sec:low_level_control_policies}

Training a single policy to generate long, continuous interaction sequences is highly inefficient. To improve learning efficiency, we draw inspiration from human motor learning. Human motor behavior is widely believed to be organized in terms of modular operators that are orchestrated to produce complex movements~\cite{berret_optimality_2019, gentner_modular_2006}.
A similar decomposition into discrete operators has long been used in HCI, in particular in Card’s Keystroke-Level Model~\cite{card_keystroke-level_1980} and the GOMS~\cite{john_goms_1996} models, where a task is modeled as a sequence of atomic operators. Following this principle, we decompose continuous touch interactions into motor operators, enabling more efficient and scalable policy learning.
We define motor operators based on the core gestures commonly observed in mobile device interaction~\cite{ruiz_user-defined_2011, tsai_testing_2017, zhu_phoneinvr_2024}: \textit{tapping} and \textit{swiping}. The tapping operator executes precise pointing actions to a specific target location on the screen. The swiping operator moves the finger in a specified direction over a set distance while maintaining continuous contact with the screen. Thus, it captures behaviors such as horizontal or vertical scrolling.
The tapping operator policy can be instantiated to reflect interpersonal differences in how task urgency and error tolerance shape behavior, capturing the speed–accuracy trade-off that governs tapping performance~\cite{zhai_speedaccuracy_2004}.
Specifically, we train a \textit{fast}, \textit{normal}, and \textit{accurate} policy in line with previous work in Fitts' law tasks~\cite{zhai_speedaccuracy_2004}.

Additionally, in human movement, no two trajectories are identical, even when performing the same action repeatedly. This inherent variability is due to motor noise, which reflects the stochastic properties of the human motor system.
Motor noise consists of two components: a signal-dependent component and a constant component~\cite{harris_signal-dependent_1998}. We model the signal-dependent noise by sampling from a normal distribution with mean 0 and standard deviation $k_{\text{SDN}} = 0.103$. 
We model the constant noise by sampling from a normal distribution with mean 0 and standard deviation $k_{\text{CN}} = 0.185$, following~\cite{van_beers_role_2004}. The resulting noisy control signal $a_{\text{noisy}}$ is computed as:


\begin{equation}
    \begin{split}
        a_{\mathrm{noisy}} &= (1 + \epsilon_{\mathrm{SDN}}) \cdot a + \epsilon_{\mathrm{CN}} \\
        \epsilon_{\mathrm{SDN}} &\sim \text{LogNormal}(0, k_{\mathrm{SDN}}) - 1 \\
        \epsilon_{\mathrm{CN}} &\sim \mathcal{N}(0, k_{\mathrm{CN}})
    \end{split}
    \label{eq:a_noisy}
\end{equation}

\subsection{From Motor Operators to Continuous Interaction Sequences} \label{sec:interaction_sequences}

To model arbitrarily long interaction sequences, we compose individual motor operators into smooth, continuous movements.
However, unlike the fixed, independent operators in KLM~\cite{card_keystroke-level_1980} and GOMS~\cite{john_goms_1996}, our motor operators are temporally dependent. Each action depends on the final state of the previous action, including starting position, muscle activation, and velocity. This dependency ensures fluid, human-like motion.
To this end, we introduce a control framework that leverages the modularity of motor operators while preserving temporal continuity between actions by conditioning each motor action on the final state of the preceding one.
Specifically, we embed the final state of the preceding motor action, including fingertip position $\mathbf{p}_{\text{finger}}^{t-1}$, fingertip velocity $\mathbf{v}_{\text{finger}}^{t-1}$, and muscle activation $\mathbf{a}_{\text{muscle}}^{t-1}$, into the initial state of the next motor action. This information is used both to construct the observation vector $\mathbf{o}t$ and to initialize new episodes. As a result, the agent can smoothly redirect the arm toward a new target $\mathbf{p}{\text{target}}^{t}$ and adjust muscle activations based on how the previous action ended. This mechanism enables the composition of motor operators into fluid, human-like movement sequences that scale to logs of arbitrary length and composition.

\subsection{Effort Model}\label{sec:effort-model}
The effort model quantifies perceived physical effort of movements by evaluating muscle activations~\cite{hincapie-ramos_consumed_2014}. It indicates how costly a motion is perceived in terms of physical exertion, enabling comparison of motor strategies and the optimization of control policies.

MyoSuite provides an effort model. However, while this default approach provides a general measure of activation, it does not account for differences in muscle size or strength, nor does it provide information about the perceived muscle effort. Several endurance-based models of perceived muscle effort for the arm have been proposed in the literature, including the Consumed Endurance model \cite{hincapie-ramos_consumed_2014,li_revisiting_2023}. In this work, we adopt the muscle effort model proposed by \citet{hincapie-ramos_consumed_2014}, which is widely used and has also been applied in \textsc{UitB}~\cite{ikkala_breathing_2022}. The model is expressed as follows:
\begin{equation}
    E(F) = \frac{1236.5}{\left(\frac{F}{F_{max}} \cdot 100 - 15\right)^{0.618}} - 72.5
\end{equation}
Here, \(F_{\text{max}}\) is the maximum force of each muscle as defined in MyoSuite, and \(F\) is the instantaneous force of that muscle, calculated based on its current activation as controlled by the RL agent. This formulation allows for a more accurate estimation of the perceived physical exertion when performing interaction movements. It has been shown to strongly correlate with perceived exertion as measured by the Borg CR10 scale~\cite{borg1998borg}.


\subsection{Training Routines}\label{sec:training}

The limitation in generalization and interaction fidelity of previously presented biomechanical forward simulations~\cite{ikkala_breathing_2022, fischer_sim2vr_2024} stems partly from training that is either too slow or does not converge at all. To solve this issue and enable precise dexterous interaction movements, we take advantage of multi-stage curriculum learning~\cite{miazga_increasing_2025, narvekar_curriculum_2020, narvekar_curriculum_2017} and adopt a staged reward shaping strategy.
At each stage, new reward components are added, gradually shaping the optimization landscape so that the learned policy balances more complex criteria and produces movements that better resemble human behavior.

In \textbf{Stage 1} the reward focuses solely on the target location: \( R^{1}_t = w_{\text{dist}} \cdot r_{\text{dist},t} + r_{\text{success},t} + r_{\text{timeout},t} \).
In \textbf{Stage 2}, we introduce a penalty for incorrect contacts to encourage precise interaction with the correct target: \( R^{2}_t = R^{1}_t + w_{\text{error}} \cdot r_{\text{error},t} \)
In \textbf{Stage 3}, we refine the \gls{RL} agent's behavior by incorporating biomechanical constraints that promote energy-efficient and smooth motion: \( R^{3}_t = R^{2}_t + w_{\text{activation}} \cdot r_{\text{activation},t} + w_{\text{jerk}} \cdot r_{\text{jerk},t} \)







For the swiping operator, we introduce an additional constraint to enforce continuous surface contact during the swipe. Any interruption in contact results in immediate episode termination and a negative reward: \( R_t = R_t + w_{\text{contact}} \cdot r_{\text{contact},t} \).

\begin{figure*}[t]
    \centering
    \includegraphics[width=0.9\linewidth]{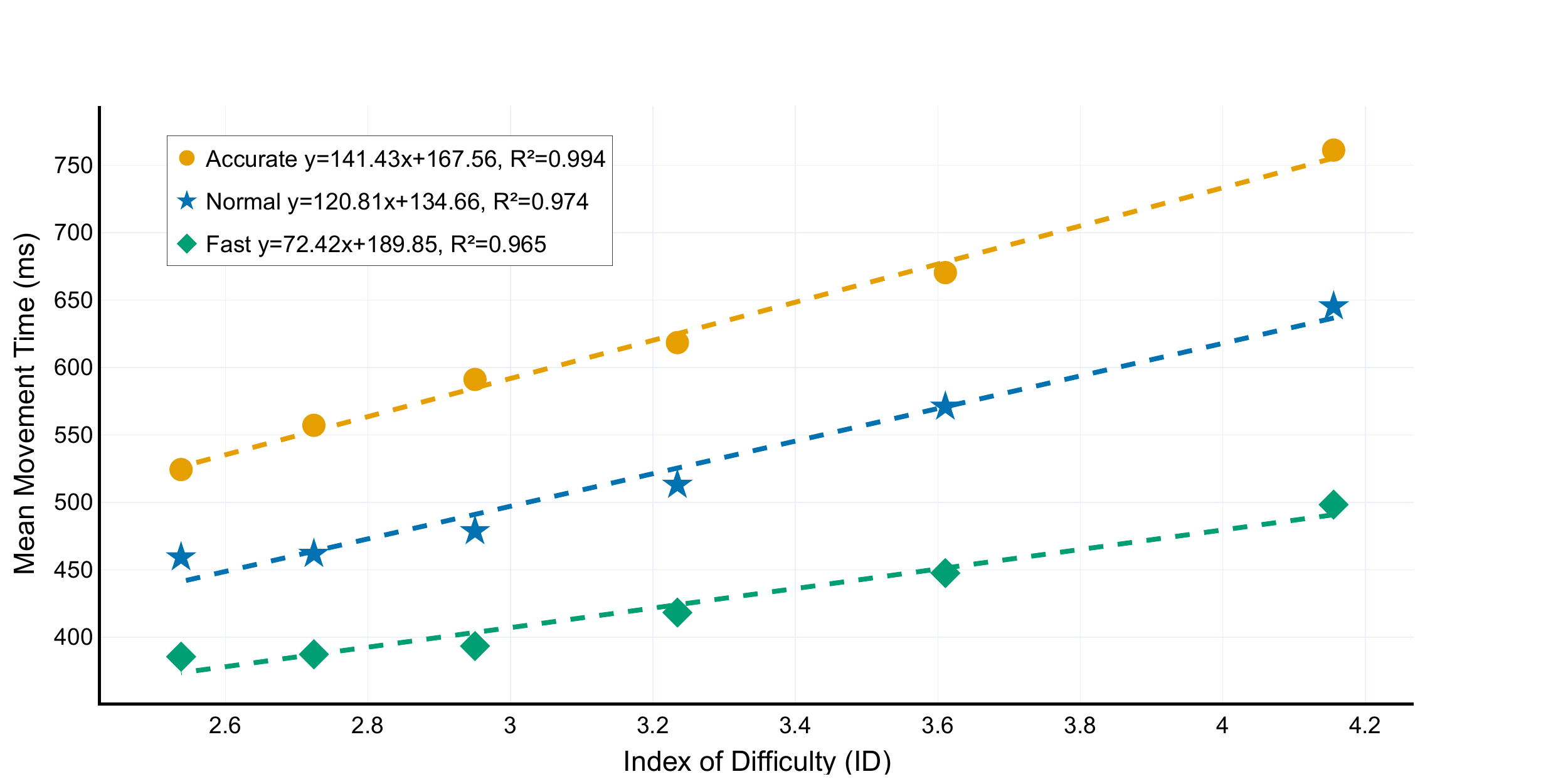}
    \caption{Simulation results showing Fitts’ law speed–accuracy trade-off for the three tapping policies \textit{accurate}, \textit{normal}, and \textit{fast}. Each point corresponds to a button size diameter (14--4 mm). The results show strong adherence to the linear relationship of movement time and task difficulty as described by Fitts' law.}
    \label{fig:fitts_tradeoff}
    \Description{A line chart of mean movement time versus index of difficulty. Three lines represent tapping policies—accurate, normal, and fast—each showing a strong linear relationship consistent with Fitts’ law.}
\end{figure*}

\section{Evaluation}

We evaluate \name by comparing its simulated movements to findings of human motor control and empirical data. Beyond verifying alignment with Fitts’ Law, we assess whether the tapping policies align with human data and compare different motor control policies in terms of endpoint accuracy. We then analyze the temporal dynamics of the simulated movements through their velocity profiles and 3D motion plots.
Finally, we analyze the difference in muscle effort between control policies positioned on different points on the speed-accuracy continuum.

\subsection{Speed-Accuracy Trade-off in Pointing Movements}\label{subsec:speed_accuracy}

Biomechanical motion synthesis should reflect the human tendency to adjust movement strategies based on task demands: reaching for larger or closer targets is typically faster, while reaching for smaller or more distant targets is slower. This trade-off between speed and accuracy is captured by Fitts’ Law~\cite{fitts_information_1954, bi_ffitts_2013}. Fitts' Law describes a linear relationship between task difficulty and the average movement time required to reach a target, and is widely used to characterize pointing and target acquisition tasks.
In addition to the task-imposed speed–accuracy trade-off, individual differences in motor strategy have been widely observed~\cite{zhai_speedaccuracy_2004}. Some users tend to prioritize speed, accepting higher error rates in exchange for faster movements. Others adopt a more cautious approach, favoring accuracy over speed by moving more slowly and precisely.

To evaluate whether \name reproduces these findings, we trained three tapping operators to represent different points along this continuum. To train these different policies, we modulate the penalty for incorrect taps (errors) to elicit varying behaviors: higher penalties encourage more precise, deliberate movements, while lower penalties favor faster but less accurate tapping. Specifically, we adjusted the $r_\text{error}$ component of the reward function introduced in \autoref{sec:reward_design}. This approach yields the following tapping operators:
\begin{itemize}
    \item \textbf{Accurate Tapping}: Control policy that prioritizes precise tapping with minimal error.
    \item \textbf{Normal Tapping}: Control policy that balances tapping speed and accuracy.
    \item \textbf{Fast Tapping}: Control policy that prioritizes speed and accepts increased error rates.
\end{itemize}

To evaluate the tapping operators, we use a range of button sizes with diameters from 4\,mm to 14\,mm, in 2\,mm increments. For context, the recommended diameter for keystroke targets is 6\,mm, while the common icon size on mobile devices is typically around 12\,mm~\cite{zhu_phoneinvr_2024}. Each trial involved tapping a target of fixed diameter from a constant starting distance, allowing us to isolate the effect of target size. For each button size and motor operator, we ran 200 simulations and measured the time from finger lift-off to screen contact.

Figure~\ref{fig:fitts_tradeoff} presents the movement time as a function of the index of difficulty using the Shannon Formulation~\cite{mackenzie_note_1989}. The results show strong adherence to the linear relationship described by Fitts' Law ($R^2 > 0.96$ for all three control policies), resulting in slower movements with an increasing index of difficulty.
We further observe a clear difference in the movement time between the different control policies (\textit{accurate}, \textit{normal}, \textit{fast}).
The slight variability in slope across the three control policies likely stems from modeled motor noise~\cite{van_beers_role_2004}, as the muscle activations that determine signal-dependent noise must vary to produce different movement speeds.


\begin{figure*}[]
    \centering
    \includegraphics[width=0.8\linewidth, trim=0 0 60 60, clip]{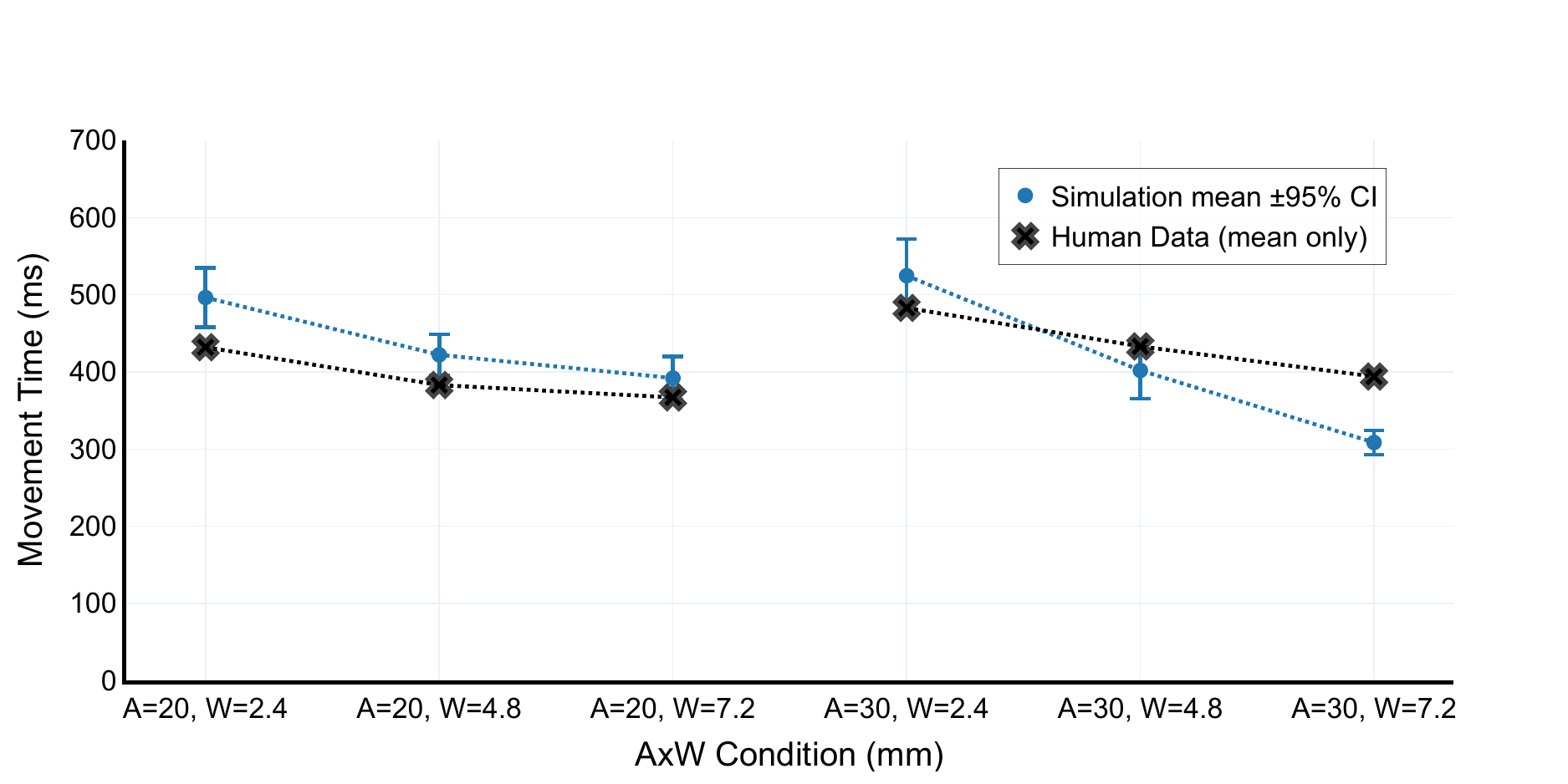}
    \caption{Movement times for the \textit{fast} tapping operator in a 1D task at distances of 20\,mm and 30\,mm with different button widths (2.4\,mm, 4.8\,mm, 7.2\,mm). 
    Simulation results are presented as mean movement times with 95\% confidence intervals based on 200 repeated runs. 
    For the human data, only mean values were available upon request.  
    The simulated mean movement times align closely with the human data across all conditions, consistent with results reported in~\cite{bi_ffitts_2013}.}
    \Description{Chart showing mean movement times with 95\% confidence intervals for a 1D tapping task. The x-axis shows six conditions based on amplitude (A) and width (W). For both A=20mm and A=30mm. The mean movement times from simulation align strongly with the available human data across all six conditions. The mean movement times are consistently higher for the A=30mm conditions than for the corresponding A=20mm conditions.}
    \label{fig:interval_movement_time_comparison}
\end{figure*}

\subsection{Pointing Performance on a Mobile Device}

We compare the tapping policy to human performance data on a mobile device from \citet{bi_ffitts_2013}. To enable a direct comparison, we replicated their experimental setup within the MuJoCo physics simulator. The biomechanical model performs the same  \textit{1D} pointing task between two bars, with target width and distance systematically varied across trials. Following the original procedure, we measure movement time from the starting bar to the target bar. We chose the \textit{fast} motor operator, as participants in the original study were instructed to acquire the target as quickly and accurately as possible, which may have biased them toward faster pointing. 

Simulation results are shown in Figure~\ref{fig:interval_movement_time_comparison}. 
As the human data were reported as means in the original paper, we compare the distribution of model movement times to these reference values.
Our results show that the simulated movement times across all AxW conditions are very close to the mean human movement times as reported by \citet{bi_ffitts_2013}, with an error between 15\,ms (20x7.2) and 77\,ms (30x7.2).
Furthermore, our simulation results replicate key human trends: movement times decrease with shorter distances and wider targets~\cite{fitts_information_1954, zhai_speedaccuracy_2004}, and the standard deviation of movement times increases with an increase in the index of difficulty~\cite{bi_ffitts_2013, kvalseth_distribution_1976}.

\subsection{Endpoint Accuracy}

\begin{figure*}[]
    \centering
    \begin{subfigure}[b]{0.24\textwidth}
        \centering
        \includegraphics[width=\linewidth]{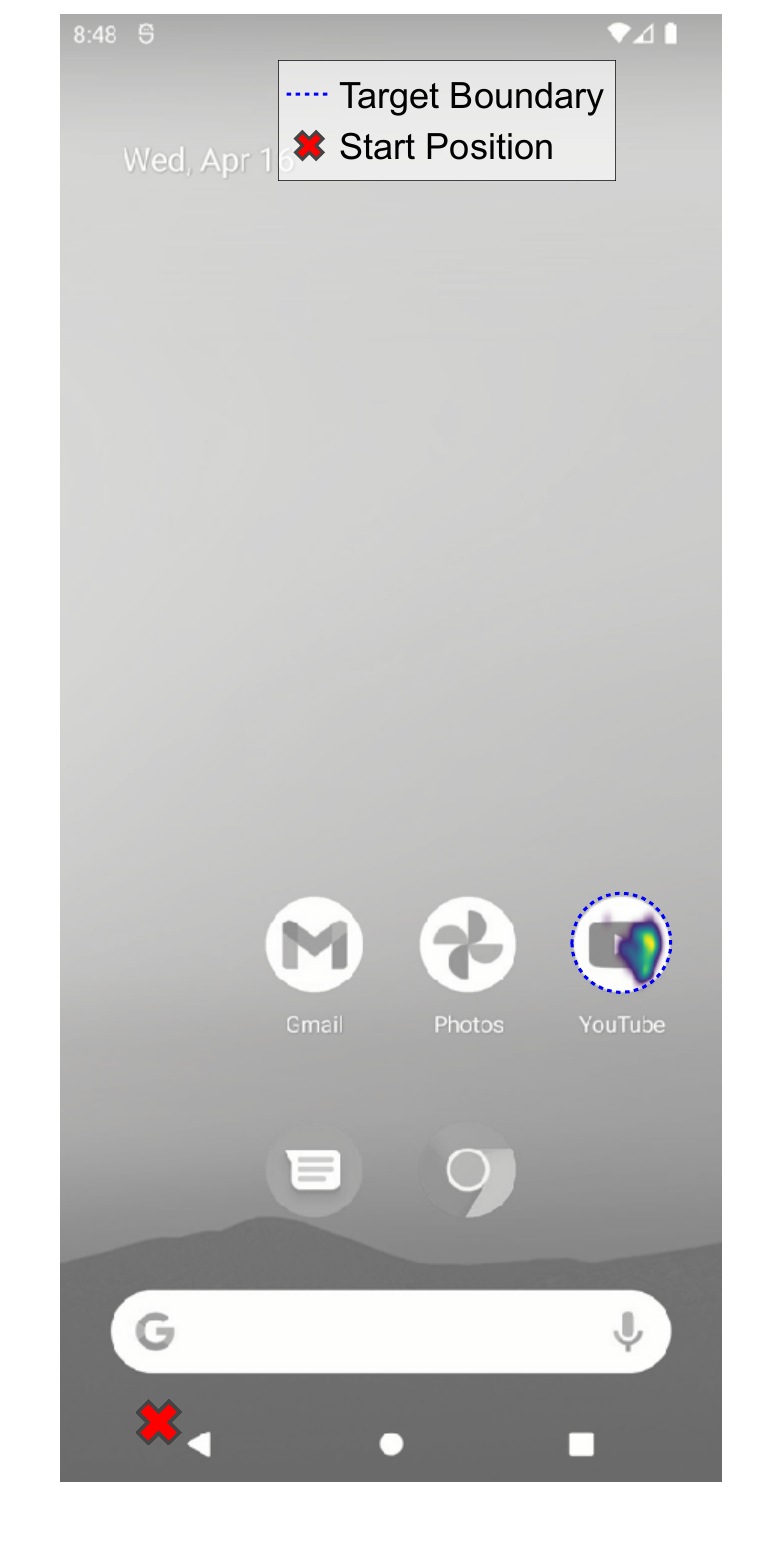}
        \caption{\textit{Accurate} tapping operator.}
    \end{subfigure}
    \begin{subfigure}[b]{0.24\textwidth}
        \centering
        \includegraphics[width=\linewidth]{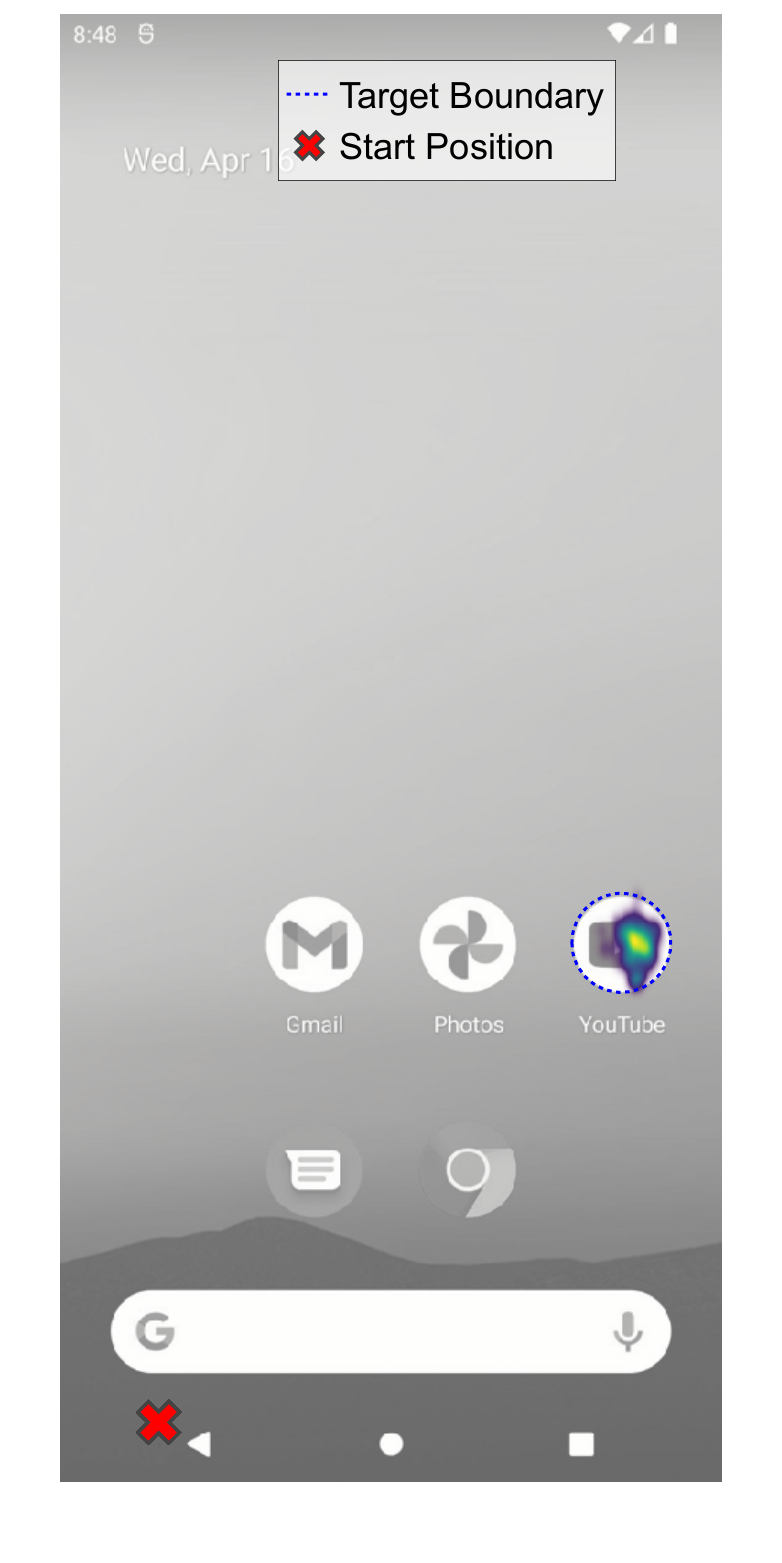}
        \caption{\textit{Normal} tapping operator.}
    \end{subfigure}
    \begin{subfigure}[b]{0.24\textwidth}
        \centering
        \includegraphics[width=\linewidth]{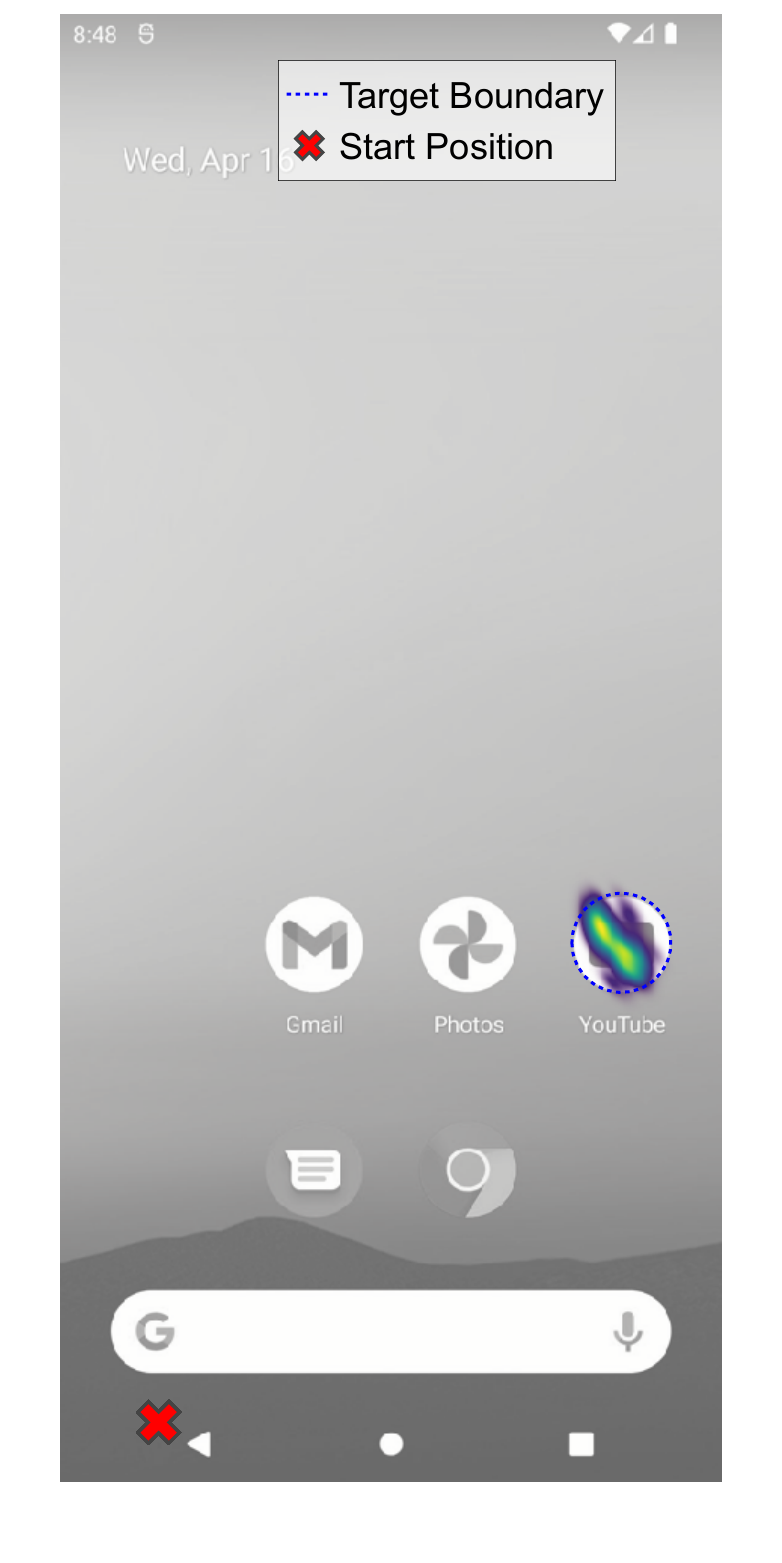}
       \caption{\textit{Fast} tapping operator.}
    \end{subfigure}
    \begin{subfigure}[b]{0.24\textwidth}
        \centering
        \includegraphics[width=\linewidth]{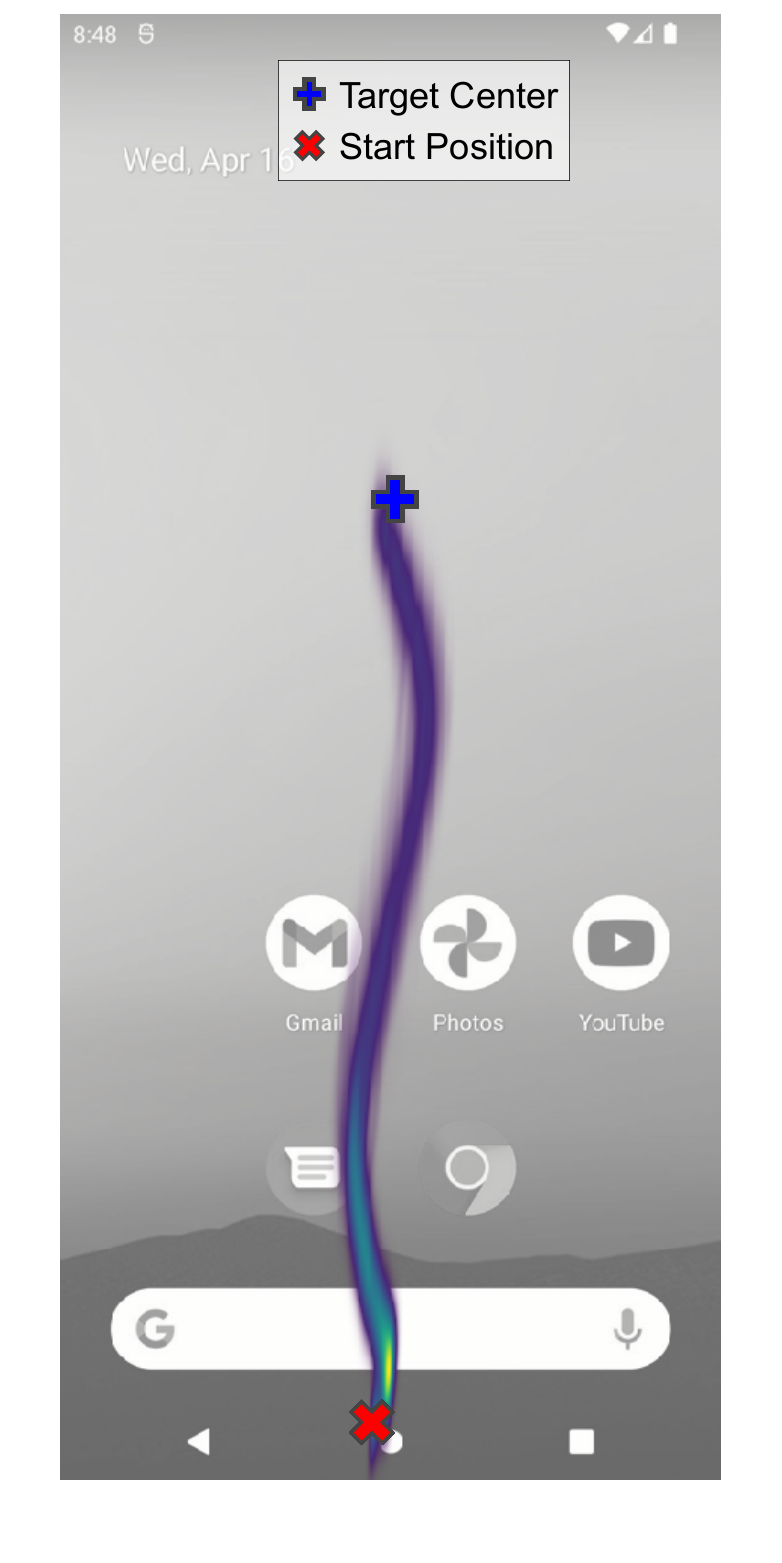}
       \caption{Swiping operator.}
    \end{subfigure}
    \caption{Heatmaps of fingertip positions during interaction with the touchscreen (200 simulation runs). For the tapping motor operators \textit{accurate}, \textit{normal}, and \textit{fast}, the heatmaps show activations within a 10\,mm diameter button, while the swipe policy represents finger movement toward the target. The heatmaps clearly show a bigger spread of endpoint contacts for the \textit{fast} operator compared to the \textit{normal} and \textit{accurate} operators.}
    \label{fig:end_point_variation}
    \Description{The figure shows four heatmaps of fingertip positions on a smartphone screen for different motor operators. Images (a), (b), and (c) display heatmaps for the accurate, normal, and fast tapping operators, respectively, on a 10 mm diameter button. The heatmap for the fast operator has a larger spread of contacts compared to the normal and accurate operators. Image (d) shows a heatmap representing the finger's movement path for a swipe policy.}
\end{figure*}

Having established that \name reproduces the speed–accuracy trade-off observed in human motor behavior, we next examine the spatial accuracy of created endpoints, which is important for matching click/touch coordinates reported in logs.
Research in typing~\cite{banovic_quantifying_2017}, for example, shows that humans reduce their typing speed to minimize typing errors. Accordingly, we found that the \textit{accurate} motor operator leads to fewer pointing errors compared to the \textit{normal} and \textit{fast} ones.
To investigate the errors made by the different control policies, \autoref{fig:end_point_variation} visualizes the fingertip contact points as a heatmap, highlighting the spatial distribution of taps across repeated trials.
Similar to the velocity profiles, we observe clear differences between the three tapping policies. The \textit{fast} policy exhibits a higher spread of endpoint interactions, indicating more errors on the screen compared to the \textit{normal} and \textit{accurate} policies.

\begin{figure*}[t]
    \centering
    \begin{subfigure}[b]{0.49\textwidth}
        \centering
        \includegraphics[width=\linewidth, trim=0 0 70 80, clip]{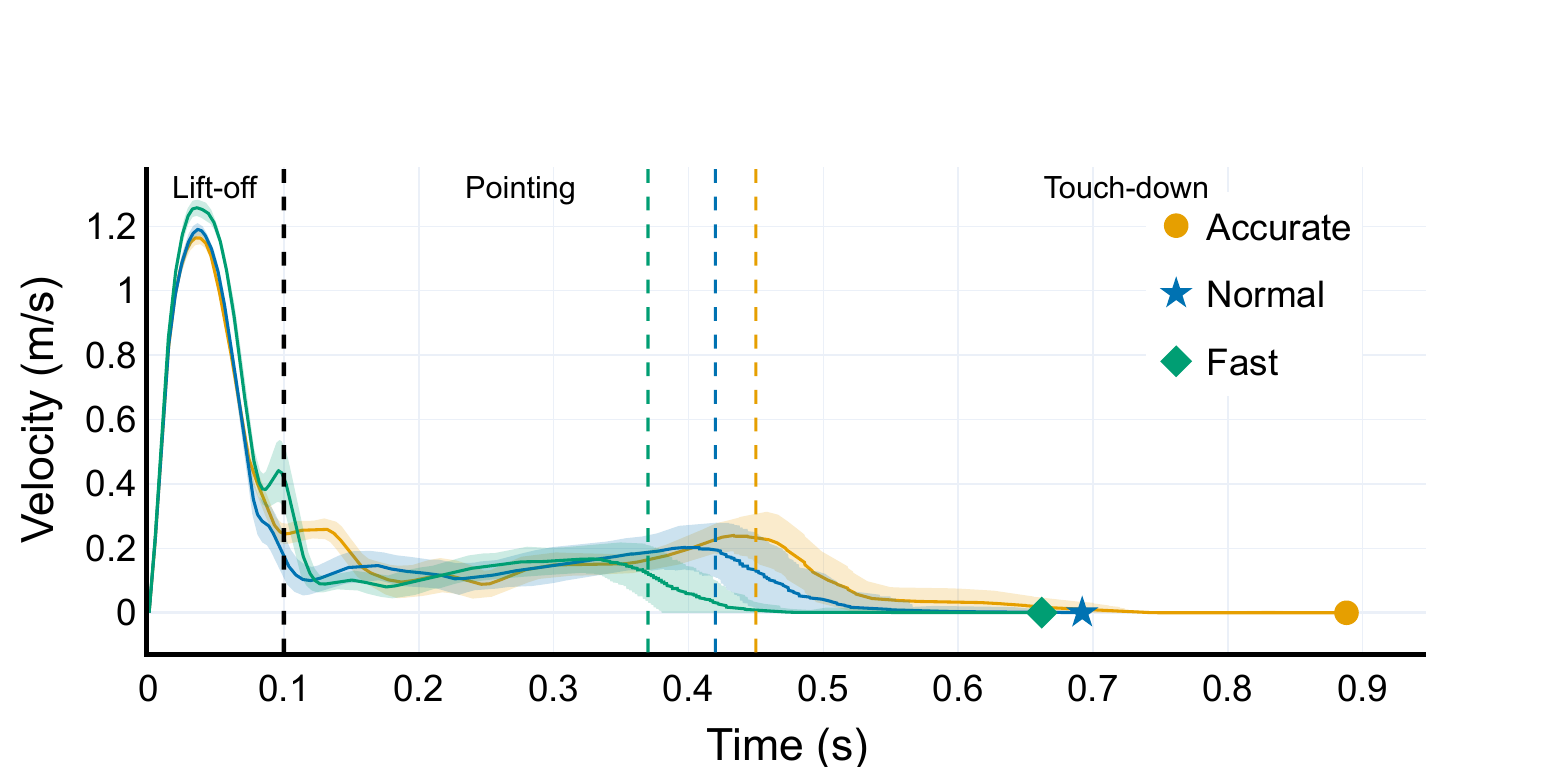}
        \caption{Velocity profiles of the tapping operators.}
        \label{fig:velocity}
    \end{subfigure}%
    \hfill
    \begin{subfigure}[b]{0.49\textwidth}
        \centering
        \includegraphics[width=\linewidth, trim=0 0 60 80, clip]{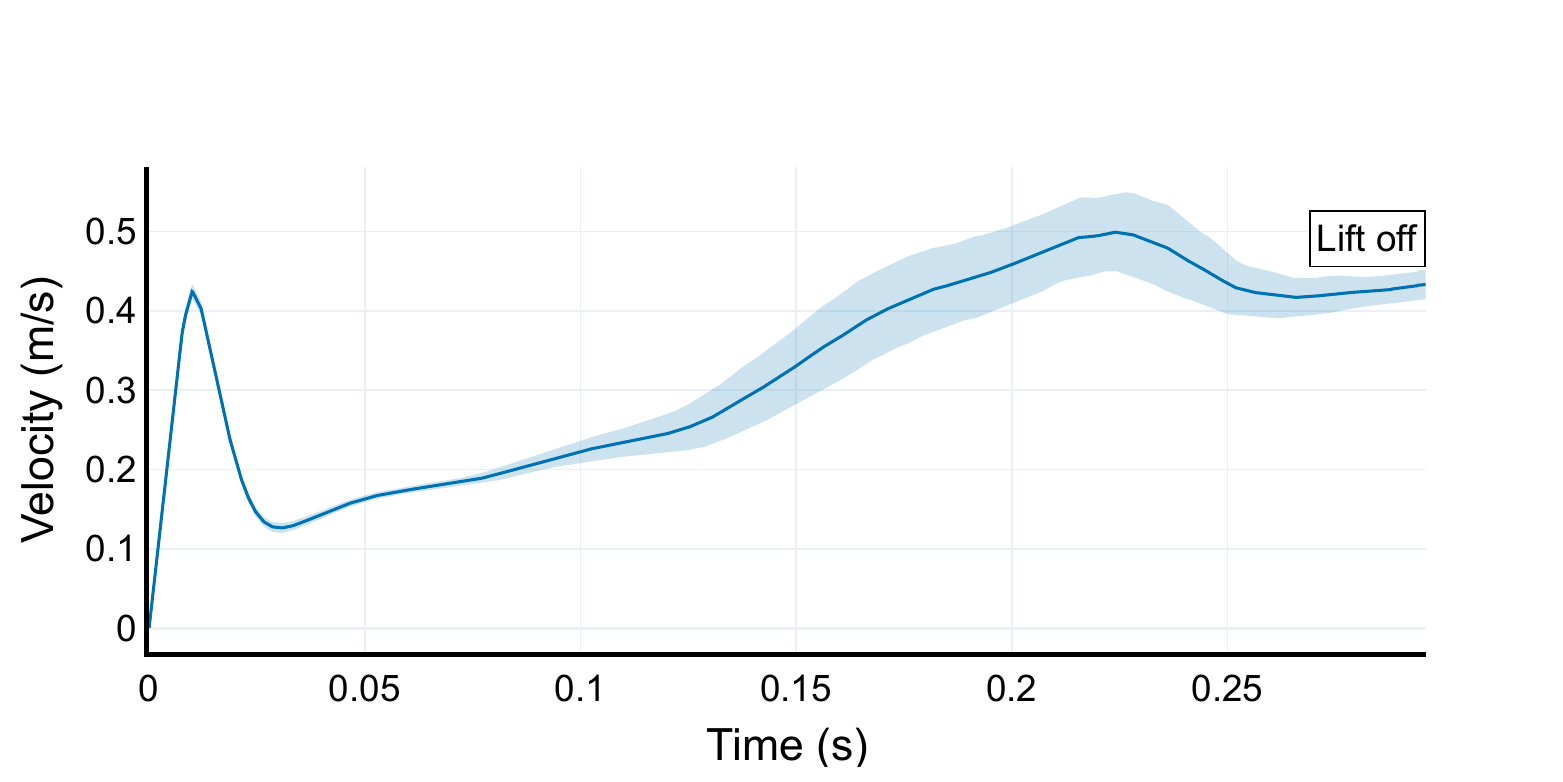}
        \caption{Velocity profiles of the swiping operator.}
        \label{fig:swiping_velocity}
    \end{subfigure}
    \caption{Velocity profiles of the \textit{accurate}, \textit{normal}, and \textit{fast} tapping operator and the swiping operator during interaction with a target of diameter 10~mm. 
    Left: overall velocity profile. Right: velocity profile measured on the display surface for \textit{Normal Swipe} policy, ending with the lift off the finger. The velocity profiles show strong similarities to the ballistic movements observed in human pointing behavior.}
    \Description{The figure consists of two line graphs. Graph (a) compares the velocity profiles of the accurate, normal, and fast tapping operators, showing that the fast operator has a higher peak velocity and a shorter movement time. Graph (b) shows a separate velocity profile for the swiping operator, which features a different curve with a lower initial peak followed by a gradual increase and a final plateau.}
    \label{fig:velocity_combined}
\end{figure*}

To evaluate the error rate, we selected button diameters of 4\,mm and 10\,mm, based on empirical measurements of current button implementations in Android running on a Google Pixel 6. While prior work recommends a diameter of approximately 6\,mm for keystrokes and 12\,mm for icons~\cite{zhu_phoneinvr_2024}, our chosen sizes reflect the smaller targets commonly encountered in real-world mobile interfaces.

The resulting error rates for the \textit{accurate}, \textit{normal}, and \textit{fast} policies are approximately 8.5\,\%, 12.5\,\%, and 42.5\,\% for the 4\,mm diameter button, and 1.5\,\%, 1.5\,\%, and 2\,\% for the 10\,mm diameter button. These values fall within the range observed in human performance~\cite{bi_ffitts_2013}.

For the \textit{swiping} motor operator, we define the starting position approximately at the home button and the ending position at three-quarters of the screen length. We observe slight variations of the fingertip trajectory on the phone surface, which result from the implemented motor noise and the friction of the screen surface, set to represent typical glass surface characteristics. The curved form of the trajectory and increased spread toward the end of the swipe motion align with findings from studies with human data~\cite{torres_tap_2023}.

\subsection{Movement Dynamics}

Synthesized movements should reflect typical features of human aimed movements, such as bell-shaped velocity profiles and smooth acceleration and deceleration~\cite{morasso_spatial_1981}. Aimed movements are often ballistic and follow minimum-jerk trajectories, minimizing sudden changes in acceleration to produce efficient and natural motion~\cite{viviani_minimum-jerk_1995}. To assess whether simulated movements exhibit these characteristics, we evaluate velocity profiles for all motor operators over 200 simulation runs (\autoref{fig:velocity_combined}), using the same start and target positions as in \autoref{subsec:speed_accuracy} and a 10\,mm button diameter.

The velocity profiles (see \autoref{fig:velocity_combined}) show strong similarities to human movements and related works~\cite{ikkala_breathing_2022, fischer_reinforcement_2021, viviani_minimum-jerk_1995, morasso_spatial_1981}. 
To better compare the three \textit{tapping} operators, we divide the movement into three distinct phases: (1) lift-off, (2) pointing, and (3) touch-down.
During the \textit{lift-off} phase, the \textit{fast} operator reaches the highest peak velocity, followed by the \textit{normal} and \textit{accurate} operators.
In the \textit{pointing} phase, where the agent moves the fingertip toward the target, velocity profiles are more similar across operators. However, the phase is noticeably longer for the \textit{accurate} operator, reflecting a more deliberate movement strategy.
The most distinct differences appear in the \textit{touch-down} phase. The \textit{fast} operator maintains higher velocities, resulting in shorter movement times but a greater likelihood of overshooting the target. In contrast, the \textit{accurate} operator slows down considerably before contact, as indicated by a long right tail in the velocity profile. This deceleration leads to a lower error rate and more precise target acquisition.

For the \textit{swiping} operator, the recorded velocity profile captures the movement while the fingertip remains in contact with the screen surface. As shown in Figure~\ref{fig:swiping_velocity}, the swipe begins with a rapid acceleration phase, followed by a brief deceleration and a second phase of sustained acceleration. The fingertip lifts off once the target swipe length is reached.

\begin{figure*}[]
    \centering
    \begin{subfigure}[b]{0.42\linewidth} 
        \centering
        \includegraphics[width=\linewidth]{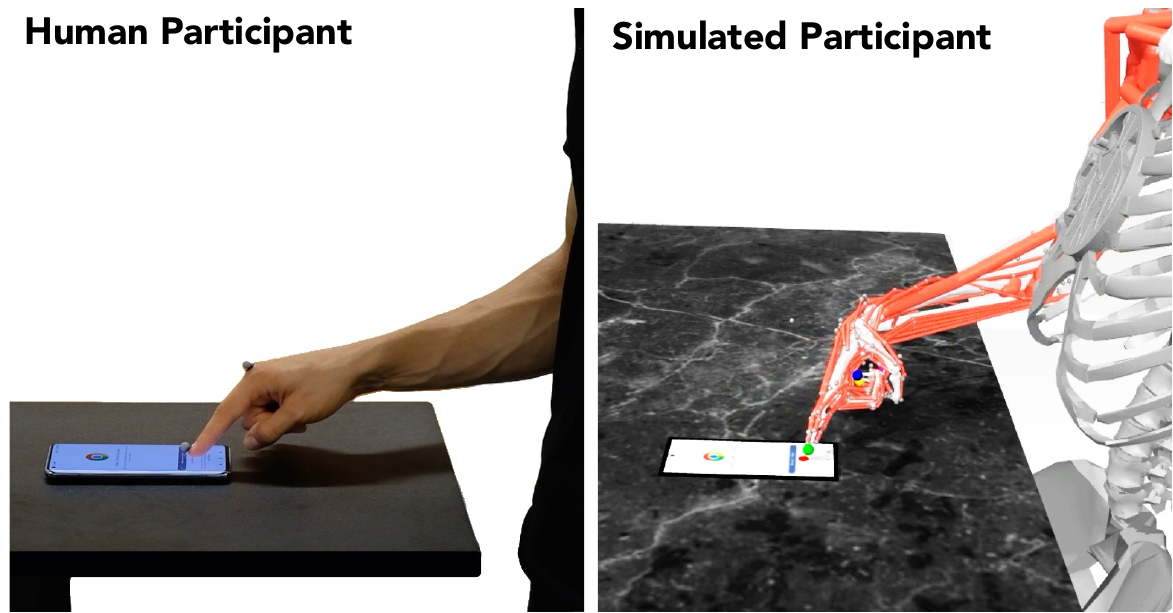}
        \caption{Study setup vs. simulation setup}
        \label{fig:sub:study_setup}
    \end{subfigure}
    \hfill
    \begin{subfigure}[b]{0.42\linewidth} 
        \centering
        \includegraphics[width=\linewidth, trim=10pt 10pt 330.0pt 330.0pt, clip]{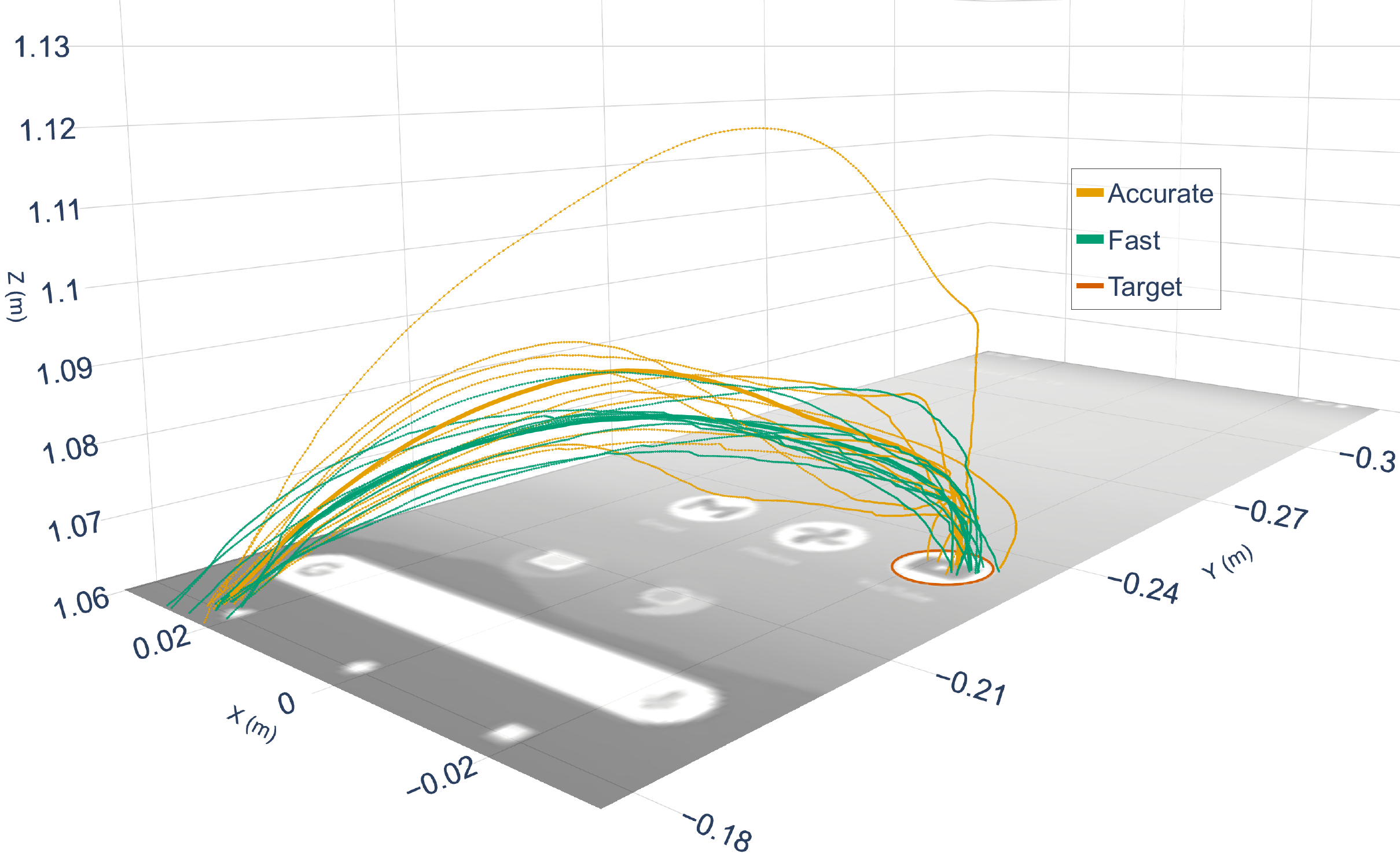}
        \caption{Movement trajectories of P1}
        \label{fig:sub:user1}
    \end{subfigure}


    \begin{subfigure}[b]{0.42\linewidth}
        \centering
        \includegraphics[width=\linewidth, trim=10pt 10pt 330.0pt 330.0pt, clip]{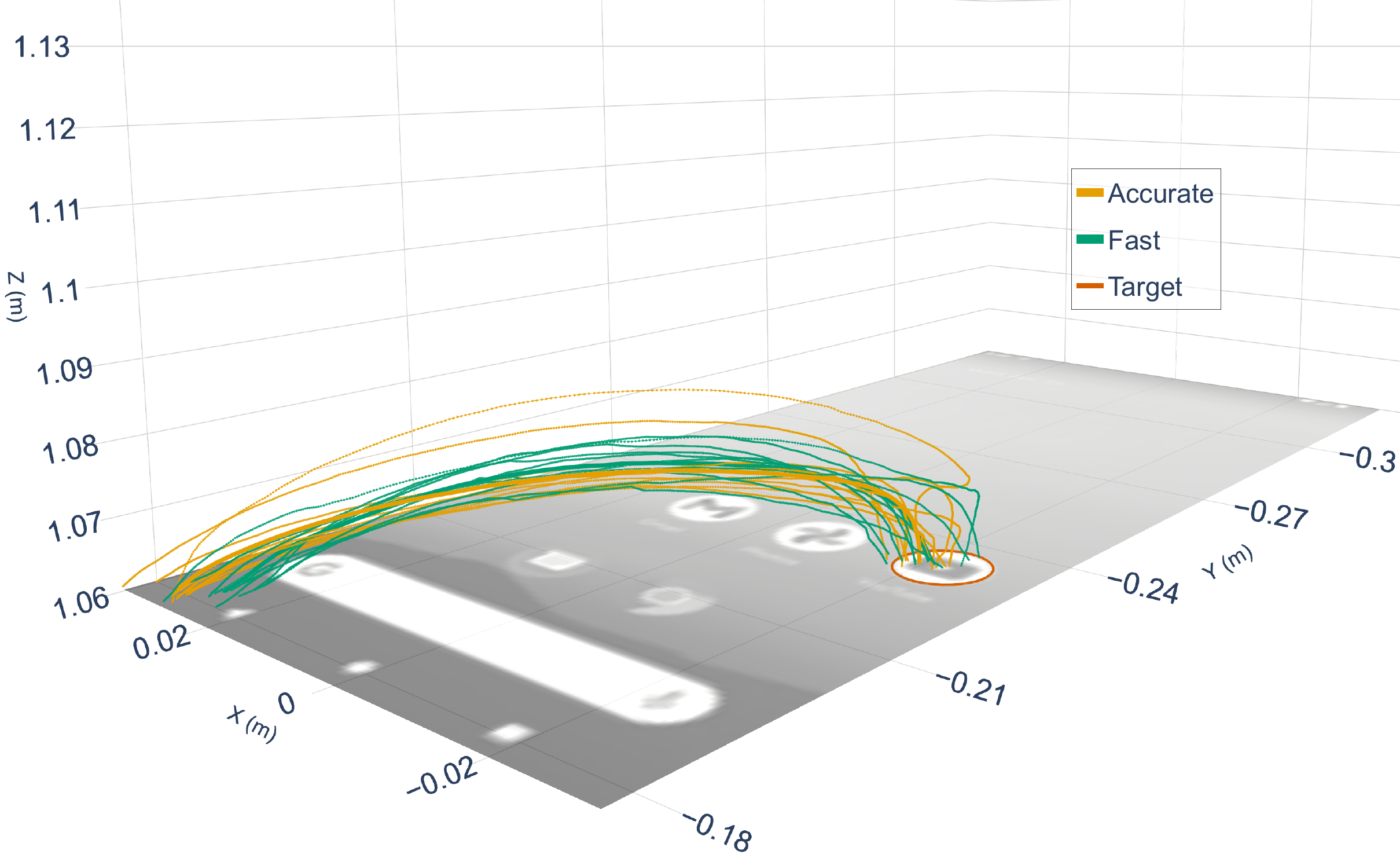}
        \caption{Movement trajectories of P2}
        \label{fig:sub:user2}
    \end{subfigure}
    \hfill
    \begin{subfigure}[b]{0.42\linewidth}
        \centering
        \includegraphics[width=\linewidth, trim=10pt 10pt 330.0pt 330.0pt, clip]{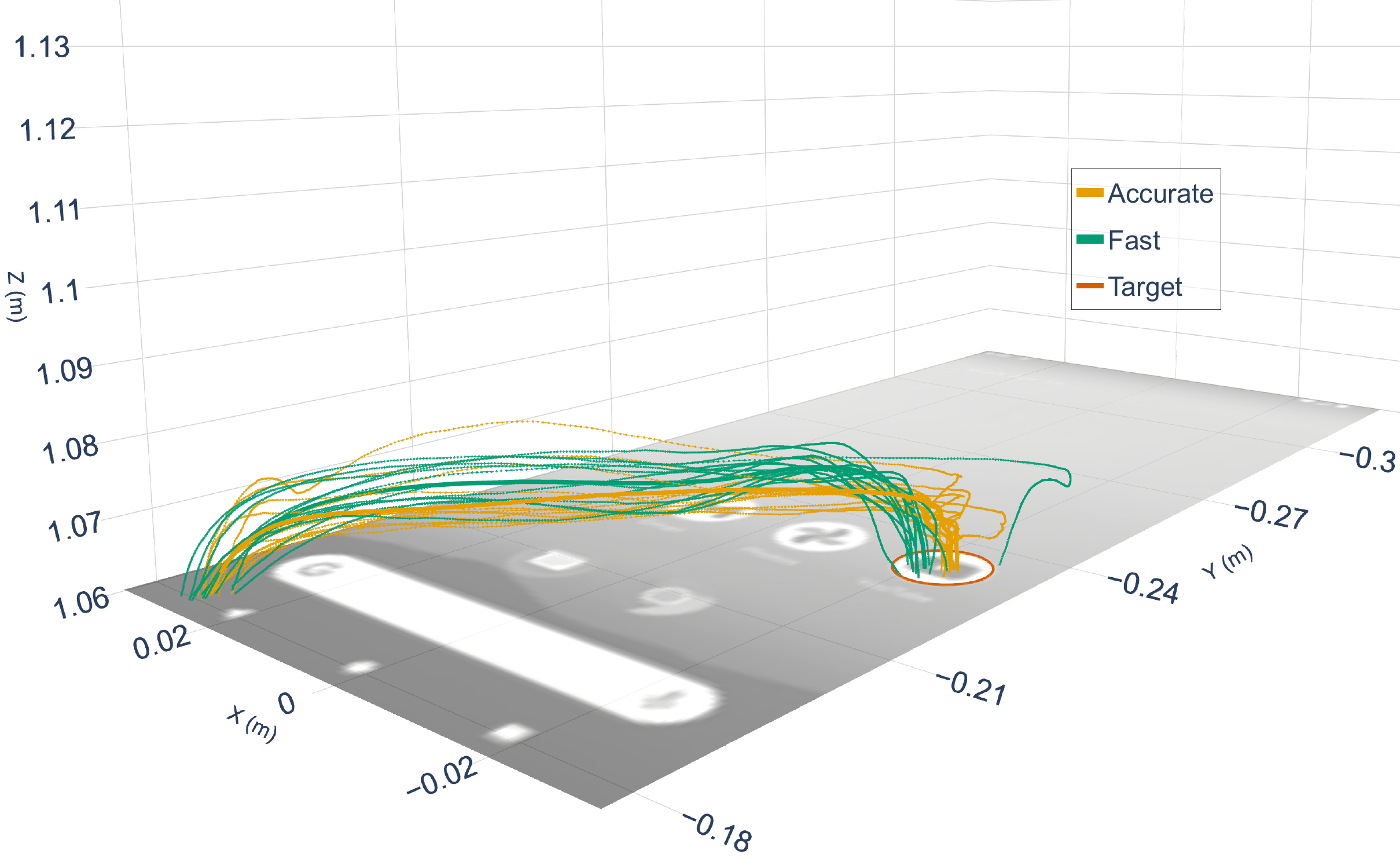}
        \caption{Movement trajectories of P3}
        \label{fig:sub:user3}
    \end{subfigure}


    \begin{subfigure}[b]{0.42\linewidth}
        \centering
        \includegraphics[width=\linewidth, trim=10pt 10pt 330.0pt 330.0pt, clip]{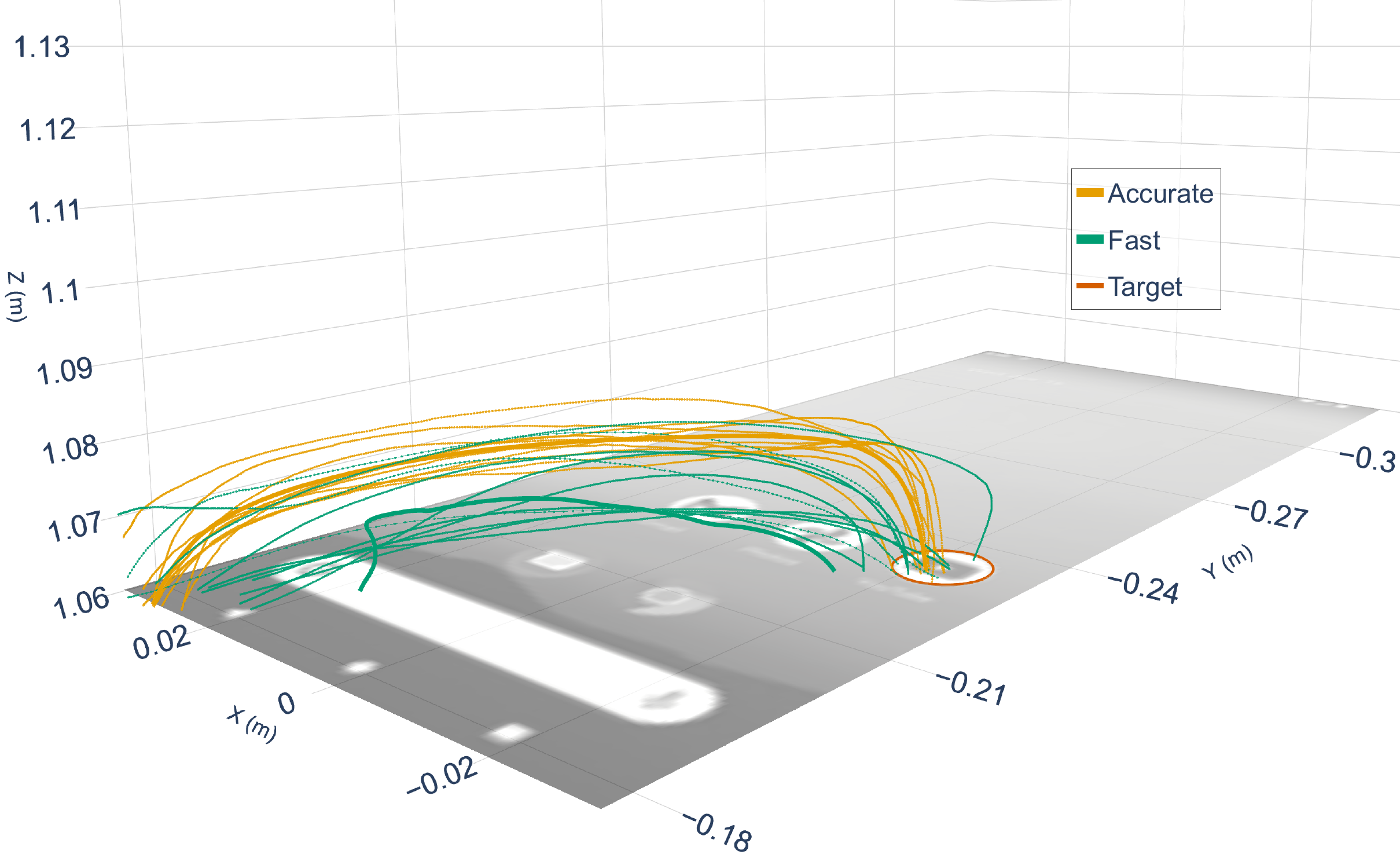}
        \caption{Movement trajectories of P4}
        \label{fig:sub:user4}
    \end{subfigure}
    \hfill
    \begin{subfigure}[b]{0.42\linewidth}
        \centering
        \includegraphics[width=\linewidth, trim=10pt 10pt 330.0pt 330.0pt, clip]{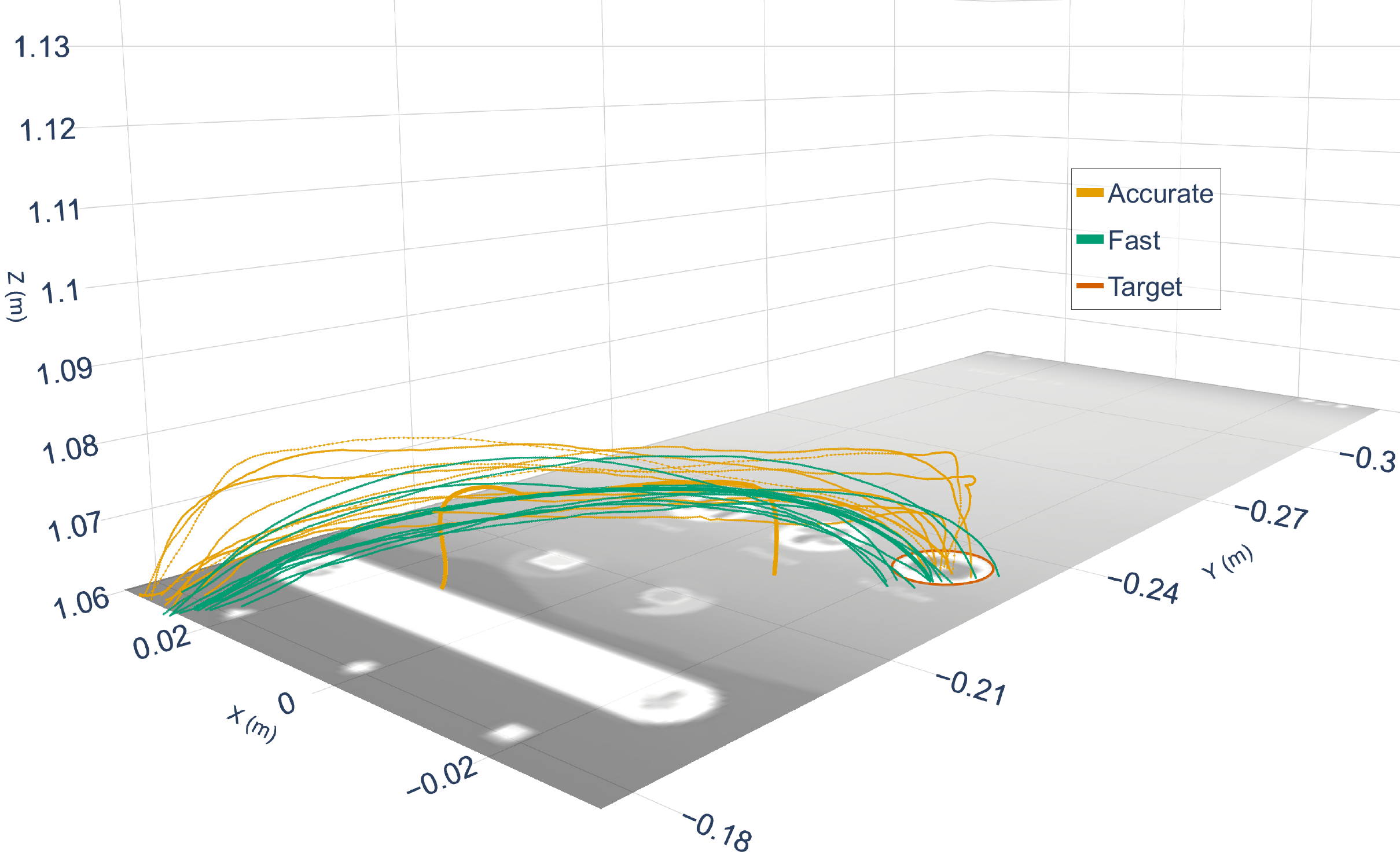}
        \caption{Movement trajectories of P5}
        \label{fig:sub:user5}
    \end{subfigure}

    \begin{subfigure}[b]{0.42\linewidth}
        \centering
        \includegraphics[width=\linewidth, trim=10pt 10pt 330.0pt 330.0pt, clip]{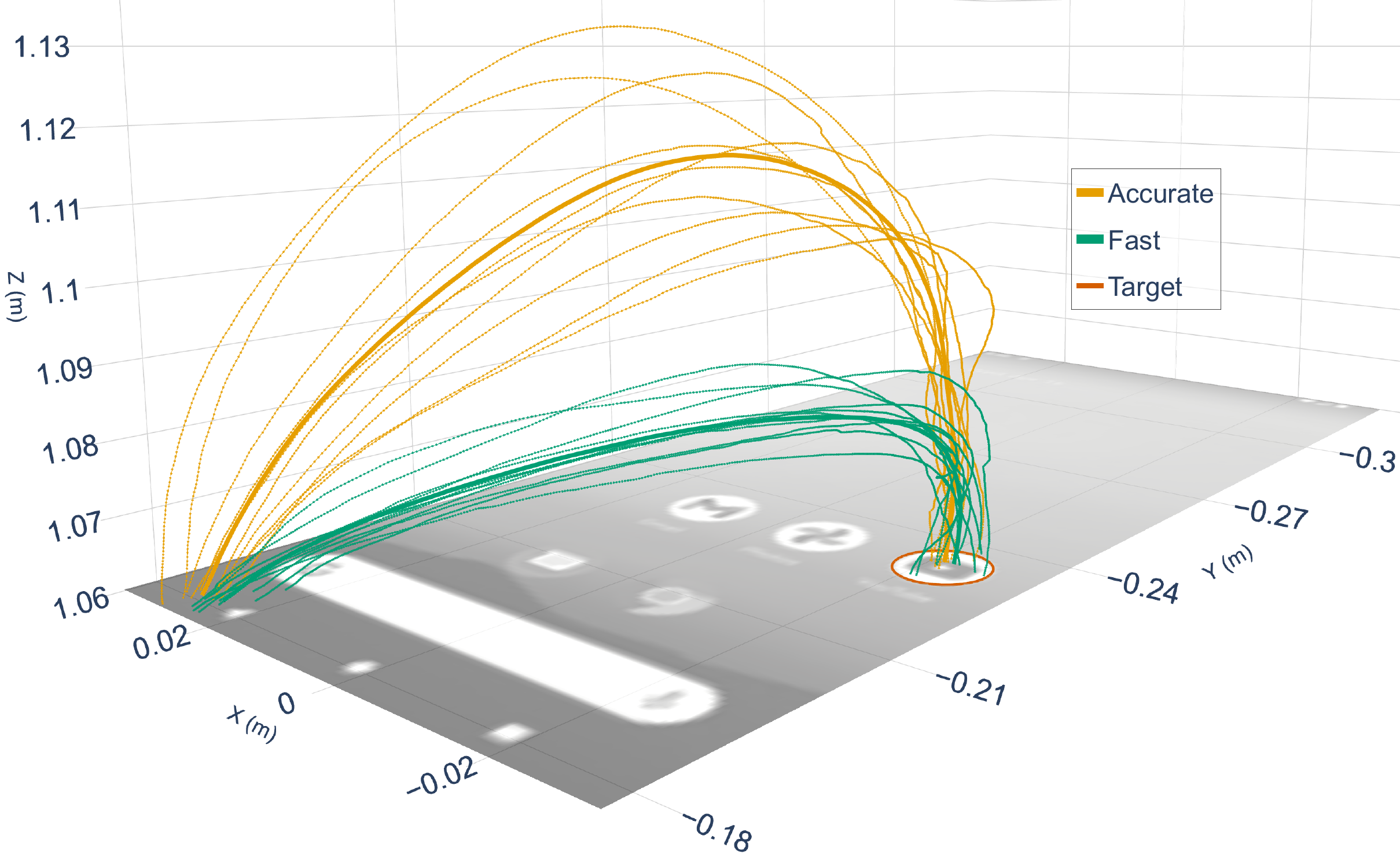}
        \caption{Movement trajectories of P6}
        \label{fig:sub:user6}
    \end{subfigure}
    \hfill
    \begin{subfigure}[b]{0.42\linewidth}
        \centering
        \includegraphics[width=\linewidth, trim=10pt 10pt 330.0pt 330.0pt, clip]{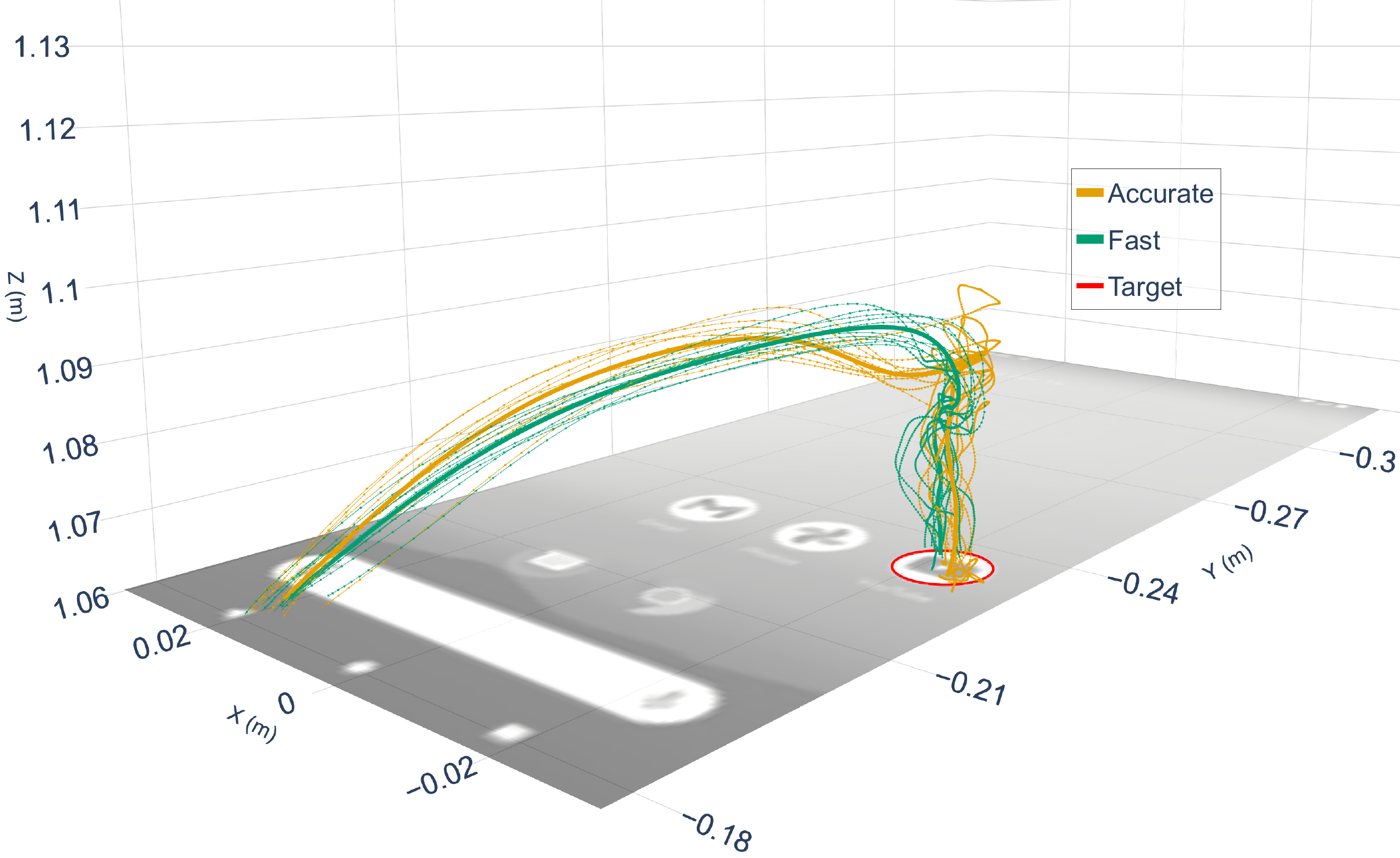}
        \caption{Movement trajectories of \name}
        \label{fig:sub:log2motion}
    \end{subfigure}

    \caption{Study setup 3D visualizations comparing participants' movement trajectories (\autoref{fig:sub:user1}-\autoref{fig:sub:user6}) and synthesized trajectories generated with \name (\autoref{fig:sub:log2motion}). Based on the DTW distances, P1 and P6 are closest to the simulation.}
    \Description{The figure contains eight panels comparing human and simulated smartphone-touch trajectories in 3D.
Panel (a) shows two images: a photo of a person reaching toward a smartphone on a table, and a simulation screenshot showing a musculoskeletal hand model performing the same task.
Panels (b) to (g) each show 3D line plots for one participant (P1 to P6). Yellow lines represent accurate movements, green lines represent fast movements, and small gray circles mark target locations on the phone surface. The trajectories rise slightly above the phone and curve toward the targets, with individual differences visible across participants.
Panel (h) displays the same style of 3D plot for trajectories generated by the Log2Motion simulation. The simulated paths closely resemble the human trajectories in curvature and range, converging on the same target positions. Across all panels, the phone is shown as a rectangular gray surface in the horizontal plane.}
    \label{fig:all_trajectories_grid_3x2}
\end{figure*}

\begin{figure*}[t]
    \centering

    \begin{subfigure}[t]{0.32\textwidth}
        \centering
        \includegraphics[width=\textwidth, trim=10 40 70 70, clip]{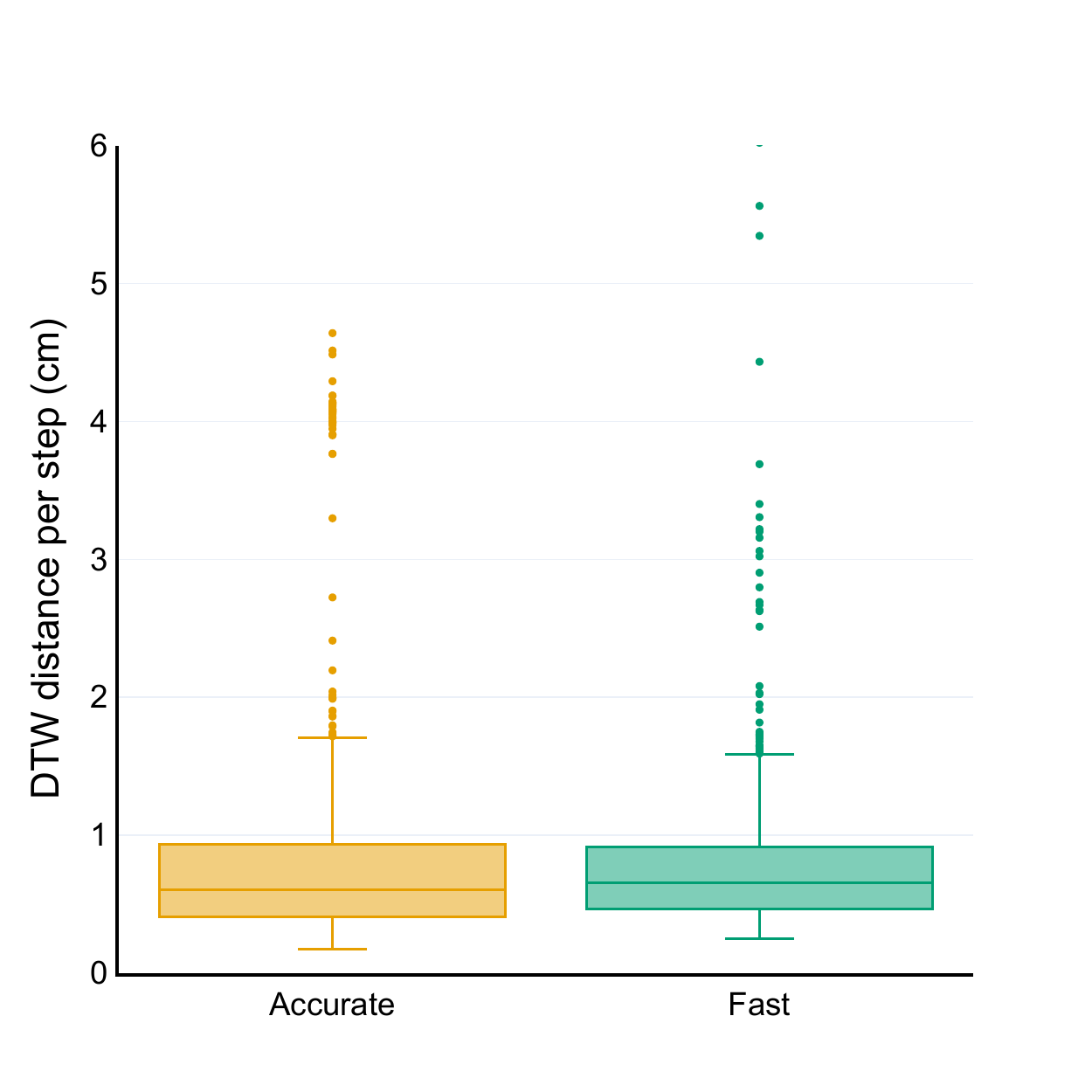}
        \caption{\small Within-participant similarity.}
        \label{fig:sub:within-DTW}
    \end{subfigure}
    \hfill
    \begin{subfigure}[t]{0.32\textwidth}
        \centering
        \includegraphics[width=\textwidth, trim=10 40 70 70, clip]{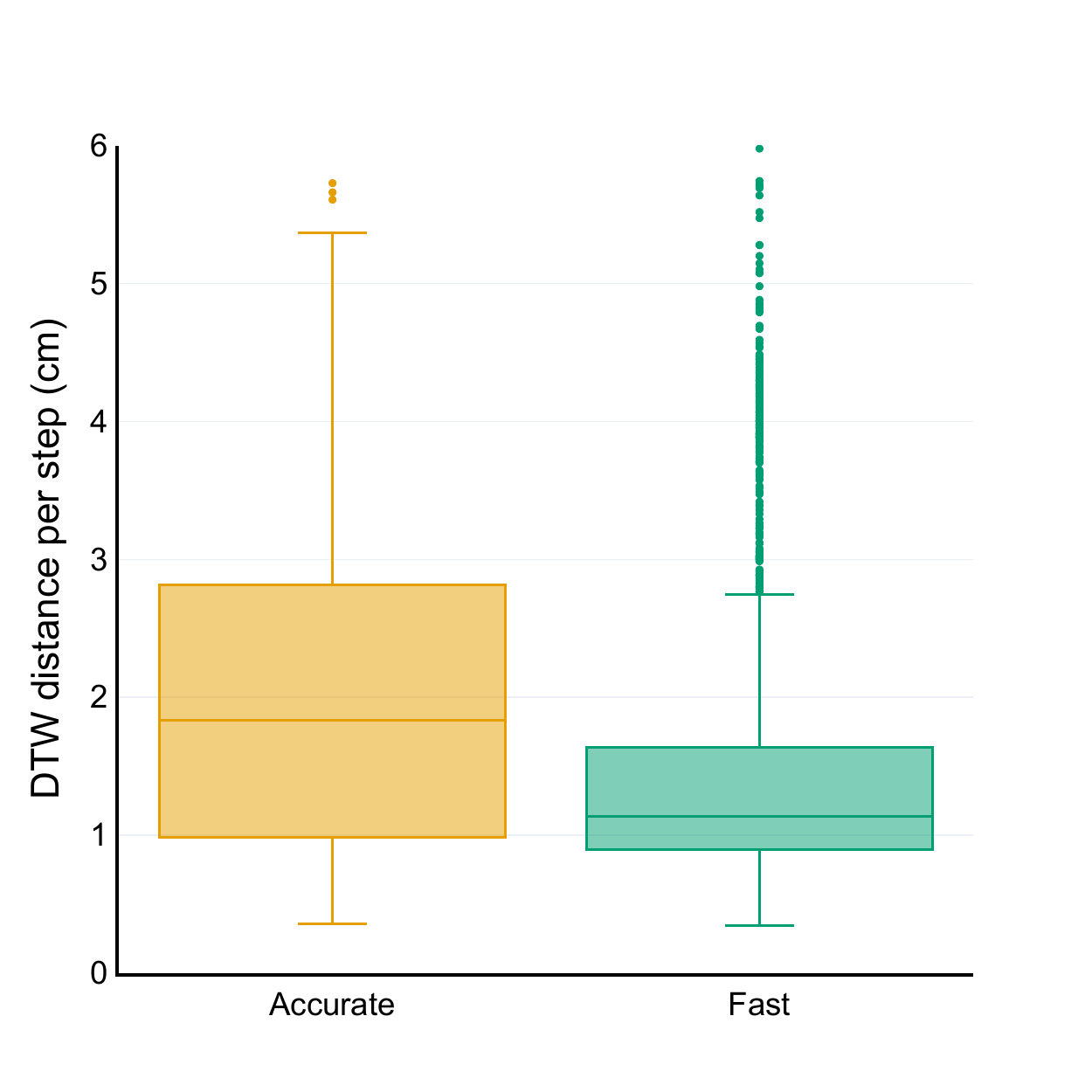}
        \caption{\small Between-participant similarity.}
        \label{fig:sub:between-DTW}
    \end{subfigure}
    \hfill
    \begin{subfigure}[t]{0.32\textwidth}
        \centering
        \includegraphics[width=\textwidth, trim=10 40 70 70, clip]{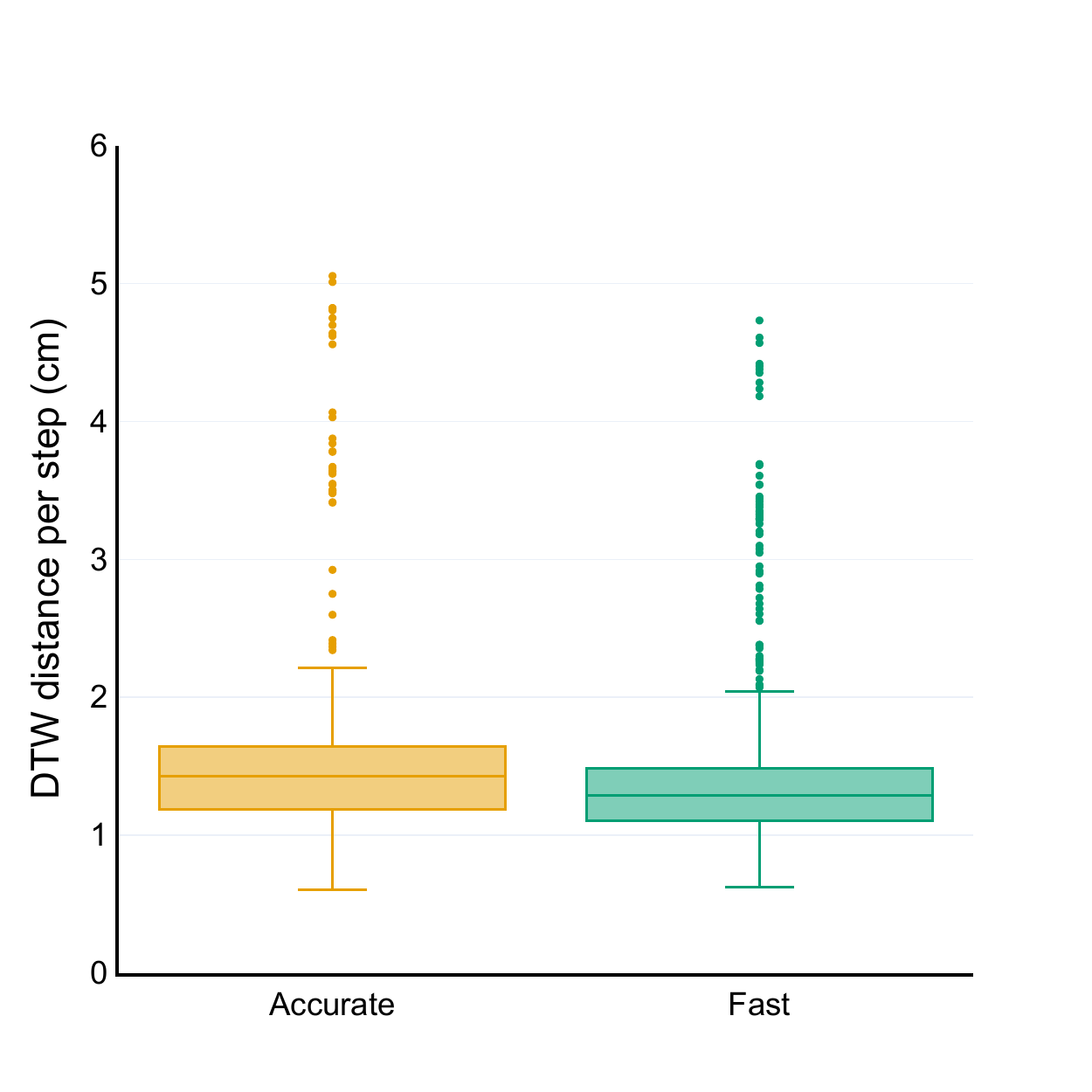}
        \caption{\small Simulation-participant similarity.}
        \label{fig:sub:DTW-user-sim}
    \end{subfigure}
\caption{
Comparison of normalized DTW distances per step for the pointing task. Distances between synthesized trajectories and all participants (c) fall within the variability observed between participants (b), indicating that \name generates human-like trajectories. Within-participant variability (a) is shown as a reference.
}
\Description{The figure consists of three side-by-side boxplots that compare DTW distance per step, measured in centimeters, for two movement styles: “Accurate” and “Fast.”
Panel (a), labeled “Within-participant similarity,” shows the smallest DTW distances, with both movement styles producing medians below 1 centimeter and several small outliers above 3 centimeters.
Panel (b), labeled “Between-participant similarity,” shows noticeably larger DTW distances, with wider boxes and higher medians for both movement styles, as well as many outliers up to about 6 centimeters.
Panel (c), labeled “Simulation-participant similarity,” shows DTW distances that are lower than in panel (b) and similar in scale and spread to panel (a).
Across all panels, accurate and fast movements are shown as separate boxplots, using gold and green colors, respectively, plotted along the same vertical axis from 0 to 6 centimeters.}
\label{fig:DTW}
\end{figure*}

\subsection{User Study: Comparing Synthesized and Real Movement Trajectories}
To evaluate whether \name synthesizes human-like trajectories beyond known motor-control regularities, we conducted a motion capture study comparing human and simulated movements. Notably, we did not fine-tune \name beyond the training procedure in \autoref{sec:policylearning}.

\subsubsection{Method}
We recruited six participants (4 male, 2 female) via opportunity sampling from the academic staff at our institute. Participants had a mean age of 29 years (\emph{SD} = 5) and a mean height of 176\,cm (\emph{SD} = 9\,cm). All participants had normal or corrected-to-normal vision and reported no motor impairments.
Participants completed three interaction tasks: (1) pointing, (2) swiping, and (3) short interaction sequences. For the pointing and swiping tasks, they performed 10 fast and 10 accurate repetitions each. For the sequential task, they performed 2 fast and 2 accurate repetitions. We placed the smartphone on a height-adjustable table in front of the participants. We counterbalanced the task order to mitigate order effects.
We recorded 3D fingertip trajectories at 300 Hz using a marker-based system consisting of 4 Qualisys Arqus A12 and 4 Qualisys Miqus cameras. Using Qualisys Track Manager, we extracted the fingertip position time series for each trial.

\subsubsection{Results}
We focus our analysis on the pointing task, as it provides isolated single-movement trajectories most suitable for direct comparison between simulated and participants' movements. 
Our results show that the trajectories synthesized by \name closely resemble human movement trajectories. Figure~\ref{fig:all_trajectories_grid_3x2} shows the 3D fingertip trajectories of all participants (\autoref{fig:sub:user1}--\autoref{fig:sub:user6}) and the trajectories synthesized by \name(\autoref{fig:sub:log2motion}). 
Human trajectories exhibited substantial inter-individual variability: some participants (P2--P5) produced relatively flat paths, whereas others (P1 and P6) showed larger vertical excursions. The separation between fast and accurate movements also varied across users, with P4 and P6 showing pronounced height differences and P2 and P3 showing minimal divergence.
The synthesized trajectories show great similarity to the human trajectories. Fast synthesized movements follow direct, low-arc paths to the target, whereas accurate movements lift the fingertip slightly higher and slow down before contact. These patterns mirror behaviors observed in the human data: overall trajectory shape best resembles that of P6, increased touchdown-phase variability aligns with P3, and the small curvature differences between fast and accurate movements align with P5.

To quantify the similarity between human and simulated movements, we compute normalized \gls{DTW} distances~\cite{sakoe_dynamic_1978, giorgino_computing_2009} (a) within users (across repetitions), (b) between users, and (c) between users and the simulation (Figure~\ref{fig:DTW}). \gls{DTW}~\cite{sakoe_dynamic_1978}, computes an optimal, non-linear temporal alignment between multivariate time series, enabling shape comparisons even when movements differ in speed or local timing.
As shown in Figure~\ref{fig:DTW}, the median DTW distance between \name and human participants is 1.43\,cm for accurate movements and 1.29\,cm for fast movements (\autoref{fig:sub:DTW-user-sim}). For comparison, distances between human participants are 1.83\,cm and 1.13\,cm, respectively (\autoref{fig:sub:between-DTW}). These results place the simulated trajectories well within between-user variability. The small distances also imply similar spatial paths and timing, as substantial speed differences would require greater temporal warping and increase DTW cost.
The similarity between simulated and human movements is evident not only in individual movements but also in the full trajectories (see \autoref{fig:task2all_trajectories_grid_3x2} in \autoref{ch:Appendix}). For swiping gestures, we observe a small systematic deviation: human swipes tend to follow slightly straighter paths than the synthesized trajectories (see \autoref{fig:end_point_users_variation} in \autoref{ch:Appendix}).

Together, the qualitative and quantitative results show that \\ \name synthesizes biomechanically plausible movement trajectories that fall well within the natural variability of human motion, supporting our claim that it produces human-like behavior. 

%

\subsection{Effort Cost}
\begin{figure*}[]
    \centering
    \begin{subfigure}[b]{0.49\textwidth}
        \centering
        \includegraphics[width=\linewidth]{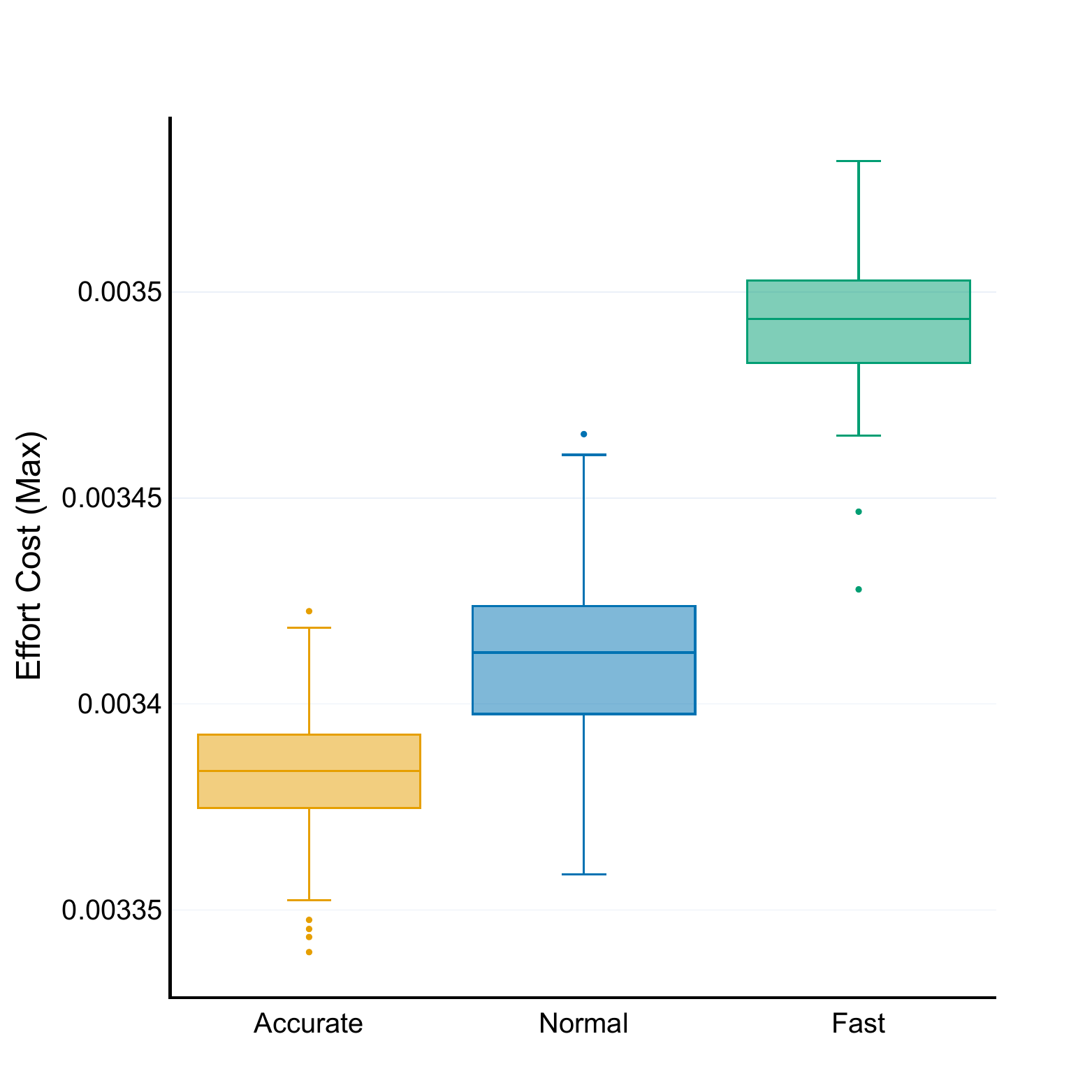}
        \caption{Maximum muscle cost.}
        \label{fig:effort_max}
    \end{subfigure}%
    \hspace{0.5em}%
    \begin{subfigure}[b]{0.49\textwidth}
        \centering
        \includegraphics[width=\linewidth]{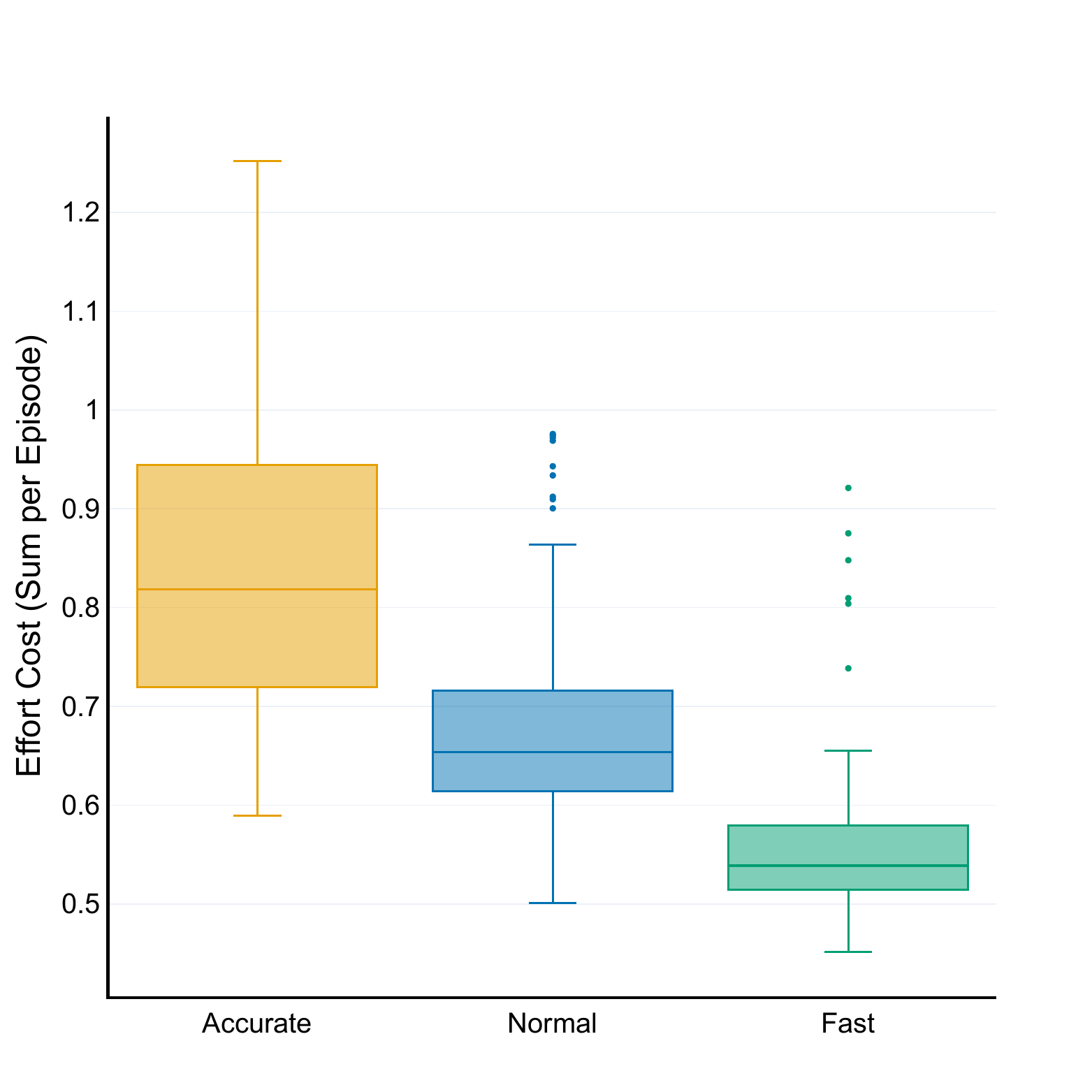}
        \caption{Total muscle cost.}
        \label{fig:effort_sum}
    \end{subfigure}
    \caption{Comparison of muscle effort metrics for the three tapping policies during interaction with a 10\,mm button over 200 attempts.}
    \label{fig:effort_comparison}
    \Description{The figure contains two box plot charts that compare muscle effort metrics for three different tapping policies: accurate, normal, and fast.

    The chart on the left (a) is titled "Maximum muscle cost." It shows that the maximum muscle cost increases from the accurate policy to the normal policy and is highest for the fast policy.

    The chart on the right (b) is titled "Total muscle cost." This chart shows that the total muscle cost decreases as the policy changes from accurate to normal to fast, with the fast policy having the lowest total muscle cost.}
\end{figure*}

We would like biomechanical motion synthesis to reveal which input interactions can be tiring when performed repeatedly.
It is known that interactions with touch surfaces can be fatiguing \cite{bachynskyi2015performance}. 

Prior research suggests that, for a given physical force, humans tend to perceive longer-duration movements as more effortful than shorter ones~\cite{morel_what_2017}.
To investigate if \name reproduces these tendencies, we compare the estimated muscle effort across the \textit{accurate}, \textit{normal}, and \textit{fast} operators. We use the same setup as in the previous sections, simulating 200 tapping interactions per policy. Muscle effort is computed as described in \autoref{sec:effort-model}.
%
%
As shown in \autoref{fig:effort_max}, the slower the movement, the lower the peak muscle effort. Peak effort is lowest for the \textit{accurate} tap gestures, followed by \textit{normal} and \textit{fast} ones. However, when considering the accumulated muscle effort across the entire movement (see \autoref{fig:effort_sum}), we observe that \textit{accurate} tap gestures require the most energy, followed by \textit{normal} and \textit{fast} ones. These results indicate that shorter, high-intensity movements are more energy-efficient for this task than slower, more controlled ones, which aligns with human perception~\cite{morel_what_2017}.
\begin{figure*}
    \centering
    \includegraphics[width=\linewidth]{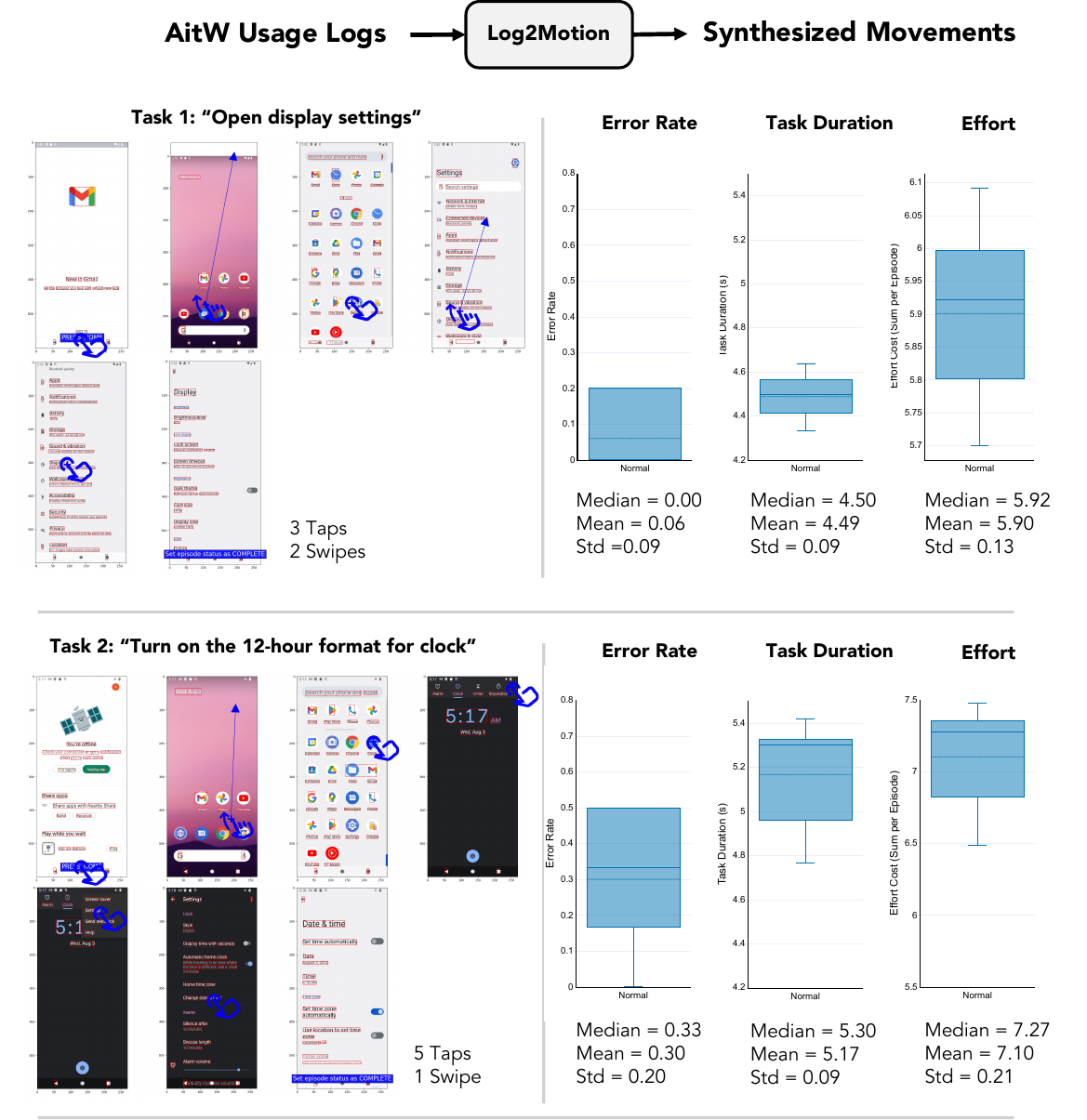}
    \caption{We take two example sequences from the Android-in-the-Wild dataset~\cite{rawles_androidinthewild_2023} (left panel) and use \name to synthesize plausible user movements. Based on 200 simulations, we derive predictions for error rates, task duration, and the effort required to perform these tasks on a smartphone.}
    \label{fig:AndroidReEnactment}
    \Description{The figure shows the results of biomechanical motion synthesis of two tasks from the "Android-in-the-Wild" dataset using the Log2Motion. The left panel shows the user's interaction flow for two tasks: "Open display settings" and "Turn on the 12-hour format for clock." The right panel presents box plots showing the predicted error rate, task duration, and effort for each task, based on 200 simulations. For the "Open display settings" task, the median error rate is 0.00, the median duration is 4.50 seconds, and the median effort is 5.92. For the "Turn on the 12-hour format for clock" task, the median error rate is 0.33, the median duration is 5.30 seconds, and the median effort is 7.27.}
\end{figure*}
\section{Demonstrations}

We demonstrate how \name can augment existing datasets with new insights and how it can be adapted to account for different interaction postures.

\subsection{Synthesizing Interaction Movements from Large Scale Logs}

\name can be applied to real usage logs to obtain rich and plausible synthesized movements that users might have carried out to perform the logged interactions. 
To illustrate this, we run it on cases from the Android-in-the-Wild dataset \cite{rawles_androidinthewild_2023}. 
This dataset was developed for mobile device control, where systems learn to interpret natural language instructions and execute them by directly manipulating the user interface. The dataset contains human demonstrations of device interactions performed on a desktop device using an Android Emulator. 

\autoref{fig:AndroidReEnactment} presents performance and physiological metrics derived from synthesized movements based on tasks from the dataset~\footnote{Videos of these synthesized movements are available in the supplemental materials of this paper.}. The motion synthesis is based solely on the usage logs recorded during user demonstrations (see \autoref{fig:teaser} for an example log structure). For both tasks, we simulated 200 motion sequences and derived key metrics such as task duration, error rates, and interaction effort.

What can we learn from these synthesized movements, and how do the two tasks compare?
Most importantly, as shown in the right panel of \autoref{fig:AndroidReEnactment}, \name produces distributions of key performance metrics not contained in the usage logs themselves. 
Comparing the two tasks, our simulations reveal systematic differences in error rates, task duration, and effort.  For Task~1 (``Open display settings''), which involves three taps and two swipes, the mean error rate is 6\,\%. In contrast, Task~2 (``Turn on the 12-hour format for clock'') shows a mean error rate of 30\,\%. Analyzing the generated motion reveals three reasons for this difference. First, Task~2 requires interactions that are spatially farther apart, leading to larger movements and therefore reduced tapping accuracy. Second, three of the six interactions are very close to the border of the screen and small in target size. Third, swipes are less sensitive to inaccuracies as success is less dependent on precise targeting; the higher share of swipes in Task~1 thus lowers its error rate.
We also observe higher task duration and effort for Task 2, partly due to the additional interaction needed to finalize Task 2, but also due to the increased number of errors that need correction. Notably, both measures show higher variance. In line with our findings in \autoref{subsec:speed_accuracy} and established results from Fitts’ Law~\cite{bi_ffitts_2013, kvalseth_distribution_1976}, we attribute this variability to the greater distances between individual interactions in Task~2.

To sum up, the key value of \name lies in generating insights that go beyond what usage logs alone can provide. At scale, such analyses make it possible to identify subtle ergonomic issues, estimate places where errors are more likely, compare alternative designs, and anticipate performance bottlenecks that would remain invisible in log-based studies alone.

\subsection{Changing the Task Environment to Adopt New Postures}\label{ch:postures}

\begin{figure*}[]
    \centering
        \includegraphics[width=0.9\linewidth, trim=0 0 0 0, clip]{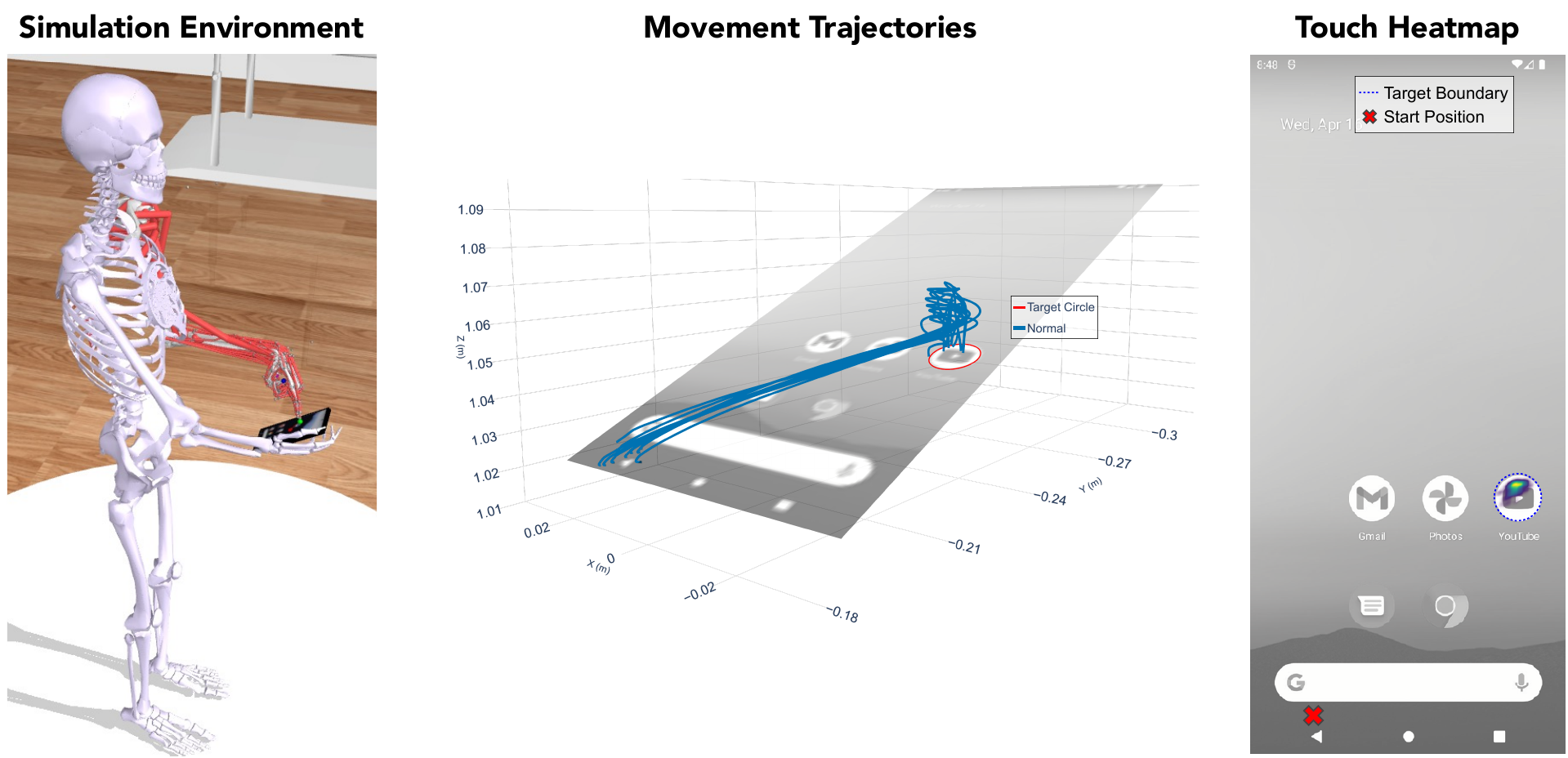}
    \caption{Left: Screenshot of the simulation environment showing the agent interacting with a slightly tilted smartphone held in the opposite hand. Middle: Movement trajectories generated for the \textit{normal} tapping operator. Right: Heatmap illustrating the spatial distribution of touch interactions. }
    \label{fig:tilted_posture}
    \Description{The figure contains three panels. The rightmost panel shows a screenshot from MuJoCo in which a standing skeleton interacts with a handheld smartphone. The middle panel shows simulated smartphone-touch trajectories in 3D. The rightmost panel shows a heatmap indicating where on the screen the interaction occurred.}

\end{figure*}

In the baseline configuration \name assumes a setting in which the mobile device lies flat on a table. While this setup offers experimental control, real-world usage often involves holding the device in one hand while interacting with the other, resulting in a tilted surface and a different posture.
We demonstrate how \name can be adapted to such alternative postures.
First, we adjust the task environment by modifying the MuJoCo XML reference to tilt and reposition the phone, reflecting the orientation of a handheld device (see left panel of \autoref{fig:tilted_posture}). In alignment with prior work~\cite{jonghun_baek_posture_2010}, we tilt the phone by $33^\circ$.
Second, we retrain the model in this modified task environment using the previously introduced training pipeline, which includes the same reward design as described in \autoref{sec:training}. We initialize the model from an existing checkpoint of the \textit{normal} operator. Since the new device position remains in the state space already explored, the existing operators largely generalize to the new orientation of the mobile device. Accordingly, only minimal fine-tuning is necessary to adapt the existing operator to the new device position.
The middle panel of \autoref{fig:tilted_posture} shows ten randomly sampled movement trajectories, while the right panel depicts a heatmap of fingertip positions during touchscreen contact. Compared to the flat-on-table setup, the handheld posture produces less pronounced movement arcs, similar pointing accuracy (0.5\% vs.\ 1.5\% error rate for the 10\,mm diameter button, 
and 13.5\% vs.\ 12.5\% for the 4\,mm diameter), and substantially reduced muscle effort ($\approx 0.31 \text{ vs.} \approx 0.65$), suggesting that the handheld orientation might be less fatiguing.
\section{Discussion}
This work has taken a step forward in applying biomechanical simulation to HCI problems. 
The key takeaway is that biomechanical motion synthesis offers a powerful lens for interpreting the interactions behind logs beyond what traditional data analysis methods can reveal. It also offers a new way for evaluation.
Specifically, it can illuminate core ergonomic and performance aspects of input interactions, including motion trajectories, effort, and the speed-accuracy tradeoff. Our motion-capture comparison and DTW analysis demonstrate that \name produces human-like trajectories for pointing tasks, with minor deviations for swipes.
Our demonstrations show how \name can augment existing datasets and be adapted to new postures, highlighting its practical relevance.
These findings motivate two broader points. First, biomechanical motion synthesis is shaped by assumptions (i.e., modeling choices) that define the physiological and environmental context in which movements emerge. Second, having control over these assumptions unlocks a spectrum of downstream applications.

\subsection{The Importance of Assumptions in Biomechanical Motion Synthesis}

Biomechanical motion synthesis critically relies on assumptions that guide the synthesis process. They act as constraints and priors on user physiology and the physical task environment. Without them, generating meaningful movements is fundamentally impossible. Our goal, therefore, is not to eliminate assumptions but to choose them carefully and ground them in validated models.

To this end, \name builds on physiological priors derived from a well-established biomechanical model~\cite{saul_benchmarking_2015, bachynskyi_is_2014}. By modeling key aspects of human motor behavior like kinematics, joint limits, and motor noise, we demonstrate that \name achieves high movement plausibility. Likewise, integrating a software emulator into the physics simulator enables interaction with real applications, bridging a longstanding gap between biomechanical modeling and interface evaluation without requiring handcrafted task environments.

A central design decision in \name is the use of forward simulation rather than inverse modeling. Instead of defining a discrepancy function and optimizing model parameters to match recorded trajectories, \name directly learns a policy that generates plausible behavior in response to logged events. This makes the synthesis process scalable and efficient, but it also means that the model assumes a particular user posture and physiology.
When logs come from real-world usage, synthesis accuracy depends on how well these assumptions match reality. In \autoref{ch:postures}, we demonstrate how \name adapts to different postures by manipulating the task environment. Many assumptions could also be inferred directly from device sensors; for example, IMU data could help recover device tilt~\cite{jonghun_baek_posture_2010} or approximate user posture~\cite{goel_gripsense_2012, hinckley_pre-touch_2016}. Embedded into logs, this information would increase synthesis accuracy.
Conversely, when logs come from expert demonstrations or design tools, explicitly specifying assumptions becomes a tool. In this case, designers and researchers can specify posture, physiology, or environmental conditions to explore novel scenarios.

Accordingly, assumptions are not limitations but an essential mechanism for steering and contextualizing motion synthesis; they can be specified for targeted exploration or inferred from real-world data to improve accuracy.


\subsection{Downstream Applications of \name}

While the main application of \name is augmenting log data, having control over assumptions may enable powerful future applications. By modifying the biomechanical model or task environment, \name can be tailored to simulate a wide range of user conditions and capabilities. 
In the future, this could enable automated accessibility and ergonomic evaluation early in the design process, long before user studies are feasible.
Such applications may become possible because \name can model core principles underlying motor impairments.
These include oscillating muscle forces as observed in tremor patients~\cite{deuschl_pathophysiology_2001}, a decrease in muscle strength as in sarcopenia~\cite{evans_what_1995}, or altered kinematics arising from prosthesis use~\cite{caggiano_myochallenge_2024}. By adjusting these parameters, designers could simulate how users with different motor conditions would perform interaction tasks.
In addition, situational impairments, such as walking-induced motion artifacts~\cite{bergstrom-lehtovirta_effects_2011}, can be introduced through external forces in the physics simulator.
This highly customizable task environment also allows designers and researchers to adjust device form factors, such as display size, and evaluate how these changes affect interaction.
Going forward, \name’s interoperability and its extensible biomechanical model and task environment may enable a variety of future applications built on top of \name.

\subsection{Limitations and Future Work}

While \name demonstrates the feasibility and value of biomechanical forward simulation for synthesizing interaction movements, it also opens up several promising directions for future work. 

Currently, \name simulates single-handed, index-finger interactions under the assumption that the mobile device rests on a table. Future extensions could significantly broaden the range of supported behaviors. Adding a second arm would enable modeling of two-handed and bi-manual interactions, such as tablet usage or thumb-typing. Similarly, enabling thumb-based input would require the model to learn how to grasp and stabilize a handheld phone, which has clear implications for one-handed use and accessibility evaluation. Recent work on dexterous manipulation~\cite{berg_sar_2023} suggests that such extensions are well within reach. Our modular architecture makes it easy to add new motor operators. This process involves defining a new POMDP, specifying reward components, training the operators, and evaluating the resulting movements. Each step follows the workflow described in \autoref{sec:policylearning}, ensuring extensions remain modular and reusable.
Furthermore, if the logs do not provide information on user posture, \name assumes a posture for synthesizing user movements. Accordingly, even better results could be achieved if the user’s posture were known or inferred~\cite{bachynskyi2015performance}. Future work could explore parameter inference techniques for user simulators to estimate posture directly from touch events and timestamps \cite{kangasraasio2019parameter}.

Another opportunity is enriching \name with models of cognition and perception. Adding vision as demonstrated in \cite{ikkala_breathing_2022} would allow to model how visual input influences touch behavior and eye–hand coordination~\cite{holz_understanding_2011}.

An example from our own comparison with motion-captured user data highlights this potential: whereas the current model hovers centrally above the screen between actions, the human participant consistently hovered slightly off-center, likely to maintain an unobstructed view of the display. Integrating vision would allow \name to capture this mutual influence between sensor and motor policies, enabling more human-like behavior.
Similarly, we see strong potential for integrating \name with cognitive computational models~\cite{fleig_mind_2025}. For instance, it could augment recent models of typing~\cite{shi_crtypist_2024, shi_simulating_2025}, which currently rely on simplified representations of motor control.

Finally, training the motor operators offers room for methodological improvement. While our control policies capture Fitts’ Law, realistic velocity profiles, and human-like movements, jointly optimizing these objectives remains challenging. This suggests opportunities for multi-objective optimization. More robust training strategies, potentially informed by automated RL~\cite{parker-holder_automated_2022}, could yield policies that satisfy multiple behavioral and biomechanical constraints without manual inspection or fine-tuning.

Ultimately, we hope \name becomes a key enabler for making biomechanical models relevant to real-world HCI problems, transforming touch log analysis into scalable evaluations of ergonomics and interaction.

\subsection{Open Science Statement}
To support the open science movement in HCI~\cite{salehzadeh_niksirat_changes_2023, ebel_changing_2024}, all source code with extensive documentation is available at: \url{https://github.com/ciao-group/Log2Motion}.

\begin{acks}
We want to thank the Computational Rationality Winter Retreat, where this idea was born, Arthur Fleig, who supported us with the data collection, and Anna Maria Feit and Mark Colley for their valuable suggestions and comments.

The authors acknowledge the financial support by the Federal Ministry of Research, Technology and Space of Germany and by Sächsische Staatsministerium für Wissenschaft, Kultur und Tourismus in the programme Center of Excellence for AI-research „Center for Scalable Data Analytics and Artificial Intelligence Dresden/Leipzig“, project identification number: ScaDS.AI.
Additionally, this work was supported by the Research Council of Finland (FCAI, grant numbers 328400, 345604, 341763; Subjective Functions, grant number 357578) and the ERC Advanced Grant (grant number 101141916).
\end{acks}
\clearpage


\balance
\bibliographystyle{ACM-Reference-Format}
\bibliography{references.bib}

\clearpage
\appendix
\begin{figure*}[t] 
    \section{Appendix}\label{ch:Appendix}
    \subsection{User Study Results Task 2}

    \centering
    \begin{subfigure}[b]{0.22\textwidth}
        \centering
        \includegraphics[width=\linewidth]{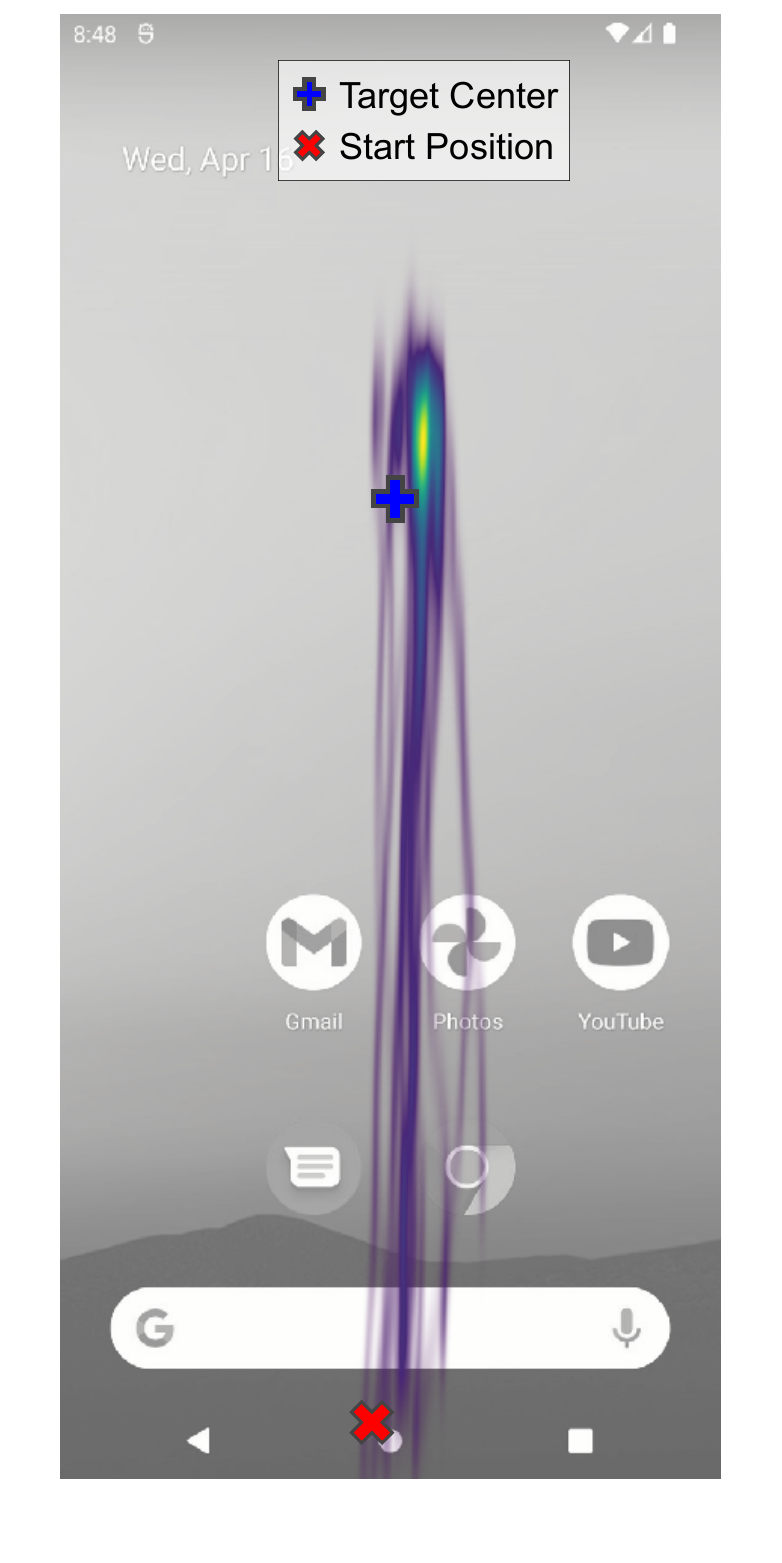}
        \caption{P1}
    \end{subfigure}
    \begin{subfigure}[b]{0.22\textwidth}
        \centering
        \includegraphics[width=\linewidth]{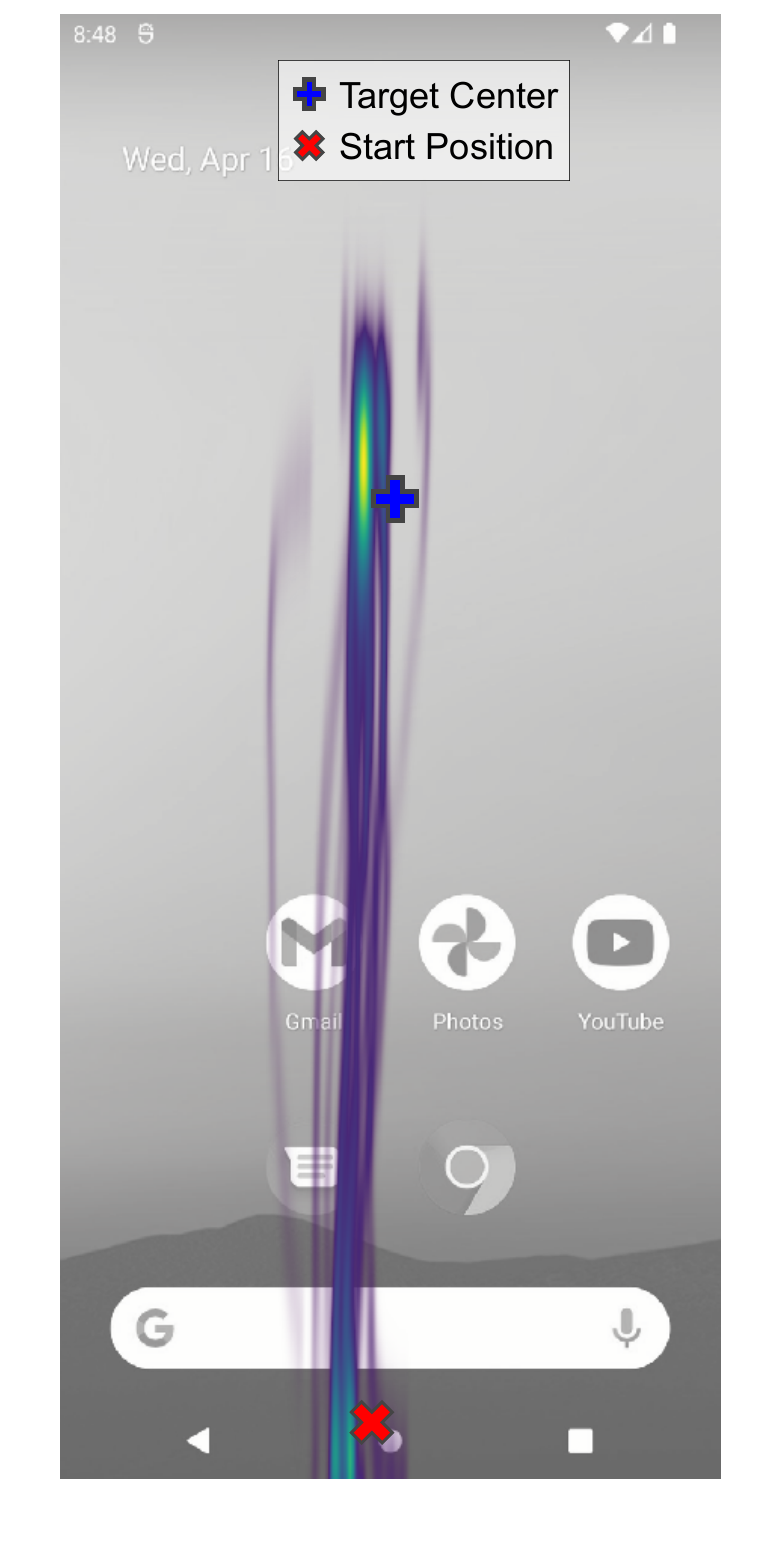}
        \caption{P2}
    \end{subfigure}
    \begin{subfigure}[b]{0.22\textwidth}
        \centering
        \includegraphics[width=\linewidth]{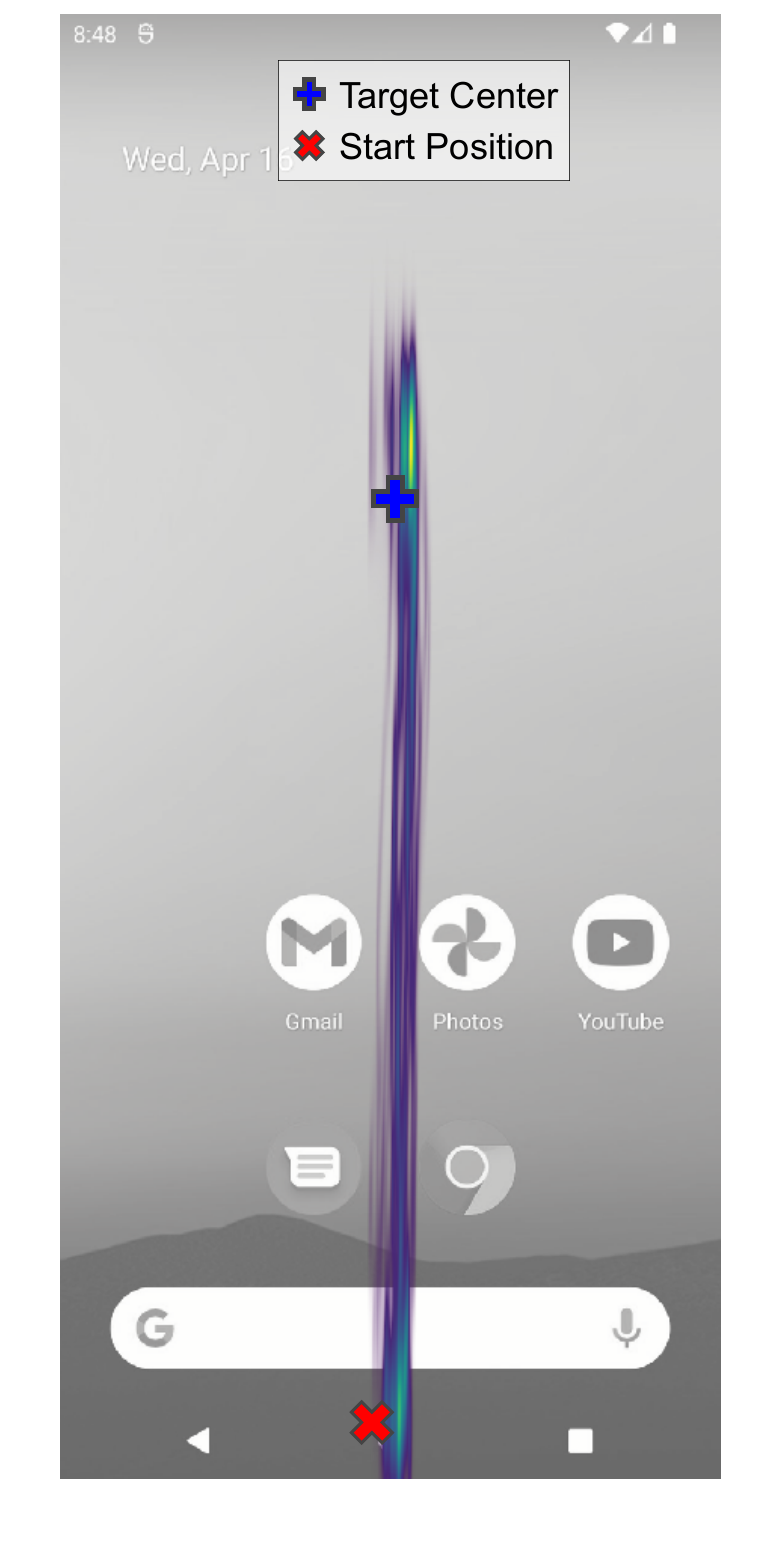}
        \caption{P3}
    \end{subfigure}
        \begin{subfigure}[b]{0.22\textwidth}
        \centering
        \includegraphics[width=\linewidth]{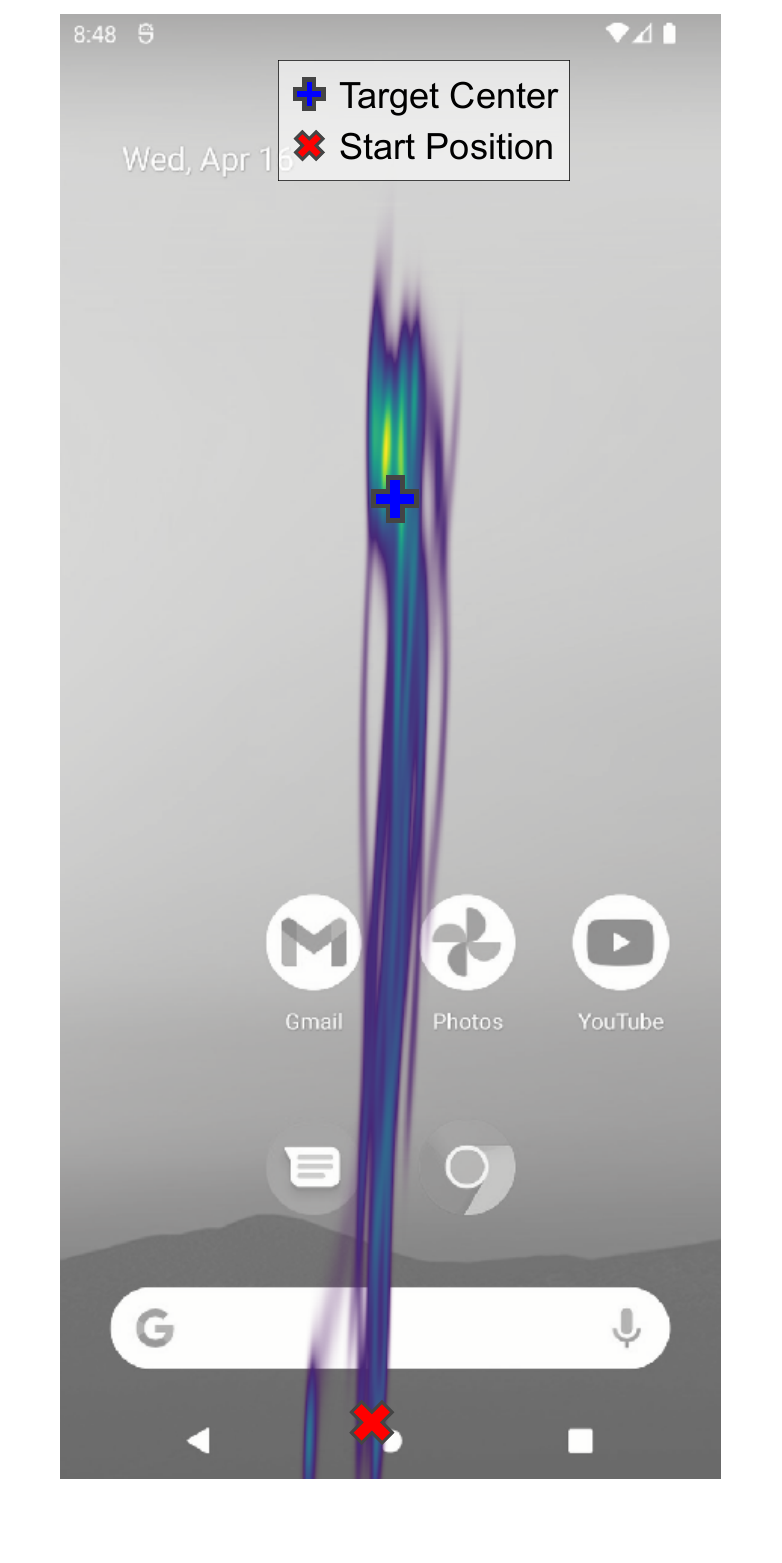}
        \caption{P4}
    \end{subfigure}


    \begin{subfigure}[b]{0.22\textwidth}
        \centering
        \includegraphics[width=\linewidth]{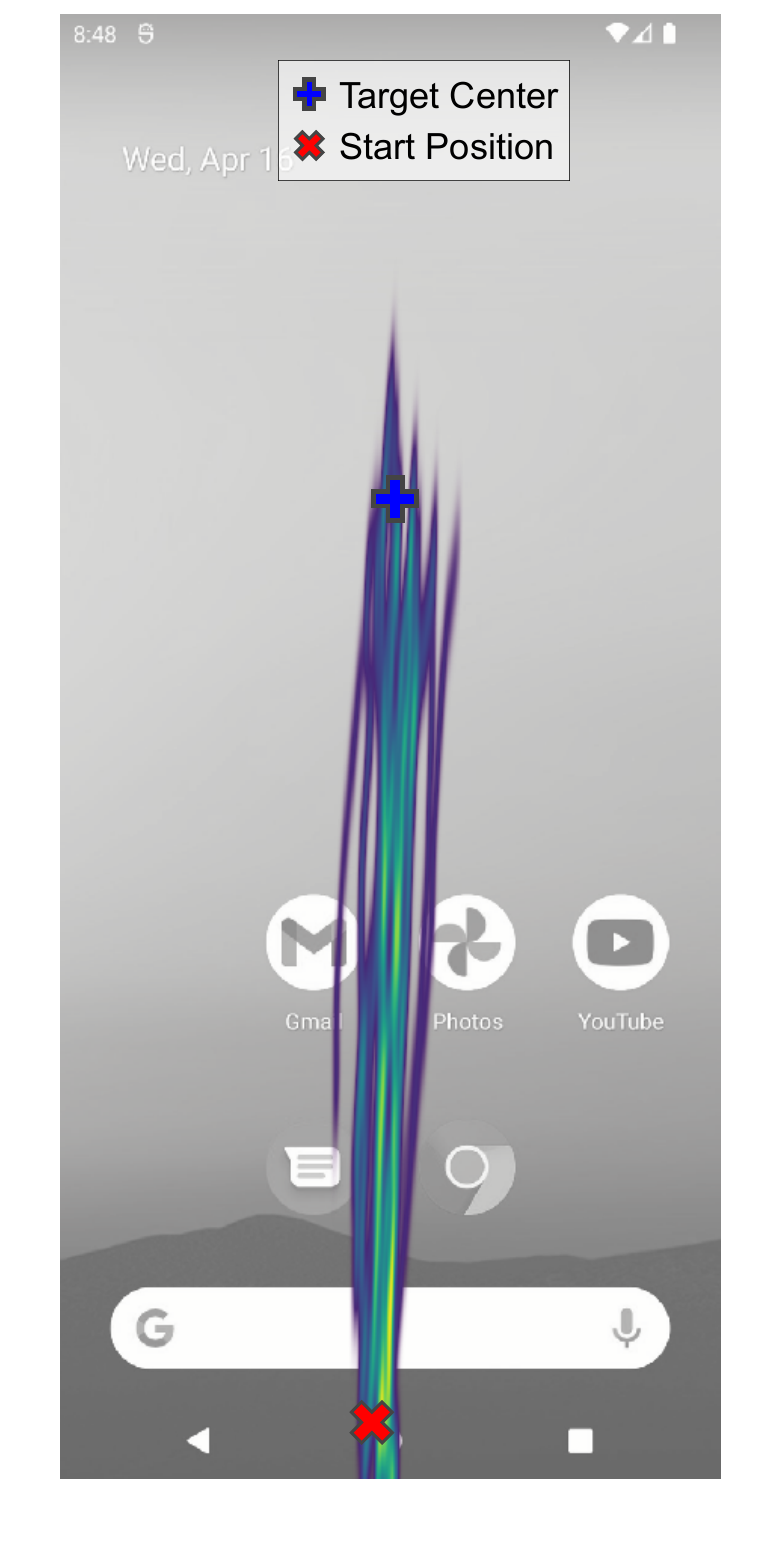}
        \caption{P5}
    \end{subfigure}
    \begin{subfigure}[b]{0.22\textwidth}
        \centering
        \includegraphics[width=\linewidth]{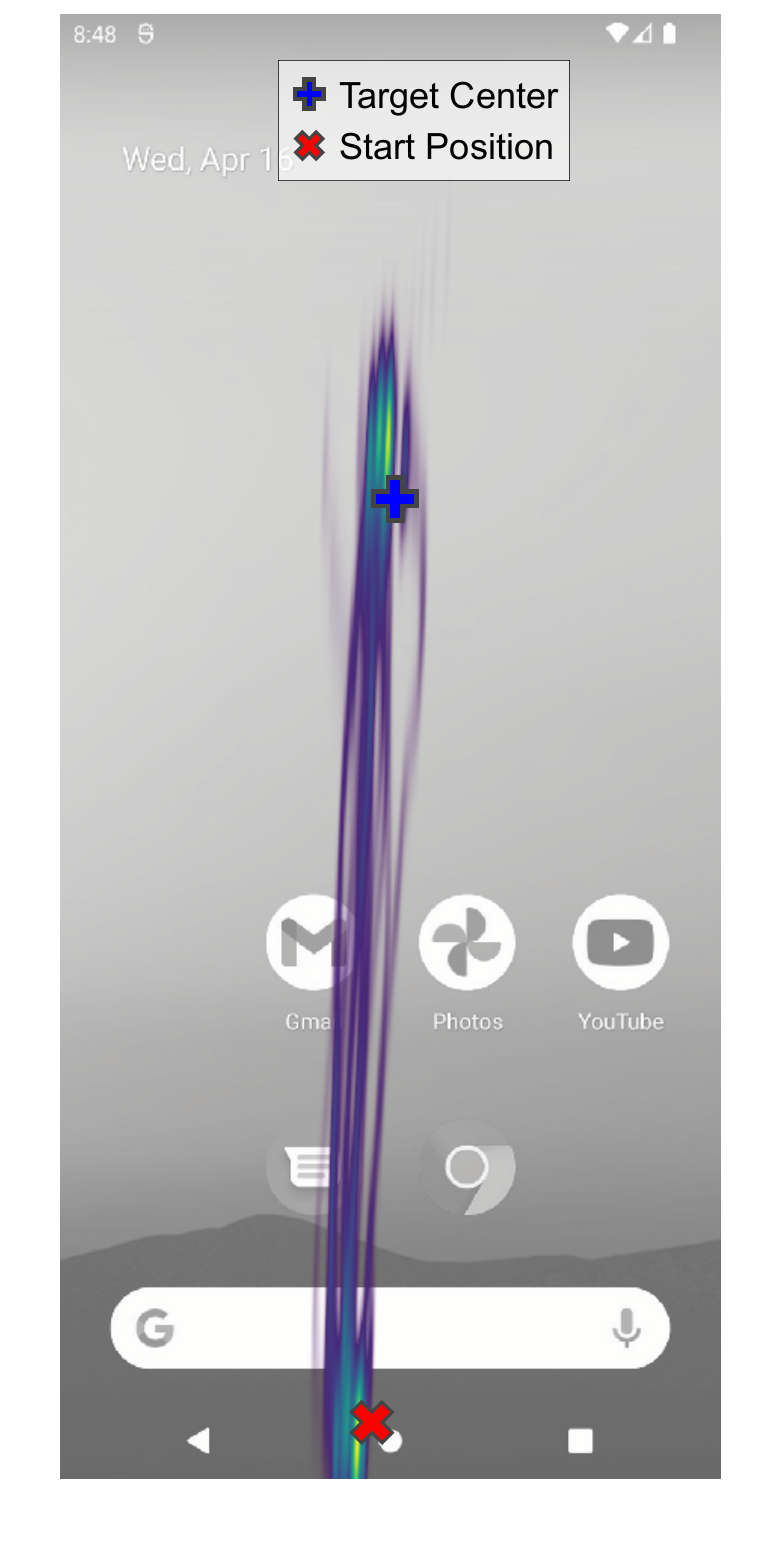}
        \caption{P6}
    \end{subfigure}
        \begin{subfigure}[b]{0.22\textwidth}
        \centering
        \includegraphics[width=\linewidth]{plots/swipe/endpoints_swipe.pdf}
        \caption{\name.}
    \end{subfigure}
    \caption{
        Heatmaps of fingertip positions during the swiping task.  
        The subfigures show individual heatmaps for Users~1--6, capturing the variability in swipe execution across participants.
    }
    \label{fig:end_point_users_variation}
    \Description{
        The figure shows heatmaps of fingertip positions on a smartphone screen during swiping interactions.
        The panels show the individual heatmaps for Participants 1–6.
    }

\end{figure*}

\begin{figure*}[]
    \subsection{User Study Results Task 3}

    \centering
    \begin{subfigure}[b]{0.40\linewidth} 
        \centering
        \includegraphics[width=\linewidth, trim=10pt 10pt 330.0pt 330.0pt, clip]{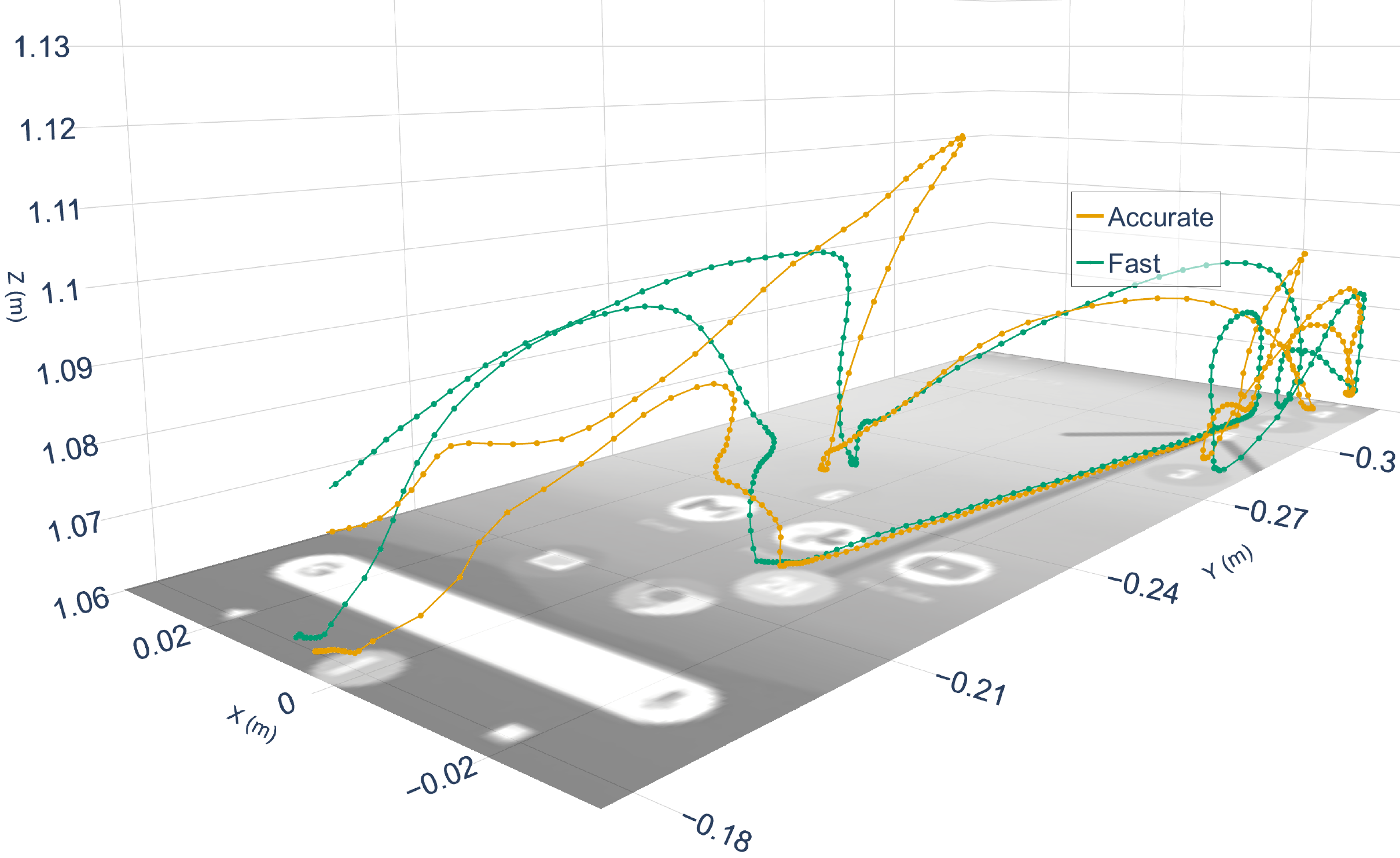}
        \caption{Movement trajectories of P1}
        \label{fig:sub:task2user1}
    \end{subfigure}
    \hfill
    \begin{subfigure}[b]{0.40\linewidth}
        \centering
        \includegraphics[width=\linewidth, trim=10pt 10pt 330.0pt 330.0pt, clip]{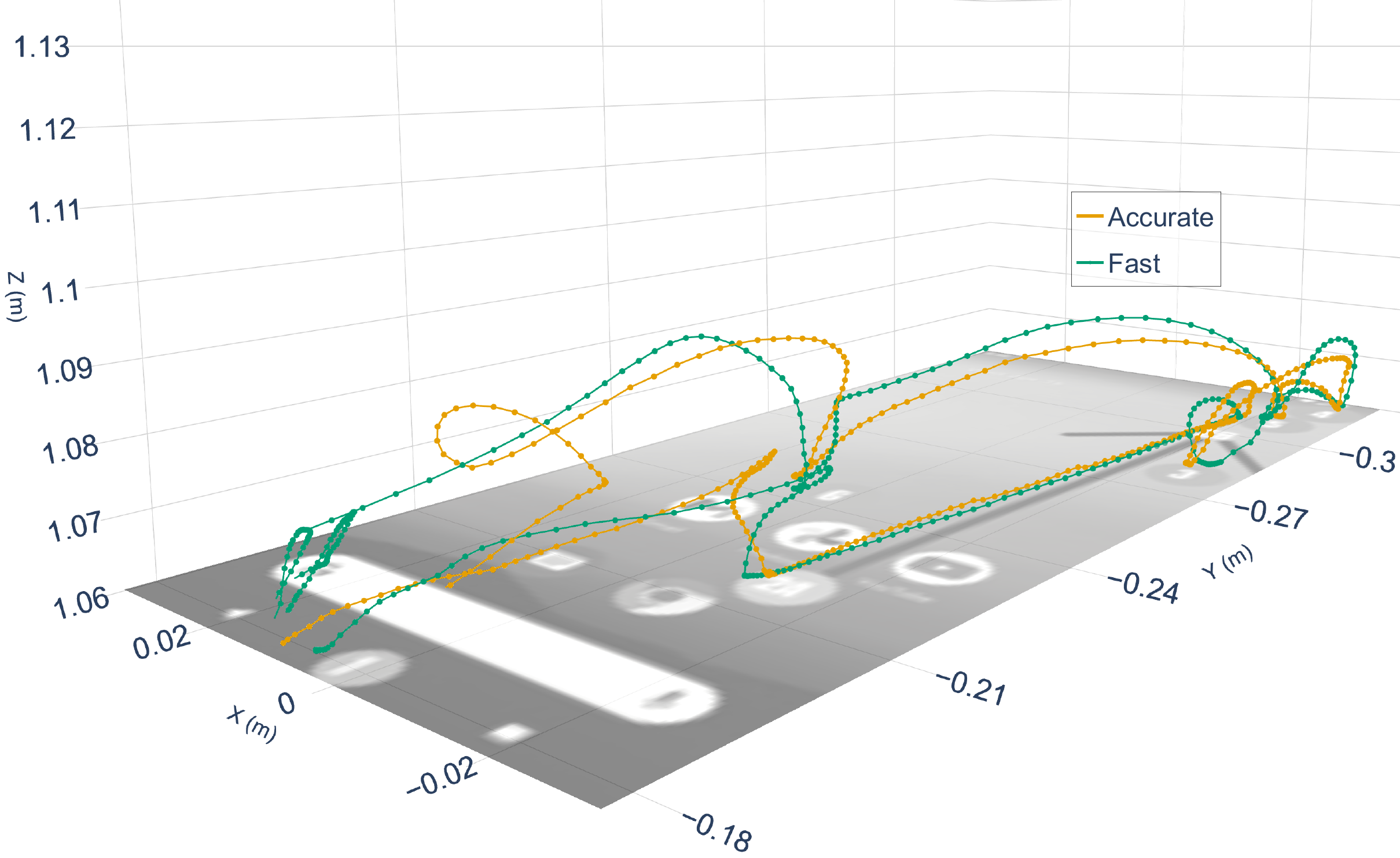}
        \caption{Movement trajectories of P2}
        \label{fig:sub:task2user2}
    \end{subfigure}


    \begin{subfigure}[b]{0.40\linewidth}
        \centering
        \includegraphics[width=\linewidth, trim=10pt 10pt 330.0pt 330.0pt, clip]{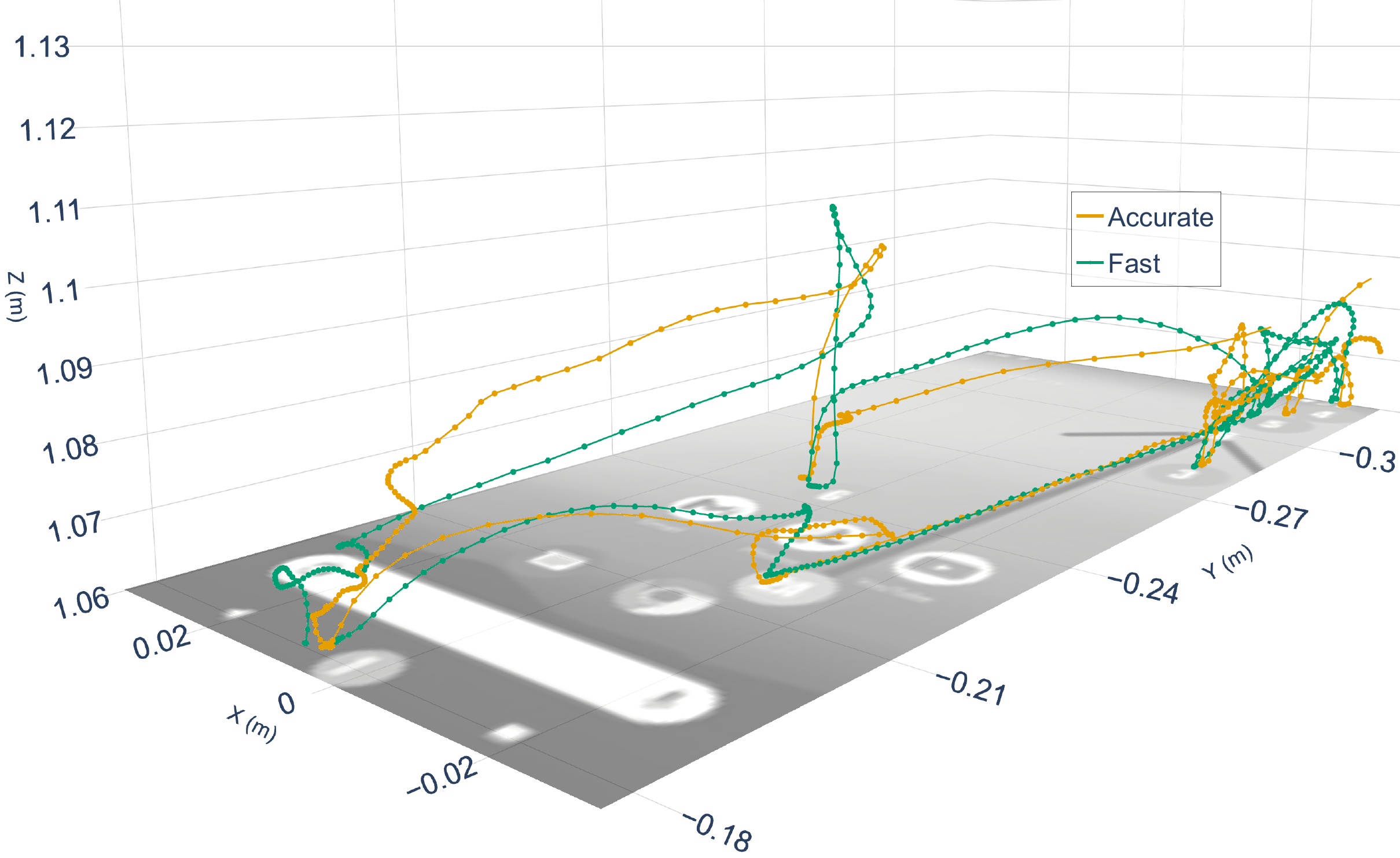}
        \caption{Movement trajectories of P3}
        \label{fig:sub:task2user3}
    \end{subfigure}
    \hfill
    \begin{subfigure}[b]{0.40\linewidth}
        \centering
        \includegraphics[width=\linewidth, trim=10pt 10pt 330.0pt 330.0pt, clip]{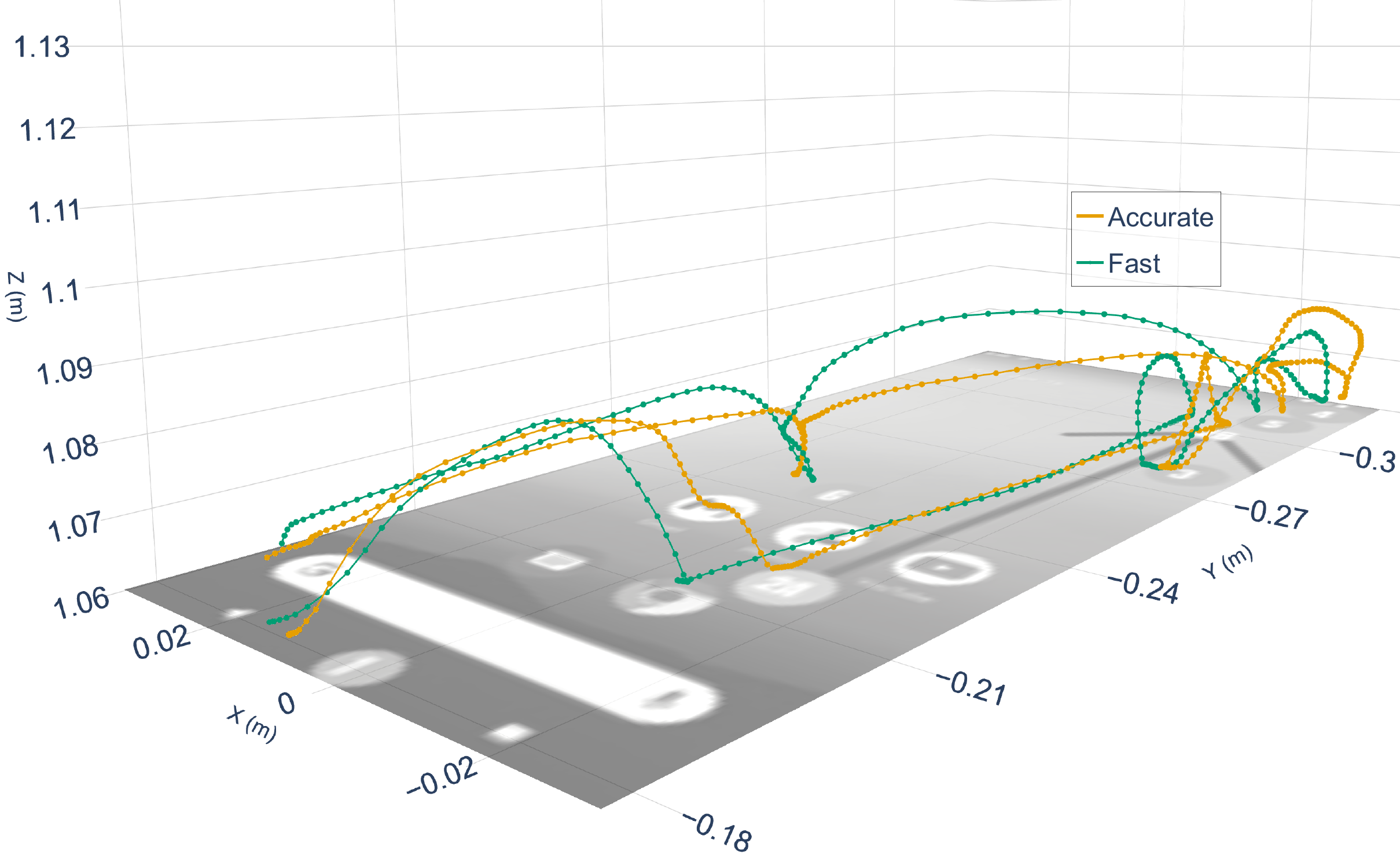}
        \caption{Movement trajectories of P4}
        \label{fig:sub:task2user4}
    \end{subfigure}


    \begin{subfigure}[b]{0.40\linewidth}
        \centering
        \includegraphics[width=\linewidth, trim=10pt 10pt 330.0pt 330.0pt, clip]{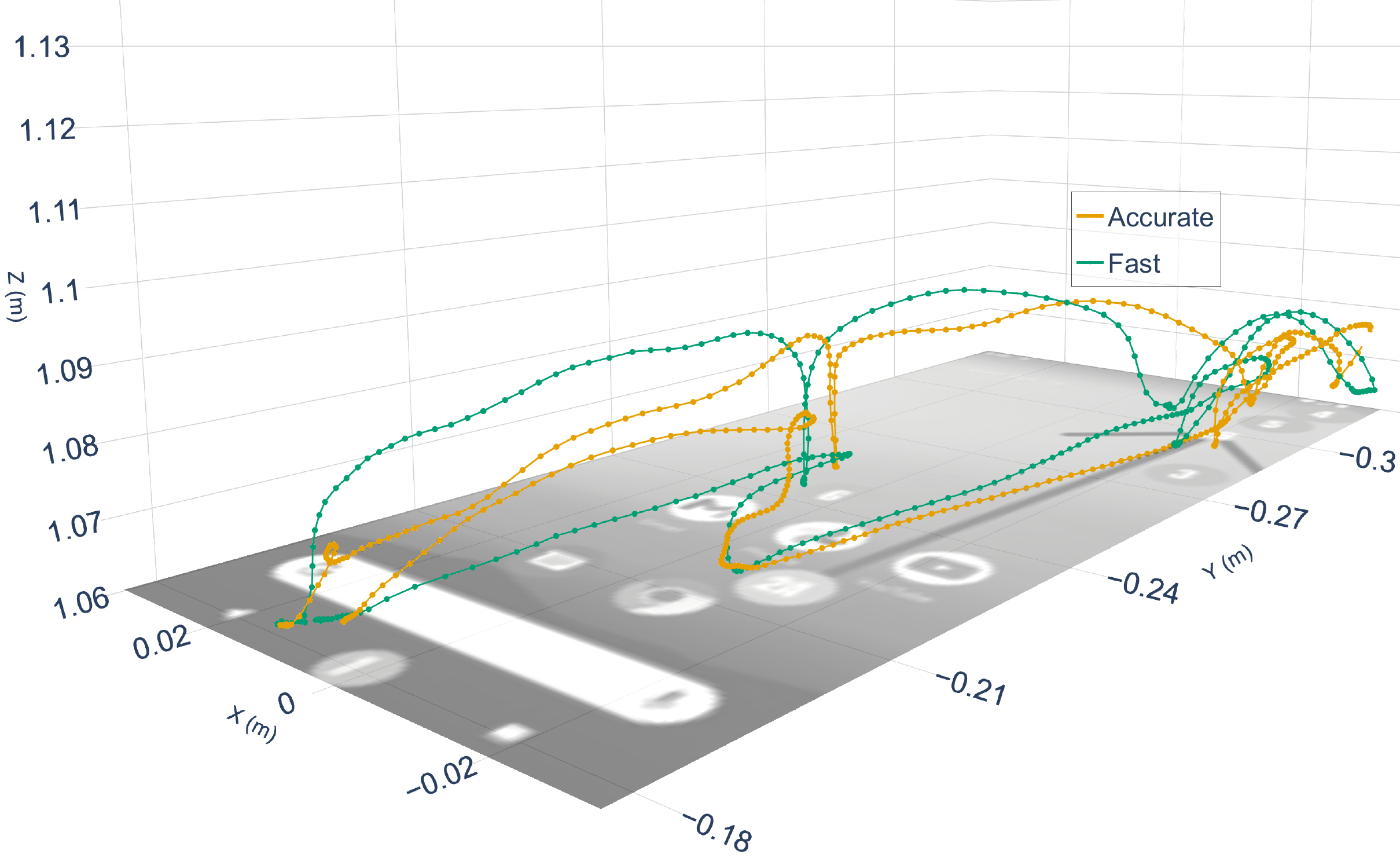}
        \caption{Movement trajectories of P5}
        \label{fig:sub:task2user5}
    \end{subfigure}
    \hfill
    \begin{subfigure}[b]{0.40\linewidth}
        \centering
        \includegraphics[width=\linewidth, trim=10pt 10pt 330.0pt 330.0pt, clip]{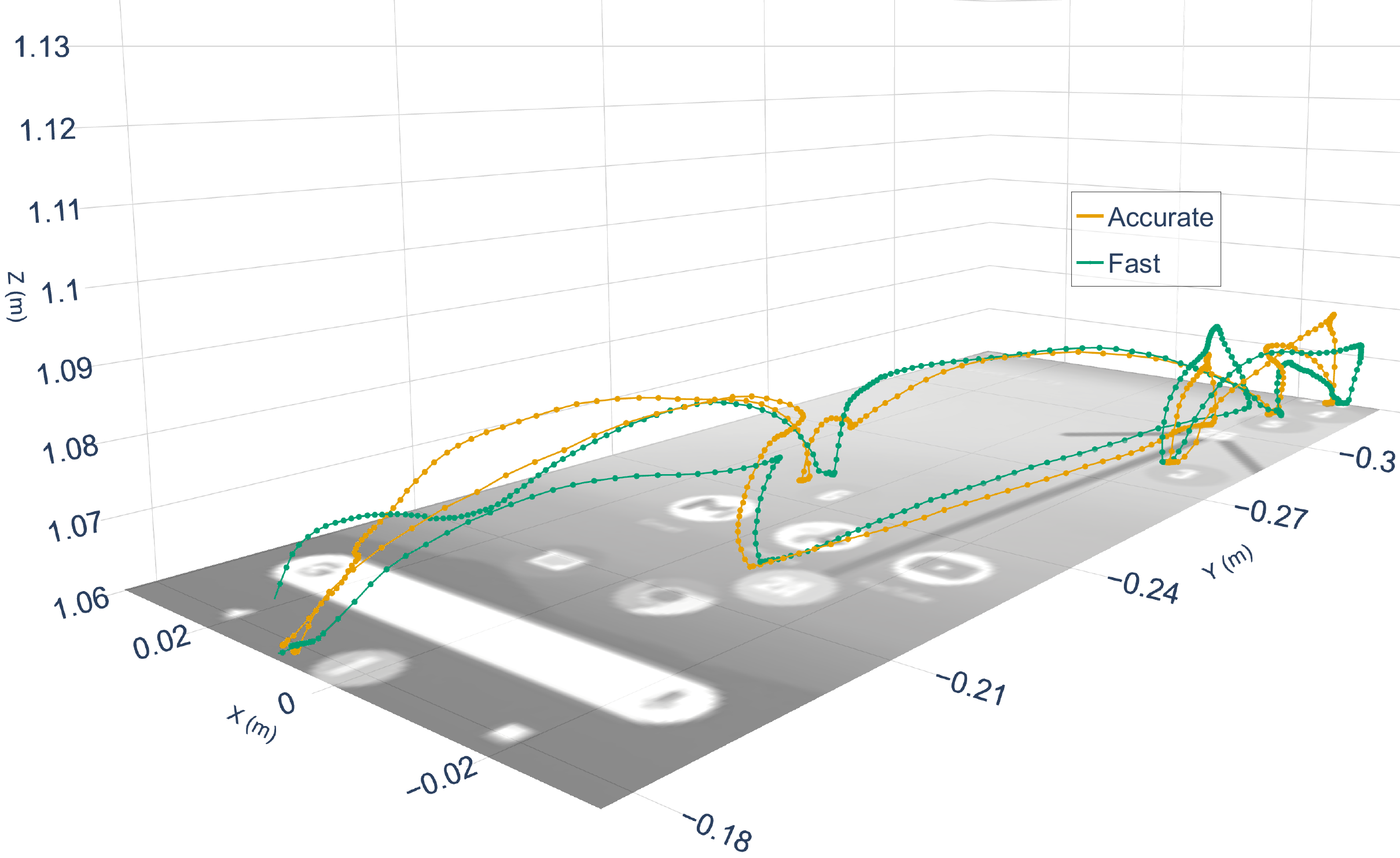}
        \caption{Movement trajectories of P6}
        \label{fig:sub:task2user6}
    \end{subfigure}

    \begin{subfigure}[b]{0.40\linewidth}
        \centering
        \includegraphics[width=\linewidth, trim=10pt 10pt 330.0pt 330.0pt, clip]{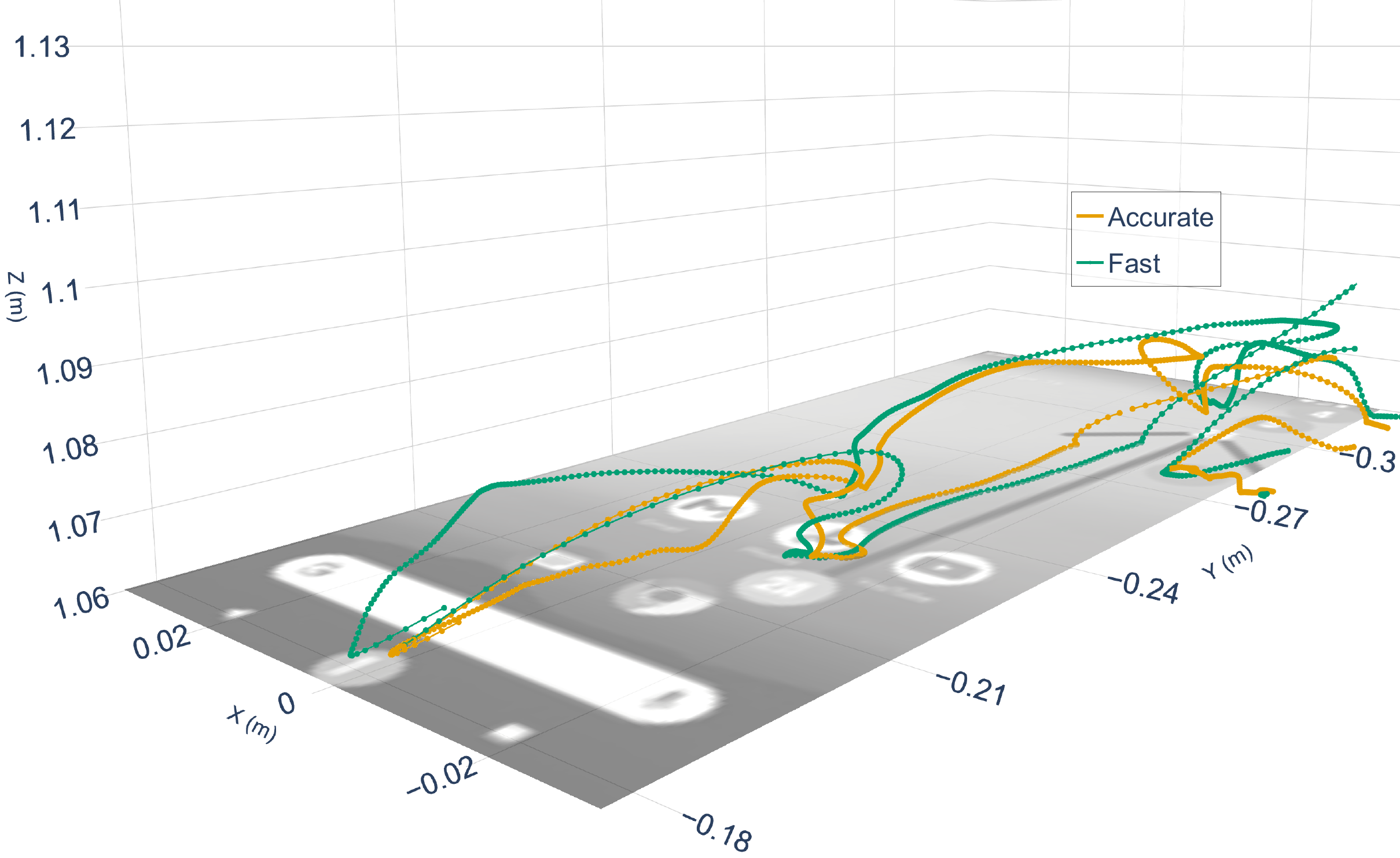}
        \caption{Movement trajectories of \name}
        \label{fig:sub:task2log2motion}
    \end{subfigure}
    \caption{Motion gallery comparing participants' movement trajectories (\autoref{fig:sub:task2user1}-\autoref{fig:sub:task2user6}) and synthesized trajectories generated with \name (\autoref{fig:sub:task2log2motion}) for the second AitW task as shown in \autoref{fig:AndroidReEnactment}.}
    \Description{The figure contains eight panels comparing human (a-f) and simulated (g) movement trajectories for the second task of the AitW demonstration}
    \label{fig:task2all_trajectories_grid_3x2}
\end{figure*}

\end{document}